\documentclass[reprint,amsmath,amssymb]{revtex4-2}

\usepackage{graphicx}
\usepackage{bm}
\usepackage{hyperref}
\usepackage{color}
\usepackage{siunitx}
\usepackage{mathrsfs}
\usepackage[explicit]{titlesec}

\usepackage[mathlines,pagewise]{lineno}
\modulolinenumbers[1]

\makeatletter
\newcommand*\bigcdot{\mathpalette\bigcdot@{.7}}
\newcommand*\bigcdot@[2]{%
	\mathbin{\vcenter{\hbox{\scalebox{#2}{$\m@th#1\bullet$}}}}%
}
\makeatother

\titlespacing*{\section}
{0pt}
{4.0ex plus .8ex minus 0.5ex}
{1.6ex plus .0ex}

\titlespacing*{\subsection}
{0pt}
{3.5ex plus .0ex minus .0ex}
{2.3ex plus .0ex}

\hypersetup{
	hypertex=true,
	colorlinks=true,
	linkcolor=blue,
	anchorcolor=blue,
	citecolor=blue
}

\usepackage{enumitem}
\usepackage{fancyhdr}
\usepackage{booktabs}
\usepackage{multirow}
\usepackage{longtable}
\usepackage{float}

\begin{document}
	
	\preprint{APS/123-QED}
	
	\title{Spin-Axis Dynamic Locking}
	
	\author{Zhiheng Lv$^{1}$}
	\thanks{These authors contributed equally to this work}
	
	\author{Jiangtao Cai$^{2}$}
	\thanks{These authors contributed equally to this work}
	
	\author{Dengpan Ma$^{1}$}
	
	\author{Yan Xing\textsuperscript{1}}
	\email{Corresponding author: xingy@imu.edu.cn}
	
	\author{Zhifeng Liu\textsuperscript{1}}
	\email{Corresponding author: zfliu@imu.edu.cn}
	
	\affiliation{
		$^1$Inner Mongolia Key Laboratory of Microscale Physics and Atomic Manufacturing
		$\&$ Research Center for Quantum Physics and Technologies,
		School of Physical Science and Technology,
		Inner Mongolia University, Hohhot 010021, China\\
		$^2$School of Physics and Information Science,
		Shaanxi University of Science and Technology,
		Xi'an 710021, China
	}
	
	\date{\today}
	
	\begin{abstract}
		The all-electrical realization of highly spin-polarized currents and their
		efficient conversion into pure spin currents remains a fundamental challenge
		in spintronics. Here, we report a \emph{spin-axis dynamic locking} (SADL)
		effect in altermagnets that pins the high and dynamically robust spin
		polarization to the crystalline axes: an in-plane electric field along one
		principal axis drives a highly spin-up-polarized current, whereas along the
		orthogonal axis, it generates a symmetry-enforced, equal-magnitude spin-down
		current. Consequently, applying an electric field diagonally yields a
		transverse pure spin current, reaching $100\%$ charge-to-spin conversion in
		the ideal limit. Mechanistically, SADL originates from a spin-split
		\emph{tent-state} band structure whose Lifshitz transitions delimit an
		open-Fermi-line regime. The momentum-separated Fermi lines carry orthogonal
		nonzero winding vectors, producing a pronounced velocity contrast while
		suppressing ordinary backscattering to dynamically stabilize the axial spin
		selectivity. High-throughput first-principles screening confirms SADL in
		broad materials. Notably, monolayer Cr$_2$WSe$_4$ and synthesized bulk
		(BaF)$_2$Mn$_2$Se$_2$O exhibit efficiencies close to the ideal limit, paving
		the way for ultra-low-power, reconfigurable spintronic devices where the spin
		states are governed solely by electric field orientation.
	\end{abstract}
	
	\maketitle
	
	\UseRawInputEncoding
	
	\emph{Introduction.}
	Modern microelectronics
	\cite{Razavi2021,li2012semiconductor,Sze2021}
	is built on the manipulation of charge, while its successor, spintronics
	\cite{Baibich1988,Fert2008,Zutic2004}, seeks to harness the spin of Bloch
	electrons as a means of transcending the limitations of charge-based devices. Yet, spintronics faces fundamental challenges \cite{Fert2008,Zutic2004,Dieny2020,Awschalom2007,Bader2010,Inomata2008} in generating and modulating highly spin-polarized currents and in achieving pure spin current with efficient charge-to-spin conversion. Overcoming these hurdles requires the ability to control and differentiate the carrier transport based on spin orientation. As an angular momentum with an associated magnetic moment, spin can be directly manipulated by magnetic fields via the Zeeman effect \cite{Fert2008}. For practical applications, however, this approach is ill-suited for nanoscale integration due to the difficulty in generating and confining strong magnetic fields with high energy efficiency and spatial selectivity \cite{Zutic2004}.
	
	Alternatively, nonmagnetic approaches (e.g., electrical or mechanical fields \cite{Matsukura2015,Fert2024,Song2017,lei2013strain,ma2012evidence}) have emerged that control spin by leveraging its coupling to various intrinsic degrees of freedom in crystals. For instance, the relativistic spin-orbit coupling (SOC), the fundamental mechanism behind various phenomena such as the Dresselhaus effect \cite{Dresselhaus1955}, Rashba effect \cite{Rashba1960,Manchon2015}, and spin Hall effect \cite{Hirsch1999,Sinova2015}, facilitates indirect electric field control of spin by modulating the carrier momentum. Likewise, spin-valley locking (SVL) \cite{Xiao2012} in noncentrosymmetric systems, where SOC plays an essential role, enables non-contact, high-speed spin manipulation through valley-selective excitation with circularly polarized light. Despite these capabilities, the energy scale of the underlying relativistic SOC remains substantially weaker than that of ferromagnetic exchange interactions, leading to low charge-to-spin conversion efficiencies and limited spin splitting around the valleys. Moreover, spin transport in these systems is confined to short diffusion lengths (typically nanoscale). Such fundamental constraints of relativistic approaches call for a paradigm shift toward alternative spin control mechanisms that function at a much stronger energy scale and are inherently non-relativistic. It is within this context that altermagnets (AMs) \cite{Smejkal2022a,Smejkal2022b,Bai2024,Song2025,Gonzalez2021} have recently been brought to the forefront as a promising platform.
	
	AMs are an emerging class of collinear magnets that distinctively combine eV-scale spin splitting---similar to ferromagnets (FMs)---with antiparallel magnetic order resulting in a zero net moment, akin to antiferromagnets (AFMs) \cite{Smejkal2022a,Smejkal2022b}. Their alternating spin-split bands, characterized by magnetic multipole order \cite{hayami2019momentum}, support spin manipulation mechanisms rooted in crystal symmetry and anisotropy \cite{Gonzalez2021,Chen2025,Ma2021,Zhang2024}, rather than relativistic SOC effects. For instance, in 2D AMs, crystal symmetry can directly induce SVL without SOC, a mechanism known as $C$-paired SVL \cite{Ma2021}. This permits spin manipulation via the valleys, with the spin-valley polarization controlled by symmetry breaking (e.g., applied uniaxial strain). Furthermore, valley mediation in multilayer 2D AMs can give rise to spin-layer coupling (SLC) \cite{Zhang2024}, which locks the spin orientation to specific layers. This allows for the electrical control of spin-valley polarization via a vertical gate field. The operation of these spin-control schemes, however, is mediated by valley- or layer-selective channel control. Although such channel selection can in principle yield fully spin-polarized transport, it first requires external perturbations to create the necessary channel imbalance, with practical energy scales typically limited to the meV range under modest experimental conditions. By contrast, the seminal spin-splitter effect (SSE)~\cite{Gonzalez2021,Ma2021} in AMs provides an all-electrical, non-relativistic route to generating spin-polarized currents and pure spin currents without additional symmetry-breaking perturbations. Yet the attainable spin polarization and charge-to-spin conversion efficiency remain strongly material dependent, and a general principle connecting altermagnetic symmetry, Fermi-surface geometry, and transport dynamics to near-ideal polarization and conversion efficiency has remained missing. Consequently, the all-electrical realization of highly spin-polarized currents together with highly efficient pure spin-current generation remains an open challenge.
	
	In this Letter, we propose a distinct spin manipulation mechanism: the \emph{spin-axis dynamic locking} (SADL) effect. Using a 2D anisotropic $d$-wave altermagnetic model, we demonstrate that SADL enables all-electrical generation of highly spin-polarized currents with dynamically robust spin polarization locked to the crystal axes, which can be switched simply by rotating the in-plane electric field. Crucially, a diagonal field orientation directly yields a transverse pure spin current, reaching $100\%$ charge-to-spin conversion in the ideal limit. We trace the origin of SADL to a unique tent-state band structure with Lifshitz-delimited open Fermi lines carrying nontrivial winding and validate this mechanism via first-principles calculations in realistic 2D and 3D materials. Our findings thereby establish the SADL effect as a direct, efficient, and non-relativistic pathway to the electrical control of spin.
	
\begin{figure}[!htbp]
		\includegraphics[width=0.98 \linewidth]{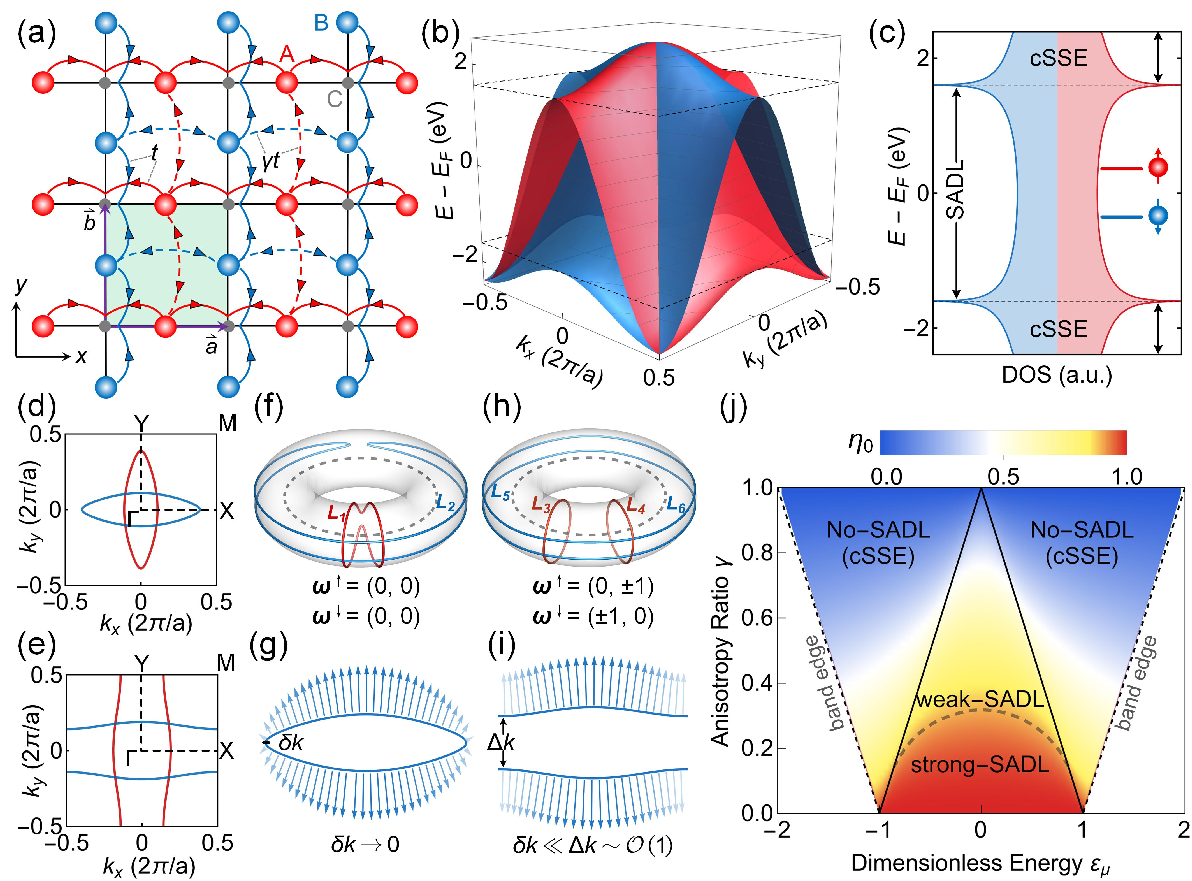}
		\caption{\label{fig:1} Lattice model, electronic structure, and the SADL effect. (a) Schematic of the N\'eel AFM lattice, with the primitive unit cell shaded in green. Arrows indicate nnn hopping pathways. (b) 3D representation of the tent-like band structure with parameters $t = -1$, $\varepsilon_0 = 0$, and $\gamma = 0.2$. (c) Spin-resolved DOS, highlighting the van Hove singularities. (d--i) Fermi contours, torus mappings, and group-velocity distributions for the closed-pocket cSSE regime (d, f, g) and open-line SADL regime (e, h, i). (j) Transport phase diagram of $\eta_0(\varepsilon_\mu,\gamma)$. Solid lines mark Lifshitz boundaries.}
	\end{figure}

	\emph{2D Altermagnetic Model and Tent State.} We introduce a square AM lattice model exhibiting N\'eel AFM order, characterized by two orthogonal crystal axes $x$ \& $y$ ($k_x$ \& $k_y$) and two spin orientations ($\uparrow$ \& $\downarrow$). As illustrated in Fig.~\ref{fig:1}a, the magnetic sites---red (A) and blue (B) spheres---possess local moments aligned along the $+z$ ($\uparrow$) and $-z$ ($\downarrow$) directions, respectively, and are bridged by non-magnetic sites (gray C spheres). In realistic materials, each site may correspond to an atom, a cluster, or a functional group. This arrangement breaks the spin-group symmetries \cite{Liu2022} of $[C_2||\tau]$ and $[C_2||T][C_2||P]$, but preserves those of $[C_2||C_{4z}]$ and $[C_2||M_\varphi]$, thus providing the necessary symmetry foundation for the $d$-wave altermagnetic characteristics \cite{Smejkal2022b,Bai2024}.
	
	A strong N\'eel AFM exchange interaction $J$ between A and B sites sets the magnetic ground state, thereby providing a stable AFM background potential for electron motion. The low-energy electron dynamics within this background are governed by hopping processes. Owing to the dynamical frustration induced by $J$, the nearest-neighbor (nn) hopping $t_1$ between A and B sites, which requires a spin flip, is strongly suppressed. In contrast, the spin-conserving next-nearest-neighbor (nnn) hopping $t_2$ ($A \leftrightarrow A$ or $B \leftrightarrow B$) becomes dominant. We therefore construct a low-energy effective model governed by symmetry-constrained nnn hopping alone.
	
	Given the strong AFM exchange interaction, in the nonrelativistic limit spin serves as a good quantum number under the emergent $[C_{\infty} || E]$ spin symmetry, which allows the Bloch Hamiltonian to be decomposed into two independent, spin-projected subspaces, resulting in a minimal model with a diagonal form (see Sec.~I of Supplementary Materials, SM~\cite{SM}): \(H = \bigoplus_{s=\uparrow,\downarrow} H_s\), where the Hamiltonian for a specific spin \emph{s} is given by: \(H_s = \sum_{\mathbf{k}} \left[ \varepsilon_0 - 2t_{2x}^s \cos(k_x a) - 2t_{2y}^s \cos(k_y a) \right] c_{\mathbf{k},s}^\dagger c_{\mathbf{k},s}\). Here, $\varepsilon_0$ is the on-site energy, $a$ is the lattice constant, $t_{2i}^s$ is the nnn hopping integral for spin-$s$ electrons along $i=x,y$, and $c_{\mathbf{k},s}^\dagger$ ($c_{\mathbf{k},s}$) is the creation (annihilation) operator. As illustrated in Fig.~\ref{fig:1}a, nonmagnetic C bridges create anisotropic hopping pathways: A-sublattice sites are bridged along $x$, and B-sublattice sites along $y$. The $[C_2||C_{4z}]$ symmetry then requires $t_{2x}^{\uparrow} = t_{2y}^{\downarrow} = t$ and $t_{2x}^{\downarrow} = t_{2y}^{\uparrow} = \gamma t$, where $t$ is the dominant C-mediated hopping and $\gamma \in [0,1)$ is the transverse-to-longitudinal hopping ratio. Within this magnetic-domain convention, spin-up and spin-down electrons preferentially hop along the orthogonal $x$ and $y$ axes, respectively; the opposite assignment corresponds to the complementary AFM domain or an interchange of spin labels and crystal axes.
	
	The resulting electronic structure is described by the dispersion $E^s(\mathbf{k})=E_0(\mathbf{k})+\sigma_s\Delta(\mathbf{k})$, with $\sigma_\uparrow=+1$ and $\sigma_\downarrow=-1$. Here $E_0(\mathbf{k})=\varepsilon_0-t(1+\gamma)[\cos(k_xa)+\cos(k_ya)]$ is the spin-independent background, whereas $\Delta(\mathbf{k})=t(1-\gamma)[\cos(k_ya)-\cos(k_xa)]$ is the $d_{x^2-y^2}$-wave altermagnetic splitting (Sec.~I of SM~\cite{SM}). The two spin-resolved bands form a symmetry-constrained tent-like band geometry when $\gamma\neq1$. In the ideal limit $\gamma=0$, $E^\uparrow$ depends only on $k_x$ and $E^\downarrow$ only on $k_y$, producing an ideal tent state with two orthogonal quasi-1D cosine sheets, dispersionless transverse directions, straight Fermi lines, and maximal $d$-wave spin splitting (Fig.~S1~\cite{SM}). Finite $\gamma$ introduces transverse dispersion, warping the tent flanks and bending the constant-energy contours while preserving the orthogonal spin-axis correspondence (Figs.~\ref{fig:1}b and S2~\cite{SM}). Thus, the spin-group symmetry fixes the tent-like skeleton, whereas $\gamma$ controls its morphology; as $\gamma\to1$, the $d$-wave splitting vanishes and the tent-like geometry collapses into a spin-degenerate band.
	
	\emph{Spin-Axis Dynamic Locking Effect.} The saddle points of the tent-like dispersion occur at $E_{\mathrm{vH}}=\varepsilon_0\pm2|t|(1-\gamma)$, producing logarithmic van Hove singularities (VHSs) in the analytically derived DOS (Fig.~\ref{fig:1}c; Sec.~II of SM~\cite{SM}). When the chemical potential crosses these energies, the Fermi contours undergo Lifshitz transitions. For $2|t|(1-\gamma)<|\mu-\varepsilon_0|\le 2|t|(1+\gamma)$, the contours form closed anisotropic pockets around the band extrema (Fig.~\ref{fig:1}d). This closed, contractible-pocket geometry corresponds to the Fermi-surface structure of the conventional spin-splitter effect (cSSE), as exemplified by $\mathrm{V_2Se_2O}$~\cite{Ma2021} and $\mathrm{RuO_2}$~\cite{Gonzalez2021}. By contrast, for $|\mu-\varepsilon_0|<2|t|(1-\gamma)$, the contours become open quasi-1D lines traversing the Brillouin zone (Fig.~\ref{fig:1}e). The spin-group operation $(k_x,k_y,\uparrow)\leftrightarrow(\pm k_y,\mp k_x,\downarrow)$ relates the two spin-resolved sets of open lines and constrains their global traversing directions to be orthogonal. To characterize this global Fermi-surface topology, we map the Brillouin zone onto a torus $\mathbb{T}^2$ and define a winding vector $\bm{\omega}$ for each contour $L_j$, with components $\omega_i=(a/2\pi)\oint_{L_j}dk_i$ (SM Sec.~III~\cite{SM}). The closed cSSE pockets $L_1$ and $L_2$ are contractible and carry trivial winding, $\bm{\omega}^{\uparrow}=\bm{\omega}^{\downarrow}=(0,0)$ (Fig.~\ref{fig:1}f), whereas the open Fermi lines are noncontractible, with $\bm{\omega}^{\uparrow}=(0,\pm1)$ for $L_3/L_4$ and $\bm{\omega}^{\downarrow}=(\pm1,0)$ for $L_5/L_6$ (Fig.~\ref{fig:1}h). 
	
	This topological distinction is accompanied by distinct transport dynamics. On closed pockets, the group velocity sweeps all directions (Fig.~\ref{fig:1}g), so small-$q$ scattering can drive angular diffusion around the connected loop, eventually linking opposite-velocity sectors. The resulting velocity-reversing scattering progressively erodes the velocity anisotropy and makes the spin polarization sensitive to momentum relaxation. In the open-line regime, the velocity is dominated by the component perpendicular to each line and is weak along it. Thus, spin-down lines winding along $k_x$ carry large $v_y$, while opposite-$v_y$ branches remain separated by a finite $\Delta k_y$ (Fig.~\ref{fig:1}i). Velocity-reversing backscattering therefore requires a large inter-branch momentum transfer and is suppressed under small-$q$, spin-conserving scattering, making the $y$-axis a dynamically robust spin-down channel. By $[C_2||C_{4z}]$ symmetry, spin-up lines winding along $k_y$ analogously establish the $x$-axis spin-up channel.
	
	We therefore define the SADL effect as a spin-transport mechanism realized in the Lifshitz-delimited open-Fermi-line regime, where $d$-wave spin-group symmetry and nontrivial Fermi-line winding jointly lock spin-selective transport channels to orthogonal crystal axes. In the present magnetic-domain convention, $\bm{\omega}^{\uparrow}$ selects the $x$-axis spin-up channel, whereas $\bm{\omega}^{\downarrow}$ selects the $y$-axis spin-down channel. This dynamically stabilized spin-axis locking distinguishes SADL from the closed-pocket cSSE regime and underlies the high axial spin polarization quantified below.
	
	Within the principal crystal-axis frame, the spin-resolved conductivity tensors are diagonal and related by the $[C_2||C_{4z}]$ symmetry. Using the Boltzmann constant-relaxation-time approximation (CRTA)~\cite{Pizzi2014,Ziman1972}, we analytically obtain $\sigma_{xx}^{\uparrow} = \sigma_{yy}^{\downarrow} \equiv \sigma_{\text{maj}}$ and $\sigma_{yy}^{\uparrow} = \sigma_{xx}^{\downarrow} \equiv \sigma_{\text{min}}$ (SM Sec.~IV~\cite{SM}). With $\tau$ treated as a constant, the resulting anisotropy directly reflects the Fermi-surface geometry and band velocities, allowing us to quantify the axial spin polarization by $\eta_0 \equiv (\sigma_{\text{maj}} - \sigma_{\text{min}})/(\sigma_{\text{maj}} + \sigma_{\text{min}})$. For an in-plane electric field applied at an angle $\theta$ relative to the principal $x$-axis, a coordinate rotation projects this axial anisotropy into tunable longitudinal and transverse responses 
	dictated by the $[C_2\Vert C_{4z}]$ symmetry
	(SM Sec.~IV and Fig.~S3~\cite{SM}): the longitudinal charge conductivity is $\sigma_{\parallel}^c = \sigma_{\text{maj}} + \sigma_{\text{min}}$, while the transverse charge currents of the two spin channels cancel, leaving a pure spin conductivity $|\sigma_{\perp}^s| = (\sigma_{\text{maj}} - \sigma_{\text{min}})|\sin 2\theta|$. Thus, the same rotation controls both the residual longitudinal spin polarization, $\eta = \eta_0 \cos 2\theta$, and the transverse charge-to-spin conversion, characterized by the unconventional spin Hall angle $\theta_{\text{SH}} \equiv |\sigma_{\perp}^s / \sigma_{\parallel}^c| = \eta_0 |\sin 2\theta|$. At the diagonal orientations $\theta = \pi/4$ (mod $\pi/2$), the longitudinal spin polarization vanishes and the transverse charge-to-spin conversion efficiency reaches its maximum value $\eta_0$.
	
	Fig.~\ref{fig:1}j summarizes the transport phase diagram of $\eta_0(\varepsilon_\mu,\gamma)$, where $\varepsilon_\mu=(\mu-\varepsilon_0)/(2|t|)$. The Lifshitz lines $|\varepsilon_\mu|=1-\gamma$ partition the diagram into an inner open-line SADL regime, where high $\eta_0$ is maintained over a broad parameter range, and outer closed-pocket cSSE regimes, where $\eta_0$ is generally lower (Figs.~\ref{fig:1}g and S3~\cite{SM}). In the ideal limit $\gamma\to0$, $\eta_0\to1$: electric fields $\mathbf{E}\parallel[100]$ and $\mathbf{E}\parallel[010]$ drive fully spin-up- and spin-down-polarized currents, respectively; notably, diagonal driving with $\mathbf{E}\parallel[110]$ generates a transverse pure spin current with $100\%$ charge-to-spin conversion. In fact, high $\eta_0$ is not confined to this ideal quasi-1D limit. At the band center, the analytic result derived in SM Sec.~IV~\cite{SM} gives $\eta_0(0)=1-\gamma^2+\mathcal{O}(\gamma^4)$, showing that finite transverse hopping only quadratically degrades the spin polarization. Taking $\eta_0\ge90\%$ as a practical strong-SADL benchmark~\cite{Huang2021}, we obtain $\gamma\lesssim0.318$ at $\varepsilon_\mu=0$, identifying a sizable strong-SADL region accessible to realistic anisotropy in materials.

	\emph{Material Exploration and Realization.} To identify viable materials for realizing the SADL effect in future spintronic applications (e.g., spin field-effect transistors (spin-FET)~\cite{Chuang2015,Jiang2019,Malik2020}, non-volatile computing-in-memory~\cite{Ielmini2018,Sebastian2020,Li2025}, and efficient magnetic devices under electrical control~\cite{Matsukura2015,Fert2024,Song2017}), we screen 2D and 3D crystals that meet the following criteria: (i) they host the key motif of anisotropic magnetic sites bridged by non-magnetic sites, analogous to the model in Fig.~\ref{fig:1}a; (ii) they are intrinsic altermagnetic semiconductors, wherein gate-tunable tent states exist in the frontier conduction ($n$-type) or valence ($p$-type) bands near the Fermi level; (iii) the tent-state bands remain well isolated in a wide energy window and robust against SOC; (iv) the hopping anisotropy is sufficiently strong, corresponding to a small effective $\gamma$, so that the material lies in the strong-SADL region with near-unity axial spin polarization; (v) they are either already synthesized or predicted to be synthesizable.	
	
\begin{figure}[!htbp]
		\includegraphics[width=0.96\linewidth]{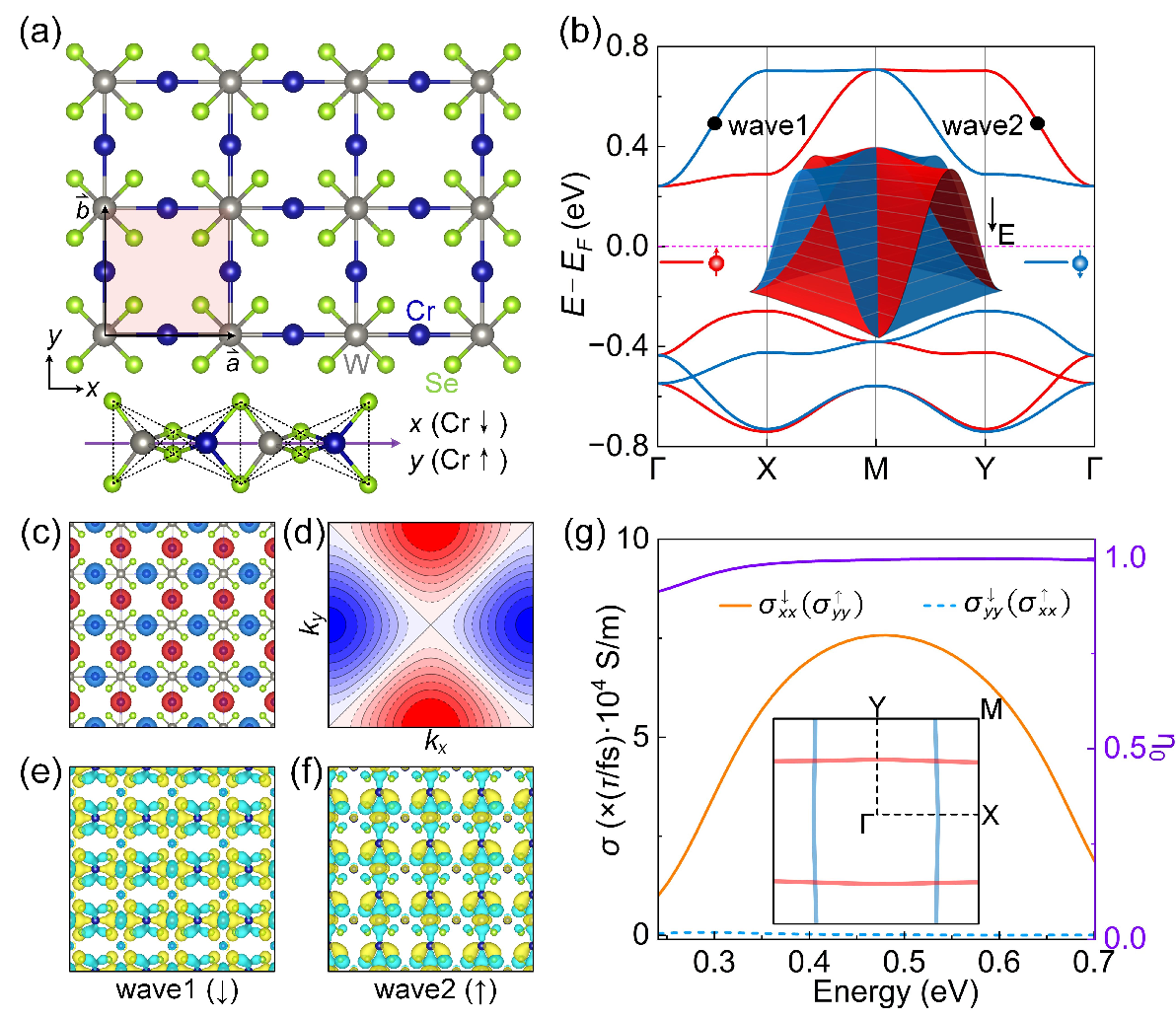}
		\caption{\label{fig:2}Crystal structure, tent state, and SADL effect in Cr$_2$WSe$_4$. (a) Top and side views of the monolayer structure. (b) Electronic band structure with an enlarged view of the tent state (inset). (c) Spin-polarized charge density distribution. (d) Alternating spin splitting. (e, f) Wave function isosurfaces corresponding to the states labeled wave-1 (spin-down) and wave-2 (spin-up) in (b), respectively. (g) Energy-dependent axial conductivity and polarization. Inset: orthogonal Fermi-surface lines of the low-energy conduction bands at 0.5 eV.}
	\end{figure}

	(1) \textit{SADL in 2D Materials.} 2D transition metal dichalcogenides (TMDs) have been extensively studied for their semiconducting properties and tunable electronic behavior~\cite{Manzeli2017,Chhowalla2013,Wang2012}. Beyond these binary compounds, recent interest has expanded to ternary monolayers (TM$_2$MX$_4$)~\cite{Li2025b,Jiang2024,Tan2025,Xu2025}. This expansion is driven by the synthesis of their layered bulk precursors (e.g., Cu$_2$MX$_4$~\cite{Crossland2005} and Ag$_2$MX$_4$~\cite{Zhan2018}) and the discovery of diverse physical phenomena (e.g., the quantum anomalous Hall effect in V$_2$MX$_4$~\cite{Jiang2024}, altermagnetic semiconducting behavior with crystal valley Hall effects~\cite{Tan2025}, and alternating piezoeffects in the Fe$_2$WX$_4$ monolayer~\cite{Xu2025}). Notably, the altermagnetic TM$_2$MX$_4$ monolayers possess a magnetic lattice analogous to the model of Fig.~\ref{fig:1}a, making them a promising platform for the SADL effect. To explore this possibility, we performed high-throughput computational screening (Fig.~S4, Sec.~VI.1 of SM~\cite{SM}) by ensuring charge balance based on the formal oxidation states of the constituent elements (Tables~S1 and S2~\cite{SM}). From an initial pool of 442 candidates, we identified eight 2D altermagnetic semiconductors (see Fig.~\ref{fig:2} and Figs.~S5-S11~\cite{SM}) that meet the preset criteria and exhibit the characteristic tent-state bands, essential for the SADL effect, near their conduction-band bottoms or valence-band tops. In the following, we focus on the Cr$_2$WSe$_4$ monolayer as a prototypical system: its small effective $\gamma$ places it in the strong-SADL regime, while the heavy W atoms provide a stringent test of SOC robustness.  	
	
	The Cr$_2$WSe$_4$ monolayer (Fig.~\ref{fig:2}a) comprises interlocking square lattices of magnetic Cr and nonmagnetic W atoms. Both Cr and W atoms are tetrahedrally coordinated by Se atoms, forming a square-planar layer encapsulated between two Se sheets. This stable tetrahedral environment, with chemically viable oxidation states (i.e., Cr$^{3+}$, W$^{2+}$ and Se$^{2-}$), confers excellent dynamical (Fig.~S12a~\cite{SM}), thermal (Fig.~S12b~\cite{SM}), and mechanical stabilities (Table~S3~\cite{SM}), making it a highly viable candidate for experimental synthesis. Its magnetic ground state is a room-temperature N\'eel AFM configuration with $T_{\text{N}} = 593$~K (Fig.~S12c~\cite{SM}), which remains robust under variations in the Hubbard $U$ parameters, strains, and carrier dopings (Figs.~S12d-f~\cite{SM}). Structurally, nearest-neighbor Cr atoms with parallel magnetic moments are bridged by nonmagnetic WSe$_4$ tetrahedra along one principal axis. In the perpendicular direction, the Cr-Cr spacing extends to $\sim 5.4$~\AA{} in the absence of bridges. This bridged/unbridged contrast suppresses transverse Cr--Cr hopping, realizing the anisotropic hopping motif in Fig.~\ref{fig:1}a. A minimal-model fit to the low-energy tent-state bands gives a small effective $\gamma=0.083$ (Fig.~S12g~\cite{SM}), placing Cr$_2$WSe$_4$ deep in the strong-SADL regime.

	The Cr$_2$WSe$_4$ monolayer is an intrinsic semiconductor (Fig.~\ref{fig:2}b) with a band gap of $0.511$~eV at the PBE level ($1.151$~eV at the HSE06 level, Fig.~S12h~\cite{SM}). It exhibits a $d$-wave altermagnetic electronic structure with $[C_2 \| C_{4z}]$ symmetry, as evidenced by its alternating spin-polarized charge density (Fig.~\ref{fig:2}c) and spin splitting (Fig.~\ref{fig:2}d). Crucially, the two lowest conduction bands form a near-ideal $n$-type tent state (see Fig.~\ref{fig:2}b), characterized by a spin-axial correspondence: the spin-down branch is strongly dispersive along $k_x$ but nearly flat along $k_y$, whereas the spin-up branch shows the opposite behavior.  Such a state is robust at the HSE06 level (Fig.~S12h~\cite{SM}) and under carrier doping/gate tuning (Fig.~S12i~\cite{SM}). The unidirectional character of the wave functions in the dispersive bands further demonstrates the locking between spin and axis (Figs.~\ref{fig:2}e, f). Owing to the extreme anisotropy, the Fermi contours form nearly perfect orthogonal open lines over a sizable energy window of $\sim 0.45$~eV, adopting a characteristic ``\texttt{\#}''-shape. All these features are in excellent agreement with our model analysis. Furthermore, our numerical transport calculations based on Wannier-function fitting within the CRTA confirm near-ideal SADL, yielding highly anisotropic spin-resolved conductivities: $\sigma_{xx}^{\downarrow} \gg \sigma_{xx}^{\uparrow} \approx 0$ and $\sigma_{yy}^{\uparrow} \gg \sigma_{yy}^{\downarrow} \approx 0$. Transport along the $x$-axis is $99.6\%$ spin-down polarized, while that along the $y$-axis is $99.6\%$ spin-up polarized (Fig.~\ref{fig:2}g). Consequently, diagonal driving generates a pure transverse spin current with a near-unity charge-to-spin conversion efficiency of $99.6\%$. SOC-included Kubo calculations and AMSET-based momentum-dependent scattering analyses (Figs.~S13 and S14; Sec.~VI.2 of SM~\cite{SM}) further confirm that the efficiency remains above the strong-SADL benchmark of $90\%$, consistent with the preceding backscattering-suppression picture.
	
	(2)\textit{SADL in 3D Materials.} Achieving low-dimensional transport within 3D materials has long been pursued to address fabrication and integration challenges of nanomaterials~\cite{Parker2013,Bilc2015,He2024,Ma2025,Zhang2025}. Although our theoretical model is based on a 2D lattice, the SADL effect can also be realized in 3D materials with quasi-2D structural motifs, when the low-energy electronic bands near the Fermi level are primarily governed by these 2D features.
	
\begin{figure}[!htbp]
		\centering
		\includegraphics[width=0.92\linewidth]{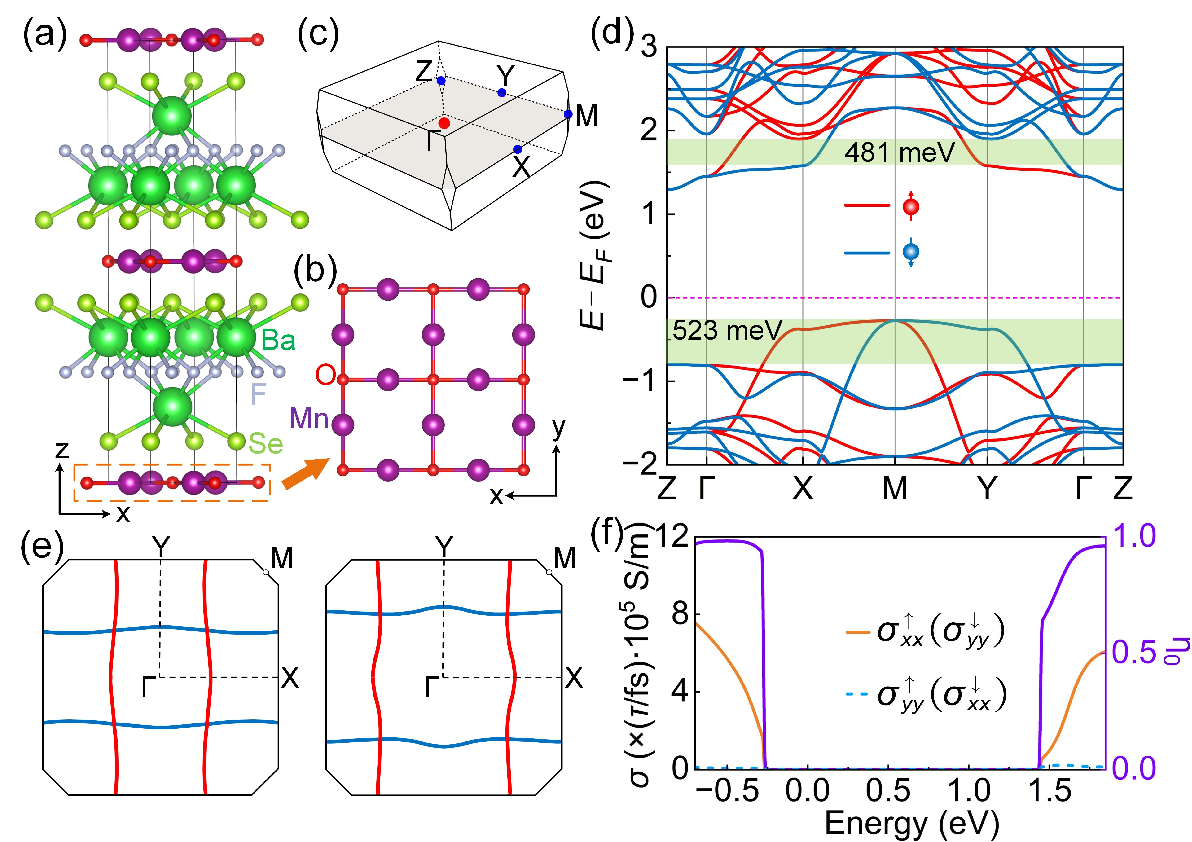}
		\caption{\label{fig:3}SADL effect in the 3D crystal (BaF)$_2$Mn$_2$Se$_2$O. (a) Crystal structure. (b) The structure of MnO layer. (c) Brillouin zone. (d) Electronic band structure. (e) Fermi surfaces of the spin-polarized valence band at $-$0.6 eV and conduction band at 1.7 eV. (f) Energy-dependent conductivities and spin polarizations along different directions.}
	\end{figure}

	From a high-throughput screening (Fig.~S15~\cite{SM}) of 3415 AFM compounds in the Materials Project database~\cite{Horton2025,Jain2013}, we identified 326 tetragonal and 110 cubic candidates. Among these, six materials with layered units analogous to Fig.~\ref{fig:1}a exhibit quasi-2D altermagnetic characteristics (Fig.~S16~\cite{SM}). Remarkably, the experimentally synthesized crystal (BaF)$_2$Mn$_2$Se$_2$O~\cite{Liu2011} emerges as a prime candidate, closely meeting our electronic structure criteria for SADL. Its crystal structure ($I4/mmm$ symmetry), depicted in Fig.~\ref{fig:3}a, features BaF and Mn$_2$Se$_2$O layers stacked along the $z$-axis. The atomic arrangement of the MnO layer (Fig.~\ref{fig:3}b) is a faithful realization of the lattice model in Fig.~\ref{fig:1}a.
	
	The electronic structure of (BaF)$_2$Mn$_2$Se$_2$O confirms its quasi-2D nature, evidenced by nearly flat, spin-degenerate band edges along the $k_z$ axis ($\Gamma \to \mathrm{Z}$, Fig.~\ref{fig:3}d), which indicates weak out-of-plane dispersion. Within the $k_x$-$k_y$ plane (Fig.~\ref{fig:3}c), both the valence and conduction band edges form tent states (see Figs.~S17a-b~\cite{SM}), whose low-energy dispersions are well captured by the minimal model with effective anisotropy ratios of $\gamma=0.042$ and $0.108$, respectively (Fig.~S17c~\cite{SM}). These states persist over substantial clean-energy windows of 523~meV (valence) and 481~meV (conduction), and remain robust under PBE+SOC (Fig.~S17d~\cite{SM}), HSE06 (Fig.~S17e~\cite{SM}), and carrier doping (Figs.~S17f-g~\cite{SM}), despite some band mixing at deeper energies. This robustness allows the Fermi level to be tuned by a gate bias into these clean-energy windows, activating the quasi-1D transport of spin-polarized carriers locked to the crystal axes (spin-up along $x$ and spin-down along $y$). Owing to the 3D nature, interlayer orbital hybridization leads to slight warping of the tent states, manifested as a minimal curvature of the Fermi contours (Fig.~\ref{fig:3}e). Nevertheless, the spin polarizations retain remarkably high values of 98.3\% for holes and 96.1\% for electrons (Fig.~\ref{fig:3}f). Consequently, a diagonal in-plane electric field generates a transverse pure spin current with similarly high charge-to-spin conversion efficiency. These results demonstrate that the core SADL phenomenology is maintained in this 3D material, enabling all-electrical control of a highly spin-polarized current. As a synthesized compound, (BaF)$_2$Mn$_2$Se$_2$O offers an experimentally accessible platform where spectroscopic and spin-transport measurements can verify the tent-state bands and SADL response (Sec.~VII of SM~\cite{SM}).
	
	\textit{Conclusions.} In summary, we propose and demonstrate a distinct spin-control mechanism in altermagnets: spin-axis dynamic locking. This effect enables the all-electrical generation of highly spin-polarized charge currents and transverse pure spin currents with $100\%$ charge-to-spin conversion efficiency in the ideal limit, controlled by the direction of an in-plane electric field. SADL is a Lifshitz-delimited spin-transport mechanism arising from a tent-like band structure: in the open-Fermi-line regime, $d$-wave spin-group symmetry and nontrivial Fermi-line winding jointly lock spin-selective transport channels to orthogonal crystal axes, while the momentum-space separation of opposite-velocity branches suppresses velocity-reversing backscattering and stabilizes high axial spin selectivity. Guided by this framework, we identify realistic altermagnetic semiconductors from the 2D $\mathrm{Cr_2WSe_4}$ monolayer to the synthesized 3D $\mathrm{(BaF)_2Mn_2Se_2O}$, both approaching the ideal SADL limit. Our results establish SADL as a nonrelativistic route toward reconfigurable, low-power spintronic devices through all-electrical spin manipulation.

	\textit{Acknowledgments.} This work is supported by the National Natural Science Foundation of China (Grants No. 12464040 and No. 12064031).
	
	\textit{Author contributions.} Z.~Lv and J.~Cai contributed equally to this work. Z.~Liu conceived the original idea and initiated the project. Z.~Liu and Y.~Xing supervised the research. All authors discussed the results and contributed to the preparation of the manuscript.
	
	\emph{Note added}: Upon completion of our work, we became aware of a study by Lai \emph{et al.}~\cite{lai2025d} on a related topic. Our work is distinguished by its establishment of the SADL effect, proposed via a minimal model and demonstrated in a separate family of materials.

% ============================================================================
% Supplemental Material
% The REVTeX main text ends above. The following local settings reproduce the
% original one-column, 12 pt Supplemental Material layout as closely as a
% single-document source permits.
% ============================================================================
\clearpage
\onecolumngrid
\begingroup
\setcounter{page}{1}
\setcounter{figure}{0}
\setcounter{table}{0}
\setcounter{equation}{0}
\renewcommand{\thefigure}{S\arabic{figure}}
\renewcommand{\thetable}{S\arabic{table}}
\renewcommand{\theHfigure}{SM.\arabic{figure}}
\renewcommand{\theHtable}{SM.\arabic{table}}
\renewcommand{\theHequation}{SM.\arabic{equation}}
\hypersetup{pageanchor=false}
\makeatletter
\renewcommand{\fnum@figure}{\textbf{Fig.\ \thefigure}}
\renewcommand{\fnum@table}{\textbf{Table \thetable.}}
\long\def\@makecaption#1#2{%
  \vskip\abovecaptionskip
  \noindent #1\space #2\par
  \vskip\belowcaptionskip}
\makeatother
\setlength{\parindent}{0pt}
\setlength{\parskip}{6pt}
\setlength{\oddsidemargin}{0in}
\setlength{\evensidemargin}{0in}
\setlength{\textwidth}{6.5in}
\setlength{\textheight}{9in}
\setlength{\columnwidth}{6.5in}
\setlength{\hsize}{6.5in}
\setlength{\vsize}{9in}
\setlength{\topmargin}{-0.5in}
\setlength{\headheight}{20pt}
\setlength{\headsep}{20pt}
\setlength{\footskip}{30pt}
\renewcommand{\baselinestretch}{1.15}\normalsize
\raggedbottom
\pagestyle{fancy}
\fancyhf{}
\fancyhead[R]{\thepage}
\renewcommand{\headrulewidth}{0pt}
\makeatletter
\renewcommand\normalsize{%
  \@setfontsize\normalsize\@xiipt{14.5}%
  \abovedisplayskip 12\p@ \@plus3\p@ \@minus7\p@
  \abovedisplayshortskip \z@ \@plus3\p@
  \belowdisplayshortskip 6.5\p@ \@plus3.5\p@ \@minus3\p@
  \belowdisplayskip \abovedisplayskip}
\renewcommand\small{\@setfontsize\small\@xipt{13.6}}
\renewcommand\large{\@setfontsize\large\@xivpt{18}}
\renewcommand\Large{\@setfontsize\Large\@xviipt{22}}
\makeatother
\normalsize\fontfamily{ptm}\selectfont
\newcommand{\journaltext}[1]{\textcolor{magenta}{#1}}
\newcommand{\boldfirstnumber}[1]{%
  \begingroup
  \def\processnumber##1,##2\relax{\textbf{##1},##2}%
  \expandafter\processnumber#1\relax
  \endgroup}
\newenvironment{reqn}[1]{\begin{equation}\tag{#1}}{\end{equation}}

% 首页不显示页码
	\thispagestyle{empty}
	
	% 自定义标题区域，减少顶部间距
	\begingroup
	\centering
	{ \textit{Supplementary Materials for}\\[18pt]
		\Large \bfseries Spin-Axis Dynamic Locking}\\[18pt]
	
	% 作者信息 - 使用小四字体
	{\normalsize
		Zhiheng Lv\textsuperscript{1,*}, Jiangtao Cai\textsuperscript{2,*}, Dengpan Ma\textsuperscript{1},  Yan Xing\textsuperscript{1,\dag}, Zhifeng Liu\textsuperscript{1,\ddag}\\[12pt]
		
		% 单位信息 - 使用小四字体
		\begin{center}
			\normalsize
			\textsuperscript{1}\textit{Inner Mongolia Key Laboratory of Microscale Physics and Atomic Manufacturing \& Research Center for Quantum Physics and Technologies, School of Physical Science and Technology, Inner Mongolia University, Hohhot 010021, China}\\[1pt]
			\textsuperscript{2}\textit{School of Physics and Information Science, Shaanxi University of Science and Technology, Xi'an 710021, China}
		\end{center}
	}
	\endgroup
	
	\vspace{24pt}
	% 目录 - 使用小四字体
	\begin{flushleft}
		\textbf{Table of Contents}
	\end{flushleft}
	
	\vspace{12pt}
	\begin{enumerate}[label=\Roman{*}. , leftmargin=*, align=left]
		\item Two-Dimensional Anisotropic $d$-Wave Altermagnetic Model.
		\item Density of States and Lifshitz Boundary.
		\item Fermi-Surface Topology and Winding Vectors.
		\item Boltzmann Transport and Charge-to-Spin Conversion.
		\item Computational Methods.
		\item Realization of SADL in Two-Dimensional Materials.
		\item Realization of SADL in Three-Dimensional Materials.
	\end{enumerate}

	% 通讯作者信息放在页面底部，使用小五字体并减少行间距
	\vspace*{\fill}
	\begin{flushleft}
		\fontsize{8}{3}\selectfont % 小五字体，行距设为10pt
		*\ These authors contributed equally to this work
		
		\dag\ Corresponding author: xingy@imu.edu.cn
		
		\ddag\ Corresponding author: zfliu@imu.edu.cn
	\end{flushleft}
	
	% 从第二页开始添加新内容
	% Section I
	\newpage
	\begin{center}
		\large\bfseries I. Two-Dimensional Anisotropic $d$-Wave Altermagnetic Model
	\end{center}	
	
	\qquad In the nonrelativistic limit considered in the main text, the effective low-energy model is obtained by projecting the N\'eel antiferromagnetic (AFM) lattice onto two spin-conserving hopping channels. For each spin channel $s$ ($s=\uparrow,\downarrow$), the real-space tight-binding Hamiltonian is written as
	\begin{equation}
		H_s = \sum_{\mathbf{R}} \varepsilon_0 c_{\mathbf{R}s}^{\dagger}c_{\mathbf{R}s} - \sum_{\mathbf{R}}\sum_{i=x,y} t_{2i}^{s} \left( c_{\mathbf{R}+\mathbf{a}_i,s}^{\dagger}c_{\mathbf{R}s} + c_{\mathbf{R}s}^{\dagger}c_{\mathbf{R}+\mathbf{a}_i,s} \right). \tag{I-1} \label{I-1}
	\end{equation}
	Here, $\varepsilon_0$ is the on-site energy; $t_{2i}^{s}$ is the spin-dependent next-nearest-neighbor hopping integral along $i=x,y$; $c_{\mathbf{R}s}^{\dagger}$ and $c_{\mathbf{R}s}$ are the creation and annihilation operators at site $\mathbf{R}$, respectively; and $\mathbf{a}_x=a\hat{\mathbf{x}}$ and $\mathbf{a}_y=a\hat{\mathbf{y}}$ are the corresponding displacement vectors along the two principal axes.	
	
	\qquad Using the Fourier transformation
	\begin{equation}
		c_{\mathbf{R}s} = \frac{1}{\sqrt{N}} \sum_{\mathbf{k}} e^{-i\mathbf{k}\cdot\mathbf{R}} c_{\mathbf{k}s}, \qquad c_{\mathbf{R}s}^{\dagger} = \frac{1}{\sqrt{N}} \sum_{\mathbf{k}} e^{i\mathbf{k}\cdot\mathbf{R}} c_{\mathbf{k}s}^{\dagger},
		\tag{I-2} \label{I-2}
	\end{equation}
	together with the lattice orthogonality relation $\sum_{\mathbf{R}}e^{i(\mathbf{k}-\mathbf{k}')\cdot\mathbf{R}}=N\delta_{\mathbf{k},\mathbf{k}'}$, the Hamiltonian becomes diagonal in momentum space:
	\begin{equation}
		H_s = \sum_{\mathbf{k}} \left[ \varepsilon_0 - 2t_{2x}^{s}\cos(k_xa) - 2t_{2y}^{s}\cos(k_ya) \right] c_{\mathbf{k}s}^{\dagger}c_{\mathbf{k}s}.
		\tag{I-3} \label{I-3}
	\end{equation}
	With the canonical anticommutation relation 
	$\bigl\{c_{\mathbf{k}s},c_{\mathbf{k}'s'}^\dagger\bigr\}
	=\delta_{\mathbf{k},\mathbf{k}'}\delta_{s,s'}$, 
	the diagonal form of $H_s$ directly gives the single-particle eigenvalue
	\begin{equation}
		E^s(\mathbf{k}) = \varepsilon_0 - 2t_{2x}^{s}\cos(k_xa) - 2t_{2y}^{s}\cos(k_ya).
		\tag{I-4} \label{I-4}
	\end{equation}
	Substituting the symmetry-constrained hopping amplitudes ($t_{2x}^{\uparrow}=t_{2y}^{\downarrow}=t$ and $ t_{2y}^{\uparrow}=t_{2x}^{\downarrow}=\gamma t$), the two spin-resolved dispersions are
	\begin{equation}
		E^{\uparrow}(\mathbf{k}) = \varepsilon_0 - 2t\cos(k_xa) - 2\gamma t\cos(k_ya),
		\tag{I-5} \label{I-5}
	\end{equation}
	and
	\begin{equation}
		E^{\downarrow}(\mathbf{k}) = \varepsilon_0 - 2\gamma t\cos(k_xa) - 2t\cos(k_ya).
		\tag{I-6} \label{I-6}
	\end{equation}
	
	\qquad Moreover, they can be written in the unified form $E^{s}(\mathbf{k}) = E_0(\mathbf{k}) + \sigma_s\Delta(\mathbf{k})$. In the ideal limit $\gamma=0$, the dispersions reduce to $E^{\uparrow}(\mathbf{k})=\varepsilon_0-2t\cos(k_xa)$ and $E^{\downarrow}(\mathbf{k})=\varepsilon_0-2t\cos(k_ya)$. This produces the ideal tent state, as shown in Fig.~\ref{fig:S1}. For finite $\gamma$, transverse hopping deforms this ideal tent state, as illustrated in Fig.~\ref{fig:S2}. As $\gamma\to1$, the $d$-wave splitting vanishes and the tent-like geometry continuously collapses into a spin-degenerate band.
	
	\qquad For either spin channel, the allowed band-energy range is $E\in[\varepsilon_0-2|t|(1+\gamma),\,\varepsilon_0+2|t|(1+\gamma)]$. The saddle points at $(k_x,k_y)=(0,\pi/a)$ and $(\pi/a,0)$ occur at the energies $E=\varepsilon_0\pm2|t|(1-\gamma)$. These saddle-point energies determine the van Hove singularities and serve as the analytic markers of the Lifshitz transition discussed in Sec.~II.
	
	\begin{figure}[H] 
		\centering \includegraphics[width=0.55\linewidth]{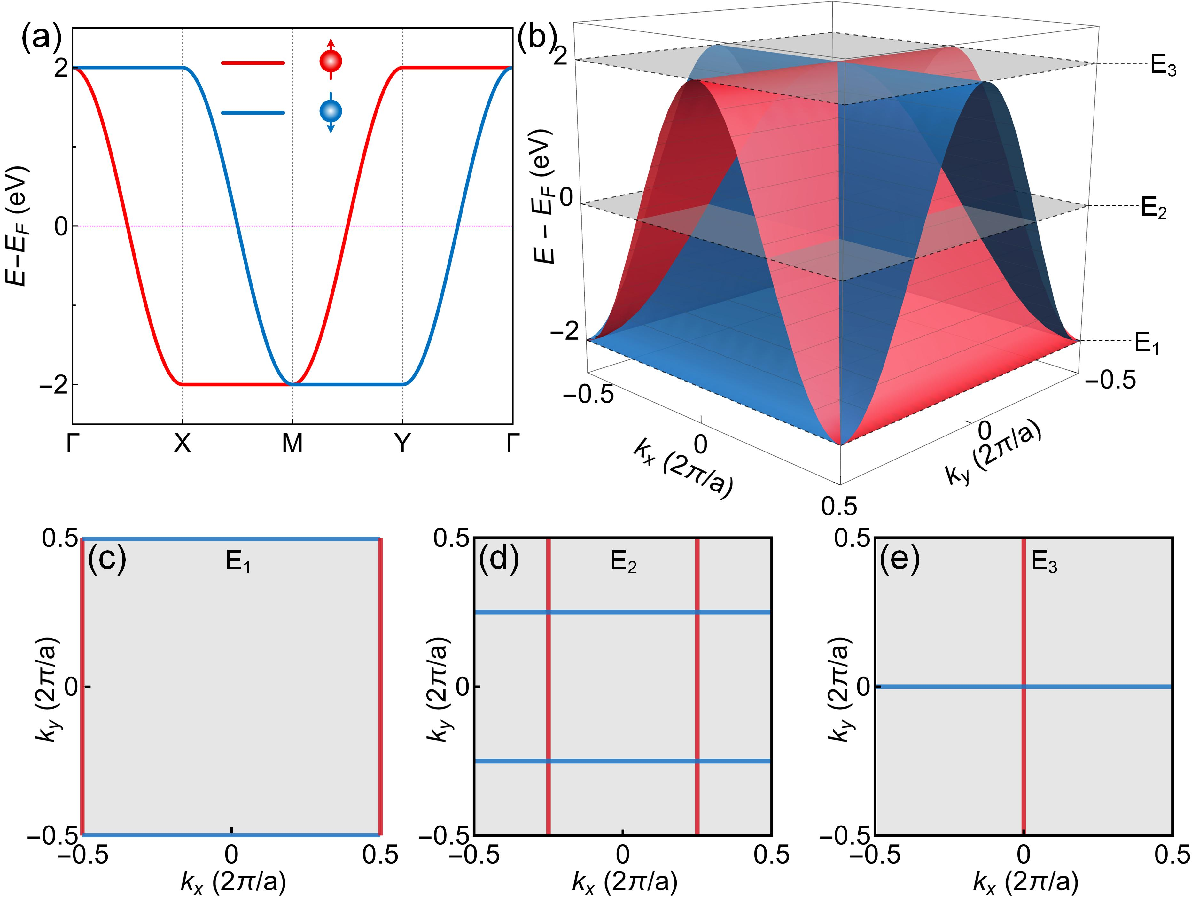} \caption{{Ideal tent state in the decoupled-chain limit at $\gamma=0$. (a) Spin-resolved band structure along the high-symmetry path. (b) 3D representation of the two orthogonal 1D cosine sheets, with three representative constant-energy cuts labeled $E_1$, $E_2$, and $E_3$. (c--e) Corresponding constant-energy contours, showing perfectly straight open lines for both spin channels.}} \label{fig:S1} 
	\end{figure}

	\begin{figure}[H]
		\centering
		\includegraphics[width=0.55\linewidth]{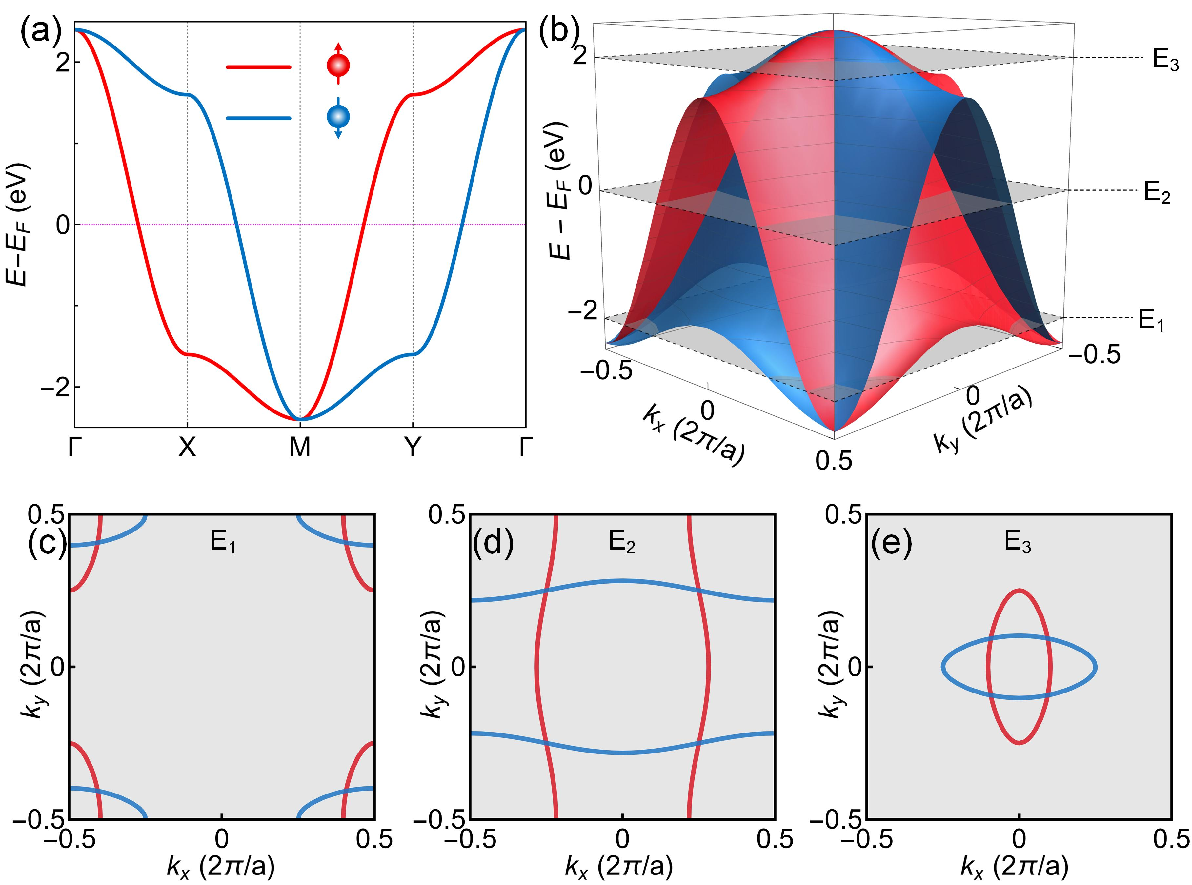}
		\caption{{Finite-anisotropy tent state at $\gamma=0.2$. 
				(a) Spin-resolved band structure along the high-symmetry path. 
				(b) 3D representation of the warped tent-state bands, with representative constant-energy cuts taken near the lower band edge, band center, and upper band edge. 
				(c--e) Corresponding constant-energy contours, illustrating the evolution from closed pockets near the band edges to open lines near the band center.}}
		\label{fig:S2}
	\end{figure}
	
	%Section II
	\clearpage
	\begin{center}
		{\large\bfseries\boldmath II. Density of States and Lifshitz Boundary}
	\end{center}
	
	\qquad In this section, we derive the density of states (DOS) for the anisotropic model introduced in Sec.~I. Because the two spin channels are related by $k_x\leftrightarrow k_y$, their Brillouin-zone-integrated DOS are identical, and it is therefore sufficient to evaluate the spin-up channel. With momenta measured in units of $1/a$ (i.e., $k_xa\to k_x$ and $k_ya\to k_y$), one has $k_x,k_y\in[-\pi,\pi]$ and $E^{\uparrow}(\mathbf{k})=\varepsilon_0-2t\cos k_x-2\gamma t\cos k_y$. The single-spin DOS per unit cell is then defined as
	\begin{equation}
		D_{\uparrow}(E) = \frac{1}{(2\pi)^2}
		\int_{-\pi}^{\pi}\int_{-\pi}^{\pi}
		\delta\left[ E-E^{\uparrow}(\mathbf{k}) \right] \, dk_x\,dk_y .
		\tag{II-1} \label{II-1}
	\end{equation}
	
	Defining the reduced energy $\varepsilon=(E-\varepsilon_0)/(2|t|)$, the $k_x$ integration over the delta-function roots and the substitution $u=\cos k_y$ give
	\begin{equation}
		D_{\uparrow}(E) = \frac{1}{2\pi^2|t|}
		\int_{-1}^{1}
		\frac{\Theta\left[1-(\varepsilon+\gamma u)^2\right]}
		{\sqrt{1-(\varepsilon+\gamma u)^2}\sqrt{1-u^2}} \, du .
		\tag{II-2} \label{II-2}
	\end{equation}
	
	The step function restricts the integral to real $k_x$ solutions. Since the denominator in Eq.~\eqref{II-2} can be written as the square root of a quartic polynomial in $u$, the integral belongs to the standard complete-elliptic-integral form and can be expressed in terms of the complete elliptic integral of the first kind, $K(m)=\int_0^{\pi/2}d\phi/\sqrt{1-m\sin^2\phi}$.
	
	\qquad For $0\le|\varepsilon|\le1-\gamma$, the single-spin DOS is
	\begin{equation}
		D_{\uparrow}(E)
		= \frac{1}{\pi^2|t|\sqrt{(1+\gamma)^2-\varepsilon^2}}
		K\left[\frac{4\gamma}{(1+\gamma)^2-\varepsilon^2}\right],
		\tag{II-3} \label{II-3}
	\end{equation}
	whereas for $1-\gamma<|\varepsilon|\le1+\gamma$, one obtains
	\begin{equation}
		D_{\uparrow}(E)
		= \frac{1}{2\pi^2|t|\sqrt{\gamma}}
		K\left[\frac{(1+\gamma)^2-\varepsilon^2}{4\gamma}\right].
		\tag{II-4} \label{II-4}
	\end{equation}
	For $|\varepsilon|>1+\gamma$, the energy lies outside the band and the DOS vanishes.
	
	\qquad In Eqs.~\eqref{II-3} and \eqref{II-4}, the elliptic-integral arguments, $m_1=4\gamma/[(1+\gamma)^2-\varepsilon^2]$ and $m_2=[(1+\gamma)^2-\varepsilon^2]/(4\gamma)$, respectively, both approach unity at $|\varepsilon|=1-\gamma$. Since $K(m)$ diverges logarithmically near $m=1$ (i.e., $K(m)\propto-\ln(1-m)$ for $m\to1^{-}$), the DOS exhibits logarithmic van Hove singularities at	
	\begin{equation}
		E_{\mathrm{vH}}=\varepsilon_0\pm2|t|(1-\gamma).
		\tag{II-5} \label{II-5}
	\end{equation}
	These are the saddle-point energies identified in Sec.~I. When the chemical potential crosses these energies, the Fermi contours undergo Lifshitz transitions, defining the Lifshitz boundary $|\mu-\varepsilon_0|=2|t|(1-\gamma)$.	For $|\mu-\varepsilon_0|<2|t|(1-\gamma)$, the Fermi contours are open lines crossing the Brillouin-zone boundary, corresponding to the open-line SADL regime. For $2|t|(1-\gamma)<|\mu-\varepsilon_0|\le2|t|(1+\gamma)$, the Fermi contours become closed pockets around the band extrema, corresponding to the closed-pocket cSSE regime. Therefore, the logarithmic VHS provides the analytic marker of the Lifshitz transition separating the open-line SADL regime from the closed-pocket cSSE regime.
	
	\qquad In the ideal decoupled-chain limit $\gamma=0$, the DOS reduces to
	\begin{equation}
		D_{\uparrow}(E) \to \frac{1}{\pi\sqrt{4t^2-(E-\varepsilon_0)^2}}.
		\tag{II-6} \label{II-6}
	\end{equation}
	In this limit, the VHS energies move to the band edges $E=\varepsilon_0\pm2|t|$, where the DOS develops the square-root divergence characteristic of a 1D tight-binding chain. At the same time, the closed-pocket cSSE interval collapses to zero width, so the entire band interior is occupied by open constant-energy lines, corresponding to the ideal SADL limit.	
	
	% Section III
	\clearpage
	\begin{center}
		{\large\bfseries\boldmath III. Fermi-Surface Topology and Winding Vectors}
	\end{center}
	
	\qquad To characterize the global Fermi-surface topology, we regard the first 2D Brillouin zone (BZ) as a torus $\mathbb{T}^2$ by gluing together the opposite edges at $k_i=-\pi/a$ and $k_i=\pi/a$ for each momentum direction $i=x,y$. Physically, this gluing reflects the equivalence between Bloch states at $\mathbf{k}$ and $\mathbf{k}+\mathbf{G}$, where $\mathbf{G}$ is a reciprocal lattice vector. Thus, both closed pockets inside the BZ and open lines crossing the BZ boundary are represented as closed contours on $\mathbb{T}^2$. For a Fermi contour $L_j$ on this torus, we define the winding vector
	\begin{equation}
		\boldsymbol{\omega}(L_j) = (\omega_x,\omega_y), \qquad
		\omega_i = \frac{a}{2\pi} \oint_{L_j} dk_i, \qquad i=x,y .
		\tag{III-1} \label{III-1}
	\end{equation}
	Here, $k_x$ and $k_y$ denote physical crystal momenta. The line integral is evaluated over one full traversal of $L_j$ on the BZ torus: starting from any point on $L_j$, one follows the contour continuously until it returns to the same point on $\mathbb{T}^2$. The resulting net displacement in the extended BZ, measured in units of $2\pi/a$, gives the winding numbers $\omega_x$ and $\omega_y$; their signs depend on the chosen traversal direction of $L_j$.
	
	\qquad Taking the spin-up channel as an example, the Fermi contour at chemical potential $\mu$ satisfies $E^\uparrow(\mathbf{k})=\mu$. Defining the reduced chemical potential $\varepsilon_\mu=(\mu-\varepsilon_0)/(2|t|)$ and $s_t=\mathrm{sgn}(t)$, this condition gives
	\begin{equation}
		\cos(k_xa)+\gamma\cos(k_ya) = -s_t\varepsilon_\mu .
		\tag{III-2} \label{III-2}
	\end{equation}
	The band edges correspond to $|\varepsilon_\mu|=1+\gamma$, while the Lifshitz boundary derived in Sec.~II is $|\varepsilon_\mu|=1-\gamma$.
	
	\qquad In the outer energy windows $1-\gamma<|\varepsilon_\mu|\le1+\gamma$, the spin-up Fermi contours form closed pockets around the band extrema. Each pocket is a contractible loop on $\mathbb{T}^2$ and can be traversed without a net displacement in the extended BZ. Using the winding definition in Eq.~\eqref{III-1}, one obtains
	\begin{equation}
		\omega_i^{\uparrow}
		= \frac{a}{2\pi}\oint_{L_j} dk_i
		= \frac{a}{2\pi}\Delta k_i
		=0.
		\tag{III-3} \label{III-3}
	\end{equation}
	Thus, the spin-up closed pockets carry trivial winding, $\boldsymbol{\omega}^{\uparrow}=(0,0)$.
	
	\qquad For $|\varepsilon_\mu|<1-\gamma$, the spin-up Fermi-contour equation can be written as $\cos(k_xa)=-s_t\varepsilon_\mu-\gamma\cos(k_ya)$. For any $k_y$ with $k_ya\in[-\pi,\pi]$, the right-hand side satisfies
	\begin{equation}
		|-s_t\varepsilon_\mu-\gamma\cos(k_ya)|\le|\varepsilon_\mu|+\gamma<1.
		\tag{III-4} \label{III-4}
	\end{equation}
	Therefore, real solutions for $k_x$ exist throughout the full $k_y$ range. By contrast, solving the same equation for $k_y$ would require $|-s_t\varepsilon_\mu-\cos(k_xa)|\le\gamma$, which is not satisfied for all $k_x$ when $\gamma<1$. Thus, the spin-up contours form open lines connecting the two opposite $k_y$ boundaries of the BZ, i.e., they wind along the $k_y$ direction on $\mathbb{T}^2$. Over one full traversal of such a line on the BZ torus, the corresponding path in the extended BZ has no net displacement along $k_x$ but advances by one reciprocal period along $k_y$, i.e., $\Delta k_x=0$ and $\Delta k_y=2\pi/a$ for one traversal direction. Using the winding definition in Eq.~\eqref{III-1}, one obtains
	\begin{equation}
		\omega_x^{\uparrow}=0,\qquad
		\omega_y^{\uparrow}=\frac{a}{2\pi}\Delta k_y=1.
		\tag{III-5} \label{III-5}
	\end{equation}
	Reversing the traversal direction changes the sign of the winding vector. Thus, the pair of spin-up open Fermi lines is characterized by $\boldsymbol{\omega}^{\uparrow}=(0,\pm1)$.
	
	\qquad By $[C_2\parallel C_{4z}]$ symmetry, which exchanges $k_x\leftrightarrow k_y$ together with $\uparrow\leftrightarrow\downarrow$, the spin-down closed pockets also have trivial winding, $\boldsymbol{\omega}^{\downarrow}=(0,0)$, whereas the spin-down open lines wind along the $k_x$ direction, giving $\boldsymbol{\omega}^{\downarrow}=(\pm1,0)$.
	
	\qquad Following the notation of Figs.~1(f) and 1(h) in the main text, the closed pockets $L_1/L_2$ carry $\boldsymbol{\omega}^{\uparrow}=\boldsymbol{\omega}^{\downarrow}=(0,0)$, while the open lines $L_3/L_4$ and $L_5/L_6$ carry $\boldsymbol{\omega}^{\uparrow}=(0,\pm1)$ and $\boldsymbol{\omega}^{\downarrow}=(\pm1,0)$, respectively. This winding change across the Lifshitz boundary provides the Fermi-surface-topological distinction between the open-line SADL and closed-pocket cSSE regimes.
	
	% Section IV
	\clearpage
	
	\begin{center}
		{\large\bfseries\boldmath IV. Boltzmann Transport and Charge-to-Spin Conversion}
	\end{center}
	
	\subsubsection*{IV.1 Spin-resolved Boltzmann conductivity and axial polarization}
	
	\qquad In the principal crystal-axis frame, each spin-resolved conductivity tensor is diagonal by symmetry, and the total charge conductivity tensor is $\bm{\sigma}^{c}=\bm{\sigma}^{\uparrow}+\bm{\sigma}^{\downarrow}=\mathrm{diag}(\sigma_{xx}^{\uparrow}+\sigma_{xx}^{\downarrow},\,\sigma_{yy}^{\uparrow}+\sigma_{yy}^{\downarrow})$. To isolate the contribution from the band anisotropy and Fermi-surface geometry, we first use semiclassical Boltzmann theory within the constant relaxation-time approximation (CRTA)~\textcolor{blue}{[\hyperlink{SM-ref-Pizzi2014}{1}, \hyperlink{SM-ref-Ziman1972}{2}]}. This provides an analytic baseline for the spin-resolved transport response, while momentum-dependent scattering effects are examined separately in the material-specific calculations. The diagonal conductivity component for spin channel $s$ at chemical potential $\mu$ is
	\begin{equation}
		\sigma_{ii}^{s}(\mu)
		=
		\frac{e^2\tau}{(2\pi)^2}
		\int_{\mathrm{BZ}}
		\left[v_i^{s}(\mathbf{k})\right]^2
		\delta\left[\mu-E^s(\mathbf{k})\right]
		d^2\mathbf{k}.
		\tag{IV-1} \label{IV-1}
	\end{equation}
	Here, $v_i^s(\mathbf{k})=(1/\hbar)\partial E^s/\partial k_i$ is the physical group velocity before momentum rescaling, and $i=x,y$.
	
	\qquad As in Sec.~II, we now rescale the momenta as $k_i a\to k_i$ ($i=x,y$), so that, from this point on, $k_x$ and $k_y$ denote dimensionless momentum variables in the interval $[-\pi,\pi]$. Under this change of variables, the momentum-space integration measure contributes a factor $a^{-2}$, while the physical group velocity becomes $v_i^s=(a/\hbar)\partial E^s/\partial k_i$, where the derivative is now taken with respect to the dimensionless variable $k_i$. These two length factors cancel exactly between the measure and the squared velocity. To isolate the remaining energy scale, we define the dimensionless band energy $\bar{E}^{s}(\mathbf{k})$ and the reduced chemical potential $\varepsilon_\mu$ as
	\begin{equation}
		\bar{E}^{s}(\mathbf{k})
		=
		\frac{E^s(\mathbf{k})-\varepsilon_0}{2|t|},
		\qquad
		\varepsilon_\mu
		=
		\frac{\mu-\varepsilon_0}{2|t|}.
		\tag{IV-2} \label{IV-2}
	\end{equation}
	Here, $E^s(\mathbf{k})$ denotes the dispersion after the same momentum rescaling. Then,
	\begin{equation}
		\frac{\partial E^s}{\partial k_i}
		=
		2|t|\frac{\partial \bar{E}^{s}}{\partial k_i},
		\qquad
		\delta\left[\mu-E^s(\mathbf{k})\right]
		=
		\frac{1}{2|t|}
		\delta\left[\varepsilon_\mu-\bar{E}^{s}(\mathbf{k})\right].
		\tag{IV-3} \label{IV-3}
	\end{equation}

	Thus, the squared-velocity term contributes an energy factor $(2|t|)^2$, while the delta function contributes $1/(2|t|)$, leaving a net factor $2|t|$. The conductivity can therefore be written as
	\begin{equation}
		\sigma_{ii}^{s}(\varepsilon_\mu)
		=
		\frac{2e^2\tau |t|}{\hbar^2}
		\Phi_{ii}^{s}(\varepsilon_\mu),
		\tag{IV-4} \label{IV-4}
	\end{equation}
	where the dimensionless transport kernel is
	\begin{equation}
		\Phi_{ii}^{s}(\varepsilon_\mu)
		=
		\frac{1}{(2\pi)^2}
		\int_{\mathrm{BZ}}
		\left(\frac{\partial \bar{E}^{s}}{\partial k_i}\right)^2
		\delta\left[\varepsilon_\mu-\bar{E}^{s}(\mathbf{k})\right]
		dk_xdk_y .
		\tag{IV-5} \label{IV-5}
	\end{equation}
	
	\qquad For the spin-up channel, the kernel reduces to
	\begin{equation}
		\Phi_{xx}^{\uparrow}(\varepsilon_\mu)
		=
		\frac{1}{\pi^2}
		\int_0^\pi
		\sqrt{1-(\varepsilon_\mu+\gamma\cos\xi)^2}\,
		\Theta\left[1-(\varepsilon_\mu+\gamma\cos\xi)^2\right] \, d\xi,
		\tag{IV-6} \label{IV-6}
	\end{equation}
	and
	\begin{equation}
		\Phi_{yy}^{\uparrow}(\varepsilon_\mu)
		=
		\frac{\gamma^2}{\pi^2}
		\int_0^\pi
		\frac{\sin^2\xi}{\sqrt{1-(\varepsilon_\mu+\gamma\cos\xi)^2}}\,
		\Theta\left[1-(\varepsilon_\mu+\gamma\cos\xi)^2\right] \, d\xi .
		\tag{IV-7} \label{IV-7}
	\end{equation}
	Here, $\xi$ is only an integration variable, and the step functions restrict the integration to the real Fermi-surface roots.
	
	\qquad The $[C_2\parallel C_{4z}]$ symmetry gives $\Phi_{xx}^{\uparrow}=\Phi_{yy}^{\downarrow}$ and $\Phi_{yy}^{\uparrow}=\Phi_{xx}^{\downarrow}$, or equivalently $\sigma_{xx}^{\uparrow}=\sigma_{yy}^{\downarrow}\equiv\sigma_{\mathrm{maj}}$ and $\sigma_{yy}^{\uparrow}=\sigma_{xx}^{\downarrow}\equiv\sigma_{\mathrm{min}}$. The axial spin polarization is therefore
	\begin{equation}
		\eta_0 =
		\frac{\sigma_{\mathrm{maj}}-\sigma_{\mathrm{min}}}
		{\sigma_{\mathrm{maj}}+\sigma_{\mathrm{min}}}
		=
		\frac{\Phi_{xx}^{\uparrow}-\Phi_{yy}^{\uparrow}}
		{\Phi_{xx}^{\uparrow}+\Phi_{yy}^{\uparrow}} .
		\tag{IV-8} \label{IV-8}
	\end{equation}
	The common prefactor in the Boltzmann conductivity cancels in this ratio, so within the CRTA, $\eta_0$ depends only on the reduced chemical potential $\varepsilon_\mu$ and the anisotropy ratio $\gamma$. This provides the analytical basis for the transport phase diagram $\eta_0(\varepsilon_\mu,\gamma)$ in Fig. 1j of the main text. Representative conductivities and axial polarizations for $\gamma=0.2$ and the ideal limit $\gamma=0$ are shown in Fig.~\ref{fig:S3}, illustrating that the Lifshitz-delimited open-line regime corresponds to the high-$\eta_0$ transport window.
		
	\qquad At the band center $\varepsilon_\mu=0$, the kernels reduce to complete elliptic integrals,
	\[
	\Phi_{xx}^{\uparrow}(0)=\frac{2}{\pi^2}E(\gamma^2),
	\qquad
	\Phi_{yy}^{\uparrow}(0)=
	\frac{2}{\pi^2}
	\left[
	E(\gamma^2)-(1-\gamma^2)K(\gamma^2)
	\right],
	\]
	where $K(m)$ and $E(m)$ denote the complete elliptic integrals of the first and second kinds, respectively. Substitution gives
	\begin{equation}
		\eta_0(0)
		=
		\frac{(1-\gamma^2)K(\gamma^2)}
		{2E(\gamma^2)-(1-\gamma^2)K(\gamma^2)} .
		\tag{IV-9} \label{IV-9}
	\end{equation}
	For $\gamma\ll1$, this reduces to
	\begin{equation}
		\eta_0(0)=1-\gamma^2+\mathcal{O}(\gamma^4),
		\tag{IV-10} \label{IV-10}
	\end{equation}
	showing that finite transverse hopping only quadratically suppresses the ideal 100\% spin polarization at the band center.
	
		\begin{figure}[H]
			\centering
			\includegraphics[width=0.75\linewidth]{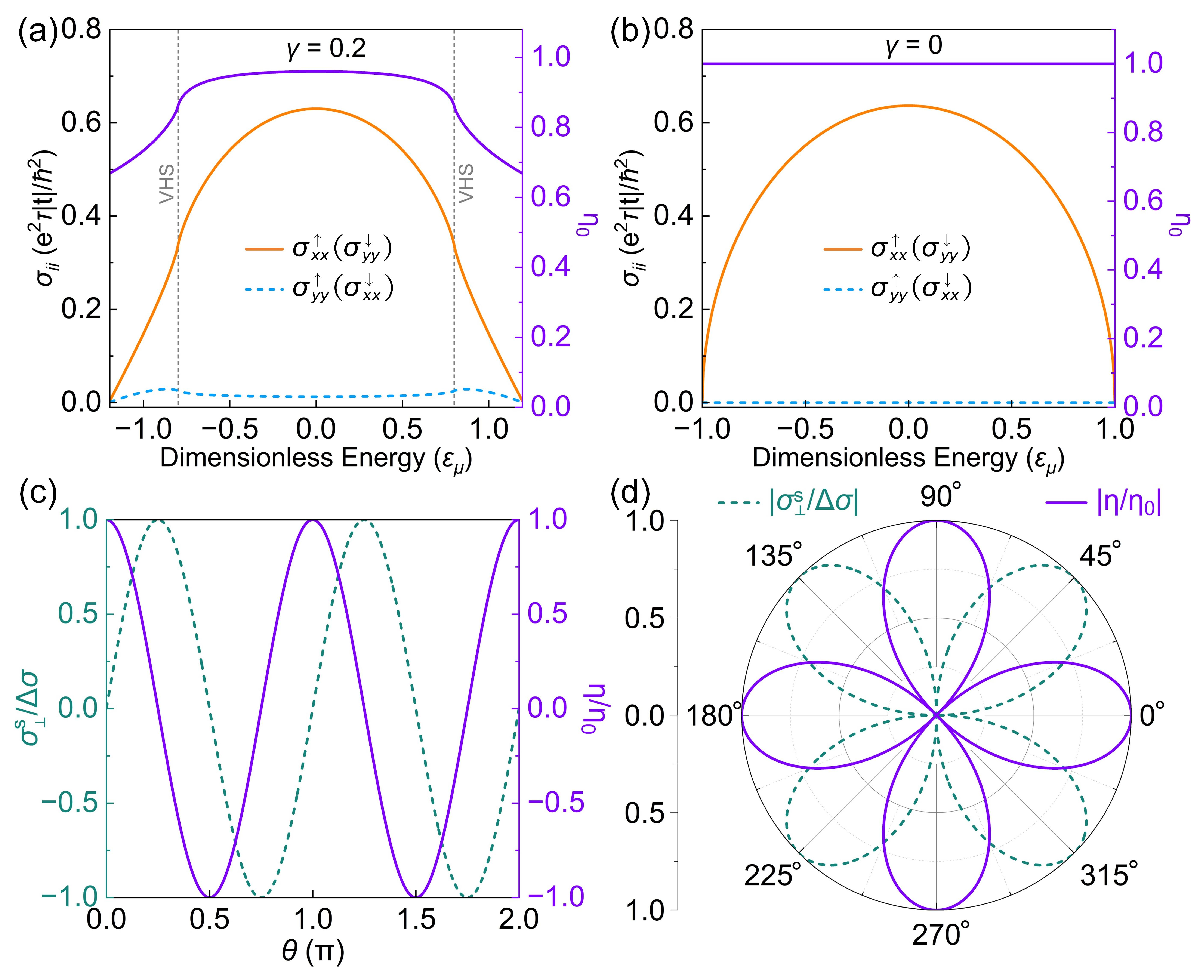}
			\caption{Energy- and angle-dependent transport responses of the anisotropic model. Spin-resolved conductivities and axial spin polarization for (a) $\gamma=0.2$ and (b) the ideal decoupled-chain limit $\gamma=0$. (c) Signed angular responses of the longitudinal spin polarization and transverse pure-spin conductivity, with their corresponding polar magnitudes shown in (d).}
			\label{fig:S3}
		\end{figure}
	
	\subsubsection*{IV.2 Tensor rotation and charge-to-spin conversion}
	
	\qquad To obtain the response for an arbitrary in-plane electric-field direction, we introduce a field-aligned frame $(x',y')$, where $x'\parallel\mathbf{E}$ and $x'$ is rotated by an angle $\theta$ from the principal $x$-axis. The tensor rotation is defined by
	\begin{equation}
		\bm{\sigma}^{s\prime}=R(\theta)\bm{\sigma}^{s}R^{T}(\theta),
		\qquad
		R(\theta)=
		\begin{pmatrix}
			\cos\theta & -\sin\theta \\
			\sin\theta & \cos\theta
		\end{pmatrix}.
		\tag{IV-11} \label{IV-11}
	\end{equation}
	Using $\bm{\sigma}^{\uparrow}=\mathrm{diag}(\sigma_{\mathrm{maj}},\sigma_{\mathrm{min}})$ and $\bm{\sigma}^{\downarrow}=\mathrm{diag}(\sigma_{\mathrm{min}},\sigma_{\mathrm{maj}})$, and defining $\Delta\sigma=\sigma_{\mathrm{maj}}-\sigma_{\mathrm{min}}$, one obtains
	\begin{equation}
		\bm{\sigma}^{\uparrow\prime}
		=
		\begin{pmatrix}
			\sigma_{\mathrm{maj}}\cos^2\theta+\sigma_{\mathrm{min}}\sin^2\theta
			&
			\Delta\sigma\sin\theta\cos\theta
			\\
			\Delta\sigma\sin\theta\cos\theta
			&
			\sigma_{\mathrm{maj}}\sin^2\theta+\sigma_{\mathrm{min}}\cos^2\theta
		\end{pmatrix},
		\tag{IV-12} \label{IV-12}
	\end{equation}
	and
	\begin{equation}
		\bm{\sigma}^{\downarrow\prime}
		=
		\begin{pmatrix}
			\sigma_{\mathrm{min}}\cos^2\theta+\sigma_{\mathrm{maj}}\sin^2\theta
			&
			-\Delta\sigma\sin\theta\cos\theta
			\\
			-\Delta\sigma\sin\theta\cos\theta
			&
			\sigma_{\mathrm{min}}\sin^2\theta+\sigma_{\mathrm{maj}}\cos^2\theta
		\end{pmatrix}.
		\tag{IV-13} \label{IV-13}
	\end{equation}	
	\qquad From the diagonal elements, the longitudinal charge conductivity is angle-independent, $\sigma_{\parallel}^{c}\equiv\sigma_{x'x'}^{\uparrow}+\sigma_{x'x'}^{\downarrow}=\sigma_{\mathrm{maj}}+\sigma_{\mathrm{min}}$. The residual longitudinal spin polarization is therefore
	\begin{equation}
		\eta(\theta)
		=
		\frac{
			\sigma_{x'x'}^{\uparrow}
			-
			\sigma_{x'x'}^{\downarrow}
		}{
			\sigma_{x'x'}^{\uparrow}
			+
			\sigma_{x'x'}^{\downarrow}
		}
		=
		\eta_0\cos2\theta .
		\tag{IV-14} \label{IV-14}
	\end{equation}
	
	\qquad From the off-diagonal elements, the transverse charge response cancels exactly, $\sigma_{\perp}^{c}\equiv\sigma_{y'x'}^{\uparrow}+\sigma_{y'x'}^{\downarrow}=0$, while the signed transverse spin conductivity is $\sigma_{\perp}^{s}\equiv \sigma_{y'x'}^{\uparrow}-\sigma_{y'x'}^{\downarrow}=\Delta\sigma \sin2\theta$. The corresponding charge-to-spin conversion efficiency is
	\begin{equation}
		\theta_{\mathrm{SH}}
		\equiv
		\left|
		\frac{\sigma_{\perp}^{s}}
		{\sigma_{\parallel}^{c}}
		\right|
		=
		\frac{{\Delta\sigma}|\sin2\theta|}
		{\sigma_{\mathrm{maj}}+\sigma_{\mathrm{min}}}
		=
		\eta_0|\sin2\theta|.
		\tag{IV-15} \label{IV-15}
	\end{equation}
	The normalized angular responses,
		$\eta/\eta_0=\cos 2\theta$ and
		$\sigma_{\perp}^{s}/\Delta\sigma=\sin 2\theta$,
		are shown in Fig.~S3(c). Their polar magnitudes in Fig.~S3(d) directly
		visualize the fourfold directional pattern imposed by the
		$[C_2\Vert C_{4z}]$ symmetry. These results show that diagonal driving (i.e., $\theta=\pi/4$ (mod $\pi/2$)) gives $\eta=0$ and $\theta_{\mathrm{SH}}=\eta_0$, whereas principal-axis driving gives ${|\eta|}=\eta_0$ and $\theta_{\mathrm{SH}}=0$. 	
	
	% Section V 
	\clearpage
	\begin{center}
		{\large\bfseries\boldmath V. Computational Methods}
	\end{center}
	
	\qquad Our first-principles calculations were performed using the Vienna \emph{
		Ab initio} Simulation Package (VASP)~\textcolor{blue}{[\hyperlink{SM-ref-vasp}{3}]}. To describe the electron-ion interactions, the projector augmented wave (PAW) ~\textcolor{blue}{[\hyperlink{SM-ref-paw}{4}]} method was employed with a plane-wave basis set and a kinetic energy cutoff of 500~eV. For the exchange-correlation functional, we adopted the Perdew-Burke-Ernzerhof (PBE)~\textcolor{blue}{[\hyperlink{SM-ref-pbe}{5}]} formulation of the generalized gradient approximation (GGA). The BZ was sampled using a Monkhorst-Pack $\mathbf{k}$-point mesh~\textcolor{blue}{[\hyperlink{SM-ref-monkhorst}{6}]} with a spacing of $2\pi\times0.01$~\AA$^{-1}$. The electronic self-consistent field calculations were converged to a tolerance of $10^{-8}$~eV. Structural relaxations were performed until the Hellmann-Feynman forces on all atoms were less than 0.01~eV/\AA. The Hubbard $U$ parameter was determined by systematically testing values from 1 to 6~eV. The optimal $U$ values for different materials were selected by comparing the GGA-PBE$+U$ electronic structures with reference HSE06 hybrid-functional calculations~\textcolor{blue}{[\hyperlink{SM-ref-hse06}{7}]}. Using the validated $U$ parameters, we first computed the electronic band structures within the GGA-PBE$+U$ framework. The corresponding electronic bands were then used as input for constructing maximally localized Wannier functions (MLWFs) via the WANNIER90 package~\textcolor{blue}{[\hyperlink{SM-ref-wannier90}{8}, \hyperlink{SM-ref-wannier90update}{9}]}. Using the Wannier-interpolated first-principles bands, spin-resolved CRTA transport ~\textcolor{blue}{[\hyperlink{SM-ref-Pizzi2014}{1}]} was evaluated to provide the material-specific band-structure baseline for the velocity anisotropy and spin-channel contrast analyzed in Sec.~VI.
	
	\qquad To include SOC in the spin-conductivity calculations, we performed SOC-included Kubo calculations based on the Wannier-interpolated Hamiltonian. Following the constant-$\Gamma$ Kubo formalism for altermagnetic spin-splitter responses~\textcolor{blue}{[\hyperlink{SM-ref-Smejkal2021}{10}, \hyperlink{SM-ref-Freimuth2014PRB}{11}]}, the spin conductivity was separated into $\mathcal{T}$-odd and $\mathcal{T}$-even components. The $\mathcal{T}$-odd component was evaluated as
	\begin{equation}
		\sigma_{bc}^{\mathcal{T}\text{-}\mathrm{odd},a}
		=
		-\frac{e\hbar}{V\pi}
		\operatorname{Re}
		\sum_{\mathbf{k},m,n}
		\frac{
			\langle u_{n}(\mathbf{k})|\hat{J}_{b}^{a}|u_{m}(\mathbf{k})\rangle
			\langle u_{m}(\mathbf{k})|\hat{v}_{c}|u_{n}(\mathbf{k})\rangle
			\Gamma^2
		}{
			\left[(E_F-E_{n}(\mathbf{k}))^2+\Gamma^2\right]
			\left[(E_F-E_{m}(\mathbf{k}))^2+\Gamma^2\right]
		}.
		\tag{V-1} \label{V-1}
	\end{equation}
	The $\mathcal{T}$-even component in the clean-limit form is
	\begin{equation}
		\sigma_{bc}^{\mathcal{T}\text{-}\mathrm{even},a}
		=
		\frac{2e\hbar}{V}
		\operatorname{Im}
		\sum_{\mathbf{k},m\ne n}^{\substack{n\,\mathrm{occ.}\\ m\,\mathrm{unocc.}}}
		\frac{
			\langle u_{n}(\mathbf{k})|\hat{J}_{b}^{a}|u_{m}(\mathbf{k})\rangle
			\langle u_{m}(\mathbf{k})|\hat{v}_{c}|u_{n}(\mathbf{k})\rangle
		}{
			\left[E_{n}(\mathbf{k})-E_{m}(\mathbf{k})\right]^2
		}.
		\tag{V-2} \label{V-2}
	\end{equation}
	Here, $|u_{n\mathbf{k}}\rangle$ is the Bloch state of band $n$ at wave vector $\mathbf{k}$, $E_{n\mathbf{k}}$ is the corresponding band energy, $E_F$ is the Fermi energy, $V$ is the crystal volume, $\hat{v}_{c}$ is the velocity operator along the electric-field direction $c$, and $e$ denotes the positive elementary charge. The spin-current operator is $\hat{J}_{b}^{a}=\frac{1}{2}\{\hat{s}_{a},\hat{v}_{b}\}$, where $a$ denotes the spin-polarization direction and $b$ denotes the spin-current-flow direction. The $\mathcal{T}$-odd component captures the altermagnetic contribution associated with SADL, whereas the $\mathcal{T}$-even component represents the conventional SOC-driven background. The Kubo sums were evaluated using the Wannier linear-response code~\textcolor{blue}{[\hyperlink{SM-ref-ZeleznyWannierLR}{12}]}. The broadening parameter $\Gamma$ in Eq.~\eqref{V-1} was varied to examine finite-lifetime effects on the SADL response in the presence of SOC.
	
	\qquad To go beyond the constant-lifetime approximations used in the CRTA and constant-$\Gamma$ Kubo calculations, we further evaluated momentum-dependent scattering using the AMSET code~\textcolor{blue}{[\hyperlink{SM-ref-amset}{13}]}. AMSET computes state-dependent scattering rates from first-principles inputs using Fermi's golden rule. For an elastic process, the scattering rate from an initial state $n\mathbf{k}$ to a final state $m,\mathbf{k}+\mathbf{q}$ is written as
	\begin{equation}
		\tau_{n\mathbf{k}\rightarrow m,\mathbf{k}+\mathbf{q}}^{-1}
		=
		\frac{2\pi}{\hbar}
		\left|g_{nm}(\mathbf{k},\mathbf{q})\right|^2
		\delta\left(
		\varepsilon_{n\mathbf{k}}
		-
		\varepsilon_{m,\mathbf{k}+\mathbf{q}}
		\right),
		\tag{V-3} \label{V-3}
	\end{equation}
	with analogous expressions for inelastic scattering processes. In this work, we included acoustic deformation-potential scattering (ADP), polar optical phonon scattering (POP), and ionized impurity scattering (IMP). The resulting calculations provide momentum- and band-dependent relaxation times beyond a single constant lifetime, and were used to assess how realistic scattering affects both the absolute conductivities and the spin-channel contrast predicted by the constant-lifetime analysis.
	
	% Section VI
	\newpage
	\begin{center}
		\large\bfseries {VI. Realization of SADL in Two-Dimensional Materials}
	\end{center}
	
	\subsubsection*{{VI.1 Screening of two-dimensional SADL candidates}}
	
	\qquad Motivated by the experimental synthesis of layered bulk materials (e.g., Cu$_2$MX$_4$ and Ag$_2$MX$_4$) and their potential for exfoliation into 2D monolayers, we focus on the broader TM$_2$MX$_4$ family, which shares the same lattice model as that shown in Fig. 1a. To systematically explore the vast chemical space, we consider a broad combination of elements: TM includes 3$d$, 4$d$, and 5$d$ transition metals; M is Mo or W; and X is selected from 15 non-metal elements (O, S, Se, Te, F, Cl, Br, I, N, P, As, Sb, B, C and Si). 
	
	\qquad All candidate compositions were screened to ensure charge neutrality according to the valence-balance rule:
	\[
	2 \times V_{\mathrm{TM}} + V_{\mathrm{M}} + 4 \times V_{\mathrm{X}} = 0,
	\]
	where $V_{\mathrm{TM}}$, $V_{\mathrm{M}}$, and $V_{\mathrm{X}}$ represent the typical oxidation states of the corresponding elements, as summarized in Table~\ref{tab:S1}. This initial screening yielded 442 charge-balanced compositions as listed in Table~\ref{tab:S2}.
	
	\begin{figure}[H]
		\centering
		\includegraphics[width=0.9\textwidth]{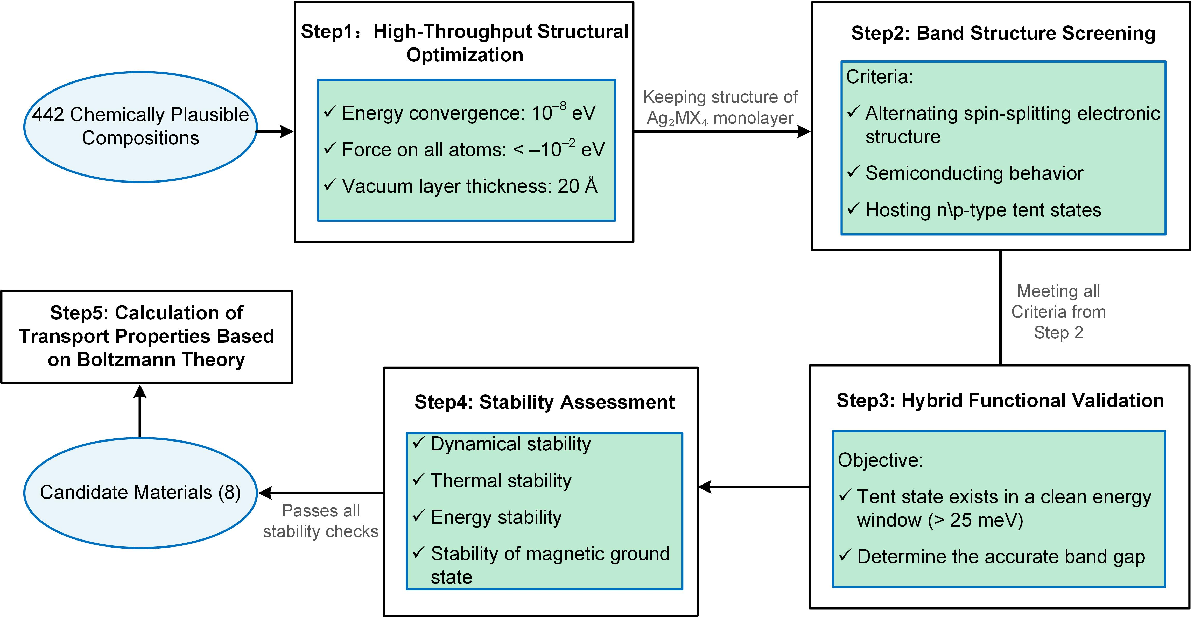}
		\caption{{Computational screening workflow for 2D tent-state materials, progressing from high-throughput structural optimization through band structure screening, hybrid functional validation, and stability assessment, culminating in the calculation of transport properties via Boltzmann theory.}}
		\label{fig:S4}
	\end{figure}
	
	\qquad Subsequently, we performed a hierarchical computational screening of the 442 initial structures, as illustrated in Fig.~\ref{fig:S4}. The workflow began with structural optimization and antiferromagnetic band structure calculations using the GGA-PBE$+U$ method. Candidate structures exhibiting tent states were then identified based on fundamental screening criteria. These promising candidates were further validated through electronic structure calculations employing the more accurate HSE06 functional. Finally, we conducted systematic stability assessments of the validated systems, including magnetic state calculations under varying strain conditions, carrier concentrations, and $U$ parameters, complemented by formation energy calculations and phonon dispersion analysis.

	\begin{table}[H]
		\centering
		\caption{Typical oxidation states of 3d, 4d, and 5d transition metals, and X elements.}
		\label{tab:S1}
		\small
		% 经过精确计算的列宽，总宽度精准等于 16.5cm (完美适配页宽)
		\begin{tabular}{|c|c|c|c|c|c|}
			\hline
			& \textbf{Elements} & \textbf{Oxidation States} & & \textbf{Elements} & \textbf{Oxidation States} \\
			\hline
			\multirow{10}{*}{3d} & Sc & $+$3 & \multirow{10}{*}{4d} & Y & $+$3 \\
			\cline{2-3}\cline{5-6}
			& Ti & $+$2, $+$3, $+$4 & & Zr & $+$3, $+$4 \\
			\cline{2-3}\cline{5-6}
			& V & $+$2, $+$3, $+$4, $+$5 & & Nb & $+$3, $+$4, $+$5 \\
			\cline{2-3}\cline{5-6}
			& Cr & $+$2, $+$3, $+$5, $+$6 & & Mo & $+$2, $+$3, $+$4, $+$5, $+$6 \\
			\cline{2-3}\cline{5-6}
			& Mn & $+$2, $+$3, $+$4, $+$5, $+$6, $+$7 & & Tc & $+$4, $+$5, $+$7 \\
			\cline{2-3}\cline{5-6}
			& Fe & $+$2, $+$3 & & Ru & $+$3, $+$4, $+$5, $+$6, $+$7, $+$8 \\
			\cline{2-3}\cline{5-6}
			& Co & $+$2, $+$3 & & Rh & $+$3, $+$4 \\
			\cline{2-3}\cline{5-6}
			& Ni & $+$2, $+$3 & & Pd & $+$2, $+$3, $+$4 \\
			\cline{2-3}\cline{5-6}
			& Cu & $+$1, $+$2 & & Ag & $+$1 \\
			\cline{2-3}\cline{5-6}
			& Zn & $+$2 & & Cd & $+$2 \\
			\hline
			\multirow{9}{*}{5d} & Hf & $+$3, $+$4 & \multirow{9}{*}{\begin{tabular}[c]{@{}c@{}}X\\ Elements\end{tabular}} & \multirow{2}{*}{O, S, Se, Te} & \multirow{2}{*}{$-$2} \\
			\cline{2-3}
			& Ta & $+$3, $+$4, $+$5 & & & \\
			\cline{2-3}\cline{5-6}
			& W & $+$4, $+$5, $+$6 & & \multirow{2}{*}{F, Cl, Br, I} & \multirow{2}{*}{$-$1} \\
			\cline{2-3}
			& Re & $+$4, $+$5, $+$6, $+$7 & & & \\
			\cline{2-3}\cline{5-6}
			& Os & $+$3, $+$4, $+$5, $+$6, $+$7, $+$8 & & \multirow{2}{*}{N, P, As, Sb, B} & \multirow{2}{*}{$-$3} \\
			\cline{2-3}
			& Ir & $+$3, $+$4, $+$5, $+$6 & & & \\
			\cline{2-3}\cline{5-6}
			& Pt & $+$2, $+$3, $+$4 & & \multirow{3}{*}{C, Si} & \multirow{3}{*}{$-$4} \\
			\cline{2-3}
			& Au & $+$1, $+$3 & & & \\
			\cline{2-3}
			& Hg & $+$1, $+$2 & & & \\
			\hline
		\end{tabular}
	\end{table}

	\begin{longtable}{|c|c|c|c|c|c|c|c|}
		\caption{Charge-balanced compositions following the TM$_2$MX$_4$ formula.} 
		\label{tab:S2} \\		
		\hline
		\textbf{No.} & \textbf{Materials} & \textbf{No.} & \textbf{Materials} & \textbf{No.} & \textbf{Materials} & \textbf{No.} & \textbf{Materials} \\
		\hline
		\endfirsthead
		\hline
		\endhead
		1 & Ag$^{+1}_2$Mo$^{+6}$O$^{-2}_4$ & 2 & Ag$^{+1}_2$Mo$^{+6}$S$^{-2}_4$ & 3 & Ag$^{+1}_2$Mo$^{+6}$Se$^{-2}_4$ & 4 & Ag$^{+1}_2$Mo$^{+6}$Te$^{-2}_4$ \\
		\hline
		5 & Ag$^{+1}_2$W$^{+6}$O$^{-2}_4$ & 6 & Ag$^{+1}_2$W$^{+6}$S$^{-2}_4$ & 7 & Ag$^{+1}_2$W$^{+6}$Se$^{-2}_4$ & 8 & Ag$^{+1}_2$W$^{+6}$Te$^{-2}_4$ \\
		\hline
		9 & Au$^{+1}_2$Mo$^{+6}$O$^{-2}_4$ & 10 & Au$^{+1}_2$Mo$^{+6}$S$^{-2}_4$ & 11 & Au$^{+1}_2$Mo$^{+6}$Se$^{-2}_4$ & 12 & Au$^{+1}_2$Mo$^{+6}$Te$^{-2}_4$ \\
		\hline
		13 & Au$^{+1}_2$W$^{+6}$O$^{-2}_4$ & 14 & Au$^{+1}_2$W$^{+6}$S$^{-2}_4$ & 15 & Au$^{+1}_2$W$^{+6}$Se$^{-2}_4$ & 16 & Au$^{+1}_2$W$^{+6}$Te$^{-2}_4$ \\
		\hline
		17 & Cd$^{+2}_2$Mo$^{+4}$O$^{-2}_4$ & 18 & Cd$^{+2}_2$Mo$^{+4}$S$^{-2}_4$ & 19 & Cd$^{+2}_2$Mo$^{+4}$Se$^{-2}_4$ & 20 & Cd$^{+2}_2$Mo$^{+4}$Te$^{-2}_4$ \\
		\hline
		21 & Cd$^{+2}_2$W$^{+4}$O$^{-2}_4$ & 22 & Cd$^{+2}_2$W$^{+4}$S$^{-2}_4$ & 23 & Cd$^{+2}_2$W$^{+4}$Se$^{-2}_4$ & 24 & Cd$^{+2}_2$W$^{+4}$Te$^{-2}_4$ \\
		\hline
		25 & Co$^{+2}_2$Mo$^{+4}$O$^{-2}_4$ & 26 & Co$^{+2}_2$Mo$^{+4}$S$^{-2}_4$ & 27 & Co$^{+2}_2$Mo$^{+4}$Se$^{-2}_4$ & 28 & Co$^{+2}_2$Mo$^{+4}$Te$^{-2}_4$ \\
		\hline
		29 & Co$^{+2}_2$W$^{+4}$O$^{-2}_4$ & 30 & Co$^{+2}_2$W$^{+4}$S$^{-2}_4$ & 31 & Co$^{+2}_2$W$^{+4}$Se$^{-2}_4$ & 32 & Co$^{+2}_2$W$^{+4}$Te$^{-2}_4$ \\
		\hline
		33 & Cr$^{+2}_2$Mo$^{+4}$O$^{-2}_4$ & 34 & Cr$^{+2}_2$Mo$^{+4}$S$^{-2}_4$ & 35 & Cr$^{+2}_2$Mo$^{+4}$Se$^{-2}_4$ & 36 & Cr$^{+2}_2$Mo$^{+4}$Te$^{-2}_4$ \\
		\hline
		37 & Cr$^{+2}_2$W$^{+4}$O$^{-2}_4$ & 38 & Cr$^{+2}_2$W$^{+4}$S$^{-2}_4$ & 39 & Cr$^{+2}_2$W$^{+4}$Se$^{-2}_4$ & 40 & Cr$^{+2}_2$W$^{+4}$Te$^{-2}_4$ \\
		\hline
		41 & Cu$^{+2}_2$Mo$^{+4}$O$^{-2}_4$ & 42 & Cu$^{+2}_2$Mo$^{+4}$S$^{-2}_4$ & 43 & Cu$^{+2}_2$Mo$^{+4}$Se$^{-2}_4$ & 44 & Cu$^{+2}_2$Mo$^{+4}$Te$^{-2}_4$ \\
		\hline
		45 & Cu$^{+2}_2$W$^{+4}$O$^{-2}_4$ & 46 & Cu$^{+2}_2$W$^{+4}$S$^{-2}_4$ & 47 & Cu$^{+2}_2$W$^{+4}$Se$^{-2}_4$ & 48 & Cu$^{+2}_2$W$^{+4}$Te$^{-2}_4$ \\
		\hline
		49 & Fe$^{+2}_2$Mo$^{+4}$O$^{-2}_4$ & 50 & Fe$^{+2}_2$Mo$^{+4}$S$^{-2}_4$ & 51 & Fe$^{+2}_2$Mo$^{+4}$Se$^{-2}_4$ & 52 & Fe$^{+2}_2$Mo$^{+4}$Te$^{-2}_4$ \\
		\hline
		53 & Fe$^{+2}_2$W$^{+4}$O$^{-2}_4$ & 54 & Fe$^{+2}_2$W$^{+4}$S$^{-2}_4$ & 55 & Fe$^{+2}_2$W$^{+4}$Se$^{-2}_4$ & 56 & Fe$^{+2}_2$W$^{+4}$Te$^{-2}_4$ \\
		\hline
		57 & Hg$^{+1}_2$Mo$^{+4}$O$^{-2}_4$ & 58 & Hg$^{+1}_2$Mo$^{+4}$S$^{-2}_4$ & 59 & Hg$^{+1}_2$Mo$^{+4}$Se$^{-2}_4$ & 60 & Hg$^{+1}_2$Mo$^{+4}$Te$^{-2}_4$ \\
		\hline
		61 & Hg$^{+1}_2$W$^{+4}$O$^{-2}_4$ & 62 & Hg$^{+1}_2$W$^{+4}$S$^{-2}_4$ & 63 & Hg$^{+1}_2$W$^{+4}$Se$^{-2}_4$ & 64 & Hg$^{+1}_2$W$^{+4}$Te$^{-2}_4$ \\
		\hline
		65 & Mn$^{+2}_2$Mo$^{+4}$O$^{-2}_4$ & 66 & Mn$^{+2}_2$Mo$^{+4}$S$^{-2}_4$ & 67 & Mn$^{+2}_2$Mo$^{+4}$Se$^{-2}_4$ & 68 & Mn$^{+2}_2$Mo$^{+4}$Te$^{-2}_4$ \\
		\hline
		69 & Mn$^{+2}_2$W$^{+4}$O$^{-2}_4$ & 70 & Mn$^{+2}_2$W$^{+4}$S$^{-2}_4$ & 71 & Mn$^{+2}_2$W$^{+4}$Se$^{-2}_4$ & 72 & Mn$^{+2}_2$W$^{+4}$Te$^{-2}_4$ \\
		\hline
		73 & Nb$^{+3}_2$Mo$^{+2}$O$^{-2}_4$ & 74 & Nb$^{+3}_2$Mo$^{+2}$S$^{-2}_4$ & 75 & Nb$^{+3}_2$Mo$^{+2}$Se$^{-2}_4$ & 76 & Nb$^{+3}_2$Mo$^{+2}$Te$^{-2}_4$ \\
		\hline
		77 & Nb$^{+3}_2$W$^{+2}$O$^{-2}_4$ & 78 & Nb$^{+3}_2$W$^{+2}$S$^{-2}_4$ & 79 & Nb$^{+3}_2$W$^{+2}$Se$^{-2}_4$ & 80 & Nb$^{+3}_2$W$^{+2}$Te$^{-2}_4$ \\
		\hline
		81 & Ni$^{+2}_2$Mo$^{+4}$O$^{-2}_4$ & 82 & Ni$^{+2}_2$Mo$^{+4}$S$^{-2}_4$ & 83 & Ni$^{+2}_2$Mo$^{+4}$Se$^{-2}_4$ & 84 & Ni$^{+2}_2$Mo$^{+4}$Te$^{-2}_4$ \\
		\hline
		85 & Ni$^{+2}_2$W$^{+4}$O$^{-2}_4$ & 86 & Ni$^{+2}_2$W$^{+4}$S$^{-2}_4$ & 87 & Ni$^{+2}_2$W$^{+4}$Se$^{-2}_4$ & 88 & Ni$^{+2}_2$W$^{+4}$Te$^{-2}_4$ \\
		\hline
		89 & Pd$^{+2}_2$Mo$^{+4}$O$^{-2}_4$ & 90 & Pd$^{+2}_2$Mo$^{+4}$S$^{-2}_4$ & 91 & Pd$^{+2}_2$Mo$^{+4}$Se$^{-2}_4$ & 92 & Pd$^{+2}_2$Mo$^{+4}$Te$^{-2}_4$ \\
		\hline
		93 & Pd$^{+2}_2$W$^{+4}$O$^{-2}_4$ & 94 & Pd$^{+2}_2$W$^{+4}$S$^{-2}_4$ & 95 & Pd$^{+2}_2$W$^{+4}$Se$^{-2}_4$ & 96 & Pd$^{+2}_2$W$^{+4}$Te$^{-2}_4$ \\
		\hline
		97 & Pt$^{+2}_2$Mo$^{+4}$O$^{-2}_4$ & 98 & Pt$^{+2}_2$Mo$^{+4}$S$^{-2}_4$ & 99 & Pt$^{+2}_2$Mo$^{+4}$Se$^{-2}_4$ & 100 & Pt$^{+2}_2$Mo$^{+4}$Te$^{-2}_4$ \\
		\hline
		101 & Pt$^{+2}_2$W$^{+4}$O$^{-2}_4$ & 102 & Pt$^{+2}_2$W$^{+4}$S$^{-2}_4$ & 103 & Pt$^{+2}_2$W$^{+4}$Se$^{-2}_4$ & 104 & Pt$^{+2}_2$W$^{+4}$Te$^{-2}_4$ \\
		\hline
		105 & Rh$^{+3}_2$Mo$^{+2}$O$^{-2}_4$ & 106 & Rh$^{+3}_2$Mo$^{+2}$S$^{-2}_4$ & 107 & Rh$^{+3}_2$Mo$^{+2}$Se$^{-2}_4$ & 108 & Rh$^{+3}_2$Mo$^{+2}$Te$^{-2}_4$ \\
		\hline
		109 & Rh$^{+3}_2$W$^{+2}$O$^{-2}_4$ & 110 & Rh$^{+3}_2$W$^{+2}$S$^{-2}_4$ & 111 & Rh$^{+3}_2$W$^{+2}$Se$^{-2}_4$ & 112 & Rh$^{+3}_2$W$^{+2}$Te$^{-2}_4$ \\
		\hline
		113 & Ru$^{+3}_2$Mo$^{+2}$O$^{-2}_4$ & 114 & Ru$^{+3}_2$Mo$^{+2}$S$^{-2}_4$ & 115 & Ru$^{+3}_2$Mo$^{+2}$Se$^{-2}_4$ & 116 & Ru$^{+3}_2$Mo$^{+2}$Te$^{-2}_4$ \\
		\hline
		117 & Ru$^{+3}_2$W$^{+2}$O$^{-2}_4$ & 118 & Ru$^{+3}_2$W$^{+2}$S$^{-2}_4$ & 119 & Ru$^{+3}_2$W$^{+2}$Se$^{-2}_4$ & 120 & Ru$^{+3}_2$W$^{+2}$Te$^{-2}_4$ \\
		\hline
		121 & Sc$^{+3}_2$Mo$^{+2}$O$^{-2}_4$ & 122 & Sc$^{+3}_2$Mo$^{+2}$S$^{-2}_4$ & 123 & Sc$^{+3}_2$Mo$^{+2}$Se$^{-2}_4$ & 124 & Sc$^{+3}_2$Mo$^{+2}$Te$^{-2}_4$ \\
		\hline
		125 & Sc$^{+3}_2$W$^{+2}$O$^{-2}_4$ & 126 & Sc$^{+3}_2$W$^{+2}$S$^{-2}_4$ & 127 & Sc$^{+3}_2$W$^{+2}$Se$^{-2}_4$ & 128 & Sc$^{+3}_2$W$^{+2}$Te$^{-2}_4$ \\
		\hline
		129 & Ta$^{+3}_2$Mo$^{+2}$O$^{-2}_4$ & 130 & Ta$^{+3}_2$Mo$^{+2}$S$^{-2}_4$ & 131 & Ta$^{+3}_2$Mo$^{+2}$Se$^{-2}_4$ & 132 & Ta$^{+3}_2$Mo$^{+2}$Te$^{-2}_4$ \\
		\hline
		133 & Ta$^{+3}_2$W$^{+2}$O$^{-2}_4$ & 134 & Ta$^{+3}_2$W$^{+2}$S$^{-2}_4$ & 135 & Ta$^{+3}_2$W$^{+2}$Se$^{-2}_4$ & 136 & Ta$^{+3}_2$W$^{+2}$Te$^{-2}_4$ \\
		\hline
		137 & Ti$^{+3}_2$Mo$^{+2}$O$^{-2}_4$ & 138 & Ti$^{+3}_2$Mo$^{+2}$S$^{-2}_4$ & 139 & Ti$^{+3}_2$Mo$^{+2}$Se$^{-2}_4$ & 140 & Ti$^{+3}_2$Mo$^{+2}$Te$^{-2}_4$ \\
		\hline
		141 & Ti$^{+3}_2$W$^{+2}$O$^{-2}_4$ & 142 & Ti$^{+3}_2$W$^{+2}$S$^{-2}_4$ & 143 & Ti$^{+3}_2$W$^{+2}$Se$^{-2}_4$ & 144 & Ti$^{+3}_2$W$^{+2}$Te$^{-2}_4$ \\
		\hline
		145 & V$^{+3}_2$Mo$^{+2}$O$^{-2}_4$ & 146 & V$^{+3}_2$Mo$^{+2}$S$^{-2}_4$ & 147 & V$^{+3}_2$Mo$^{+2}$Se$^{-2}_4$ & 148 & V$^{+3}_2$Mo$^{+2}$Te$^{-2}_4$ \\
		\hline
		149 & V$^{+3}_2$W$^{+2}$O$^{-2}_4$ & 150 & V$^{+3}_2$W$^{+2}$S$^{-2}_4$ & 151 & V$^{+3}_2$W$^{+2}$Se$^{-2}_4$ & 152 & V$^{+3}_2$W$^{+2}$Te$^{-2}_4$ \\
		\hline
		153 & Y$^{+3}_2$Mo$^{+2}$O$^{-2}_4$ & 154 & Y$^{+3}_2$Mo$^{+2}$S$^{-2}_4$ & 155 & Y$^{+3}_2$Mo$^{+2}$Se$^{-2}_4$ & 156 & Y$^{+3}_2$Mo$^{+2}$Te$^{-2}_4$ \\
		\hline
		157 & Y$^{+3}_2$W$^{+2}$O$^{-2}_4$ & 158 & Y$^{+3}_2$W$^{+2}$S$^{-2}_4$ & 159 & Y$^{+3}_2$W$^{+2}$Se$^{-2}_4$ & 160 & Y$^{+3}_2$W$^{+2}$Te$^{-2}_4$ \\
		\hline
		161 & Zn$^{+2}_2$Mo$^{+4}$O$^{-2}_4$ & 162 & Zn$^{+2}_2$Mo$^{+4}$S$^{-2}_4$ & 163 & Zn$^{+2}_2$Mo$^{+4}$Se$^{-2}_4$ & 164 & Zn$^{+2}_2$Mo$^{+4}$Te$^{-2}_4$ \\
		\hline
		165 & Zn$^{+2}_2$W$^{+4}$O$^{-2}_4$ & 166 & Zn$^{+2}_2$W$^{+4}$S$^{-2}_4$ & 167 & Zn$^{+2}_2$W$^{+4}$Se$^{-2}_4$ & 168 & Zn$^{+2}_2$W$^{+4}$Te$^{-2}_4$ \\
		\hline
		169 & Ag$^{+1}_2$Mo$^{+2}$Cl$^{-1}_4$ & 170 & Ag$^{+1}_2$Mo$^{+2}$Br$^{-1}_4$ & 171 & Ag$^{+1}_2$Mo$^{+2}$F$^{-1}_4$ & 172 & Ag$^{+1}_2$Mo$^{+2}$I$^{-1}_4$ \\
		\hline
		173 & Ag$^{+1}_2$W$^{+2}$Cl$^{-1}_4$ & 174 & Ag$^{+1}_2$W$^{+2}$Br$^{-1}_4$ & 175 & Ag$^{+1}_2$W$^{+2}$F$^{-1}_4$ & 176 & Ag$^{+1}_2$W$^{+2}$I$^{-1}_4$ \\
		\hline
		177 & Au$^{+1}_2$Mo$^{+2}$Cl$^{-1}_4$ & 178 & Au$^{+1}_2$Mo$^{+2}$Br$^{-1}_4$ & 179 & Au$^{+1}_2$Mo$^{+2}$F$^{-1}_4$ & 180 & Au$^{+1}_2$Mo$^{+2}$I$^{-1}_4$ \\
		\hline
		181 & Au$^{+1}_2$W$^{+2}$Cl$^{-1}_4$ & 182 & Au$^{+1}_2$W$^{+2}$Br$^{-1}_4$ & 183 & Au$^{+1}_2$W$^{+2}$F$^{-1}_4$ & 184 & Au$^{+1}_2$W$^{+2}$I$^{-1}_4$ \\
		\hline
		185 & Cu$^{+1}_2$Mo$^{+2}$Cl$^{-1}_4$ & 186 & Cu$^{+1}_2$Mo$^{+2}$Br$^{-1}_4$ & 187 & Cu$^{+1}_2$Mo$^{+2}$F$^{-1}_4$ & 188 & Cu$^{+1}_2$Mo$^{+2}$I$^{-1}_4$ \\
		\hline
		189 & Cu$^{+1}_2$W$^{+2}$Cl$^{-1}_4$ & 190 & Cu$^{+1}_2$W$^{+2}$Br$^{-1}_4$ & 191 & Cu$^{+1}_2$W$^{+2}$F$^{-1}_4$ & 192 & Cu$^{+1}_2$W$^{+2}$I$^{-1}_4$ \\
		\hline
		193 & Co$^{+3}_2$Mo$^{+6}$As$^{-3}_4$ & 194 & Co$^{+3}_2$Mo$^{+6}$B$^{-3}_4$ & 195 & Co$^{+3}_2$Mo$^{+6}$N$^{-3}_4$ & 196 & Co$^{+3}_2$Mo$^{+6}$P$^{-3}_4$ \\
		\hline
		197 & Co$^{+3}_2$Mo$^{+6}$Sb$^{-3}_4$ & 198 & Co$^{+3}_2$W$^{+6}$As$^{-3}_4$ & 199 & Co$^{+3}_2$W$^{+6}$B$^{-3}_4$ & 200 & Co$^{+3}_2$W$^{+6}$N$^{-3}_4$ \\
		\hline
		201 & Co$^{+3}_2$W$^{+6}$P$^{-3}_4$ & 202 & Co$^{+3}_2$W$^{+6}$Sb$^{-3}_4$ & 203 & Cr$^{+3}_2$Mo$^{+6}$As$^{-3}_4$ & 204 & Cr$^{+3}_2$Mo$^{+6}$B$^{-3}_4$ \\
		\hline
		205 & Cr$^{+3}_2$Mo$^{+6}$N$^{-3}_4$ & 206 & Cr$^{+3}_2$Mo$^{+6}$P$^{-3}_4$ & 207 & Cr$^{+3}_2$Mo$^{+6}$Sb$^{-3}_4$ & 208 & Cr$^{+3}_2$W$^{+6}$As$^{-3}_4$ \\
		\hline
		209 & Cr$^{+3}_2$W$^{+6}$B$^{-3}_4$ & 210 & Cr$^{+3}_2$W$^{+6}$N$^{-3}_4$ & 211 & Cr$^{+3}_2$W$^{+6}$P$^{-3}_4$ & 212 & Cr$^{+3}_2$W$^{+6}$Sb$^{-3}_4$ \\
		\hline
		213 & Fe$^{+3}_2$Mo$^{+6}$As$^{-3}_4$ & 214 & Fe$^{+3}_2$Mo$^{+6}$B$^{-3}_4$ & 215 & Fe$^{+3}_2$Mo$^{+6}$N$^{-3}_4$ & 216 & Fe$^{+3}_2$Mo$^{+6}$P$^{-3}_4$ \\
		\hline
		217 & Fe$^{+3}_2$Mo$^{+6}$Sb$^{-3}_4$ & 218 & Fe$^{+3}_2$W$^{+6}$As$^{-3}_4$ & 219 & Fe$^{+3}_2$W$^{+6}$B$^{-3}_4$ & 220 & Fe$^{+3}_2$W$^{+6}$N$^{-3}_4$ \\
		\hline
		221 & Fe$^{+3}_2$W$^{+6}$P$^{-3}_4$ & 222 & Fe$^{+3}_2$W$^{+6}$Sb$^{-3}_4$ & 223 & Hf$^{+3}_2$Mo$^{+6}$As$^{-3}_4$ & 224 & Hf$^{+3}_2$Mo$^{+6}$B$^{-3}_4$ \\
		\hline
		225 & Hf$^{+3}_2$Mo$^{+6}$N$^{-3}_4$ & 226 & Hf$^{+3}_2$Mo$^{+6}$P$^{-3}_4$ & 227 & Hf$^{+3}_2$Mo$^{+6}$Sb$^{-3}_4$ & 228 & Hf$^{+3}_2$W$^{+6}$As$^{-3}_4$ \\
		\hline
		229 & Hf$^{+3}_2$W$^{+6}$B$^{-3}_4$ & 230 & Hf$^{+3}_2$W$^{+6}$N$^{-3}_4$ & 231 & Hf$^{+3}_2$W$^{+6}$P$^{-3}_4$ & 232 & Hf$^{+3}_2$W$^{+6}$Sb$^{-3}_4$ \\
		\hline
		233 & Ir$^{+3}_2$Mo$^{+6}$As$^{-3}_4$ & 234 & Ir$^{+3}_2$Mo$^{+6}$B$^{-3}_4$ & 235 & Ir$^{+3}_2$Mo$^{+6}$N$^{-3}_4$ & 236 & Ir$^{+3}_2$Mo$^{+6}$P$^{-3}_4$ \\
		\hline
		237 & Ir$^{+3}_2$Mo$^{+6}$Sb$^{-3}_4$ & 238 & Ir$^{+3}_2$W$^{+6}$As$^{-3}_4$ & 239 & Ir$^{+3}_2$W$^{+6}$B$^{-3}_4$ & 240 & Ir$^{+3}_2$W$^{+6}$N$^{-3}_4$ \\
		\hline
		241 & Ir$^{+3}_2$W$^{+6}$P$^{-3}_4$ & 242 & Ir$^{+3}_2$W$^{+6}$Sb$^{-3}_4$ & 243 & Mn$^{+3}_2$Mo$^{+6}$As$^{-3}_4$ & 244 & Mn$^{+3}_2$Mo$^{+6}$B$^{-3}_4$ \\
		\hline
		245 & Mn$^{+3}_2$Mo$^{+6}$N$^{-3}_4$ & 246 & Mn$^{+3}_2$Mo$^{+6}$P$^{-3}_4$ & 247 & Mn$^{+3}_2$Mo$^{+6}$Sb$^{-3}_4$ & 248 & Mn$^{+3}_2$W$^{+6}$As$^{-3}_4$ \\
		\hline
		249 & Mn$^{+3}_2$W$^{+6}$B$^{-3}_4$ & 250 & Mn$^{+3}_2$W$^{+6}$N$^{-3}_4$ & 251 & Mn$^{+3}_2$W$^{+6}$P$^{-3}_4$ & 252 & Mn$^{+3}_2$W$^{+6}$Sb$^{-3}_4$ \\
		\hline
		253 & Nb$^{+3}_2$Mo$^{+6}$As$^{-3}_4$ & 254 & Nb$^{+3}_2$Mo$^{+6}$B$^{-3}_4$ & 255 & Nb$^{+3}_2$Mo$^{+6}$N$^{-3}_4$ & 256 & Nb$^{+3}_2$Mo$^{+6}$P$^{-3}_4$ \\
		\hline
		257 & Nb$^{+3}_2$Mo$^{+6}$Sb$^{-3}_4$ & 258 & Nb$^{+3}_2$W$^{+6}$As$^{-3}_4$ & 259 & Nb$^{+3}_2$W$^{+6}$B$^{-3}_4$ & 260 & Nb$^{+3}_2$W$^{+6}$N$^{-3}_4$ \\
		\hline
		261 & Nb$^{+3}_2$W$^{+6}$P$^{-3}_4$ & 262 & Nb$^{+3}_2$W$^{+6}$Sb$^{-3}_4$ & 263 & Ni$^{+3}_2$Mo$^{+6}$As$^{-3}_4$ & 264 & Ni$^{+3}_2$Mo$^{+6}$B$^{-3}_4$ \\
		\hline
		265 & Ni$^{+3}_2$Mo$^{+6}$N$^{-3}_4$ & 266 & Ni$^{+3}_2$Mo$^{+6}$P$^{-3}_4$ & 267 & Ni$^{+3}_2$Mo$^{+6}$Sb$^{-3}_4$ & 268 & Ni$^{+3}_2$W$^{+6}$As$^{-3}_4$ \\
		\hline
		269 & Ni$^{+3}_2$W$^{+6}$B$^{-3}_4$ & 270 & Ni$^{+3}_2$W$^{+6}$N$^{-3}_4$ & 271 & Ni$^{+3}_2$W$^{+6}$P$^{-3}_4$ & 272 & Ni$^{+3}_2$W$^{+6}$Sb$^{-3}_4$ \\
		\hline
		273 & Os$^{+3}_2$Mo$^{+6}$As$^{-3}_4$ & 274 & Os$^{+3}_2$Mo$^{+6}$B$^{-3}_4$ & 275 & Os$^{+3}_2$Mo$^{+6}$N$^{-3}_4$ & 276 & Os$^{+3}_2$Mo$^{+6}$P$^{-3}_4$ \\
		\hline
		277 & Os$^{+3}_2$Mo$^{+6}$Sb$^{-3}_4$ & 278 & Os$^{+3}_2$W$^{+6}$As$^{-3}_4$ & 279 & Os$^{+3}_2$W$^{+6}$B$^{-3}_4$ & 280 & Os$^{+3}_2$W$^{+6}$N$^{-3}_4$ \\
		\hline
		281 & Os$^{+3}_2$W$^{+6}$P$^{-3}_4$ & 282 & Os$^{+3}_2$W$^{+6}$Sb$^{-3}_4$ & 283 & Pd$^{+3}_2$Mo$^{+6}$As$^{-3}_4$ & 284 & Pd$^{+3}_2$Mo$^{+6}$B$^{-3}_4$ \\
		\hline
		285 & Pd$^{+3}_2$Mo$^{+6}$N$^{-3}_4$ & 286 & Pd$^{+3}_2$Mo$^{+6}$P$^{-3}_4$ & 287 & Pd$^{+3}_2$Mo$^{+6}$Sb$^{-3}_4$ & 288 & Pd$^{+3}_2$W$^{+6}$As$^{-3}_4$ \\
		\hline
		289 & Pd$^{+3}_2$W$^{+6}$B$^{-3}_4$ & 290 & Pd$^{+3}_2$W$^{+6}$N$^{-3}_4$ & 291 & Pd$^{+3}_2$W$^{+6}$P$^{-3}_4$ & 292 & Pd$^{+3}_2$W$^{+6}$Sb$^{-3}_4$ \\
		\hline
		293 & Pt$^{+3}_2$Mo$^{+6}$As$^{-3}_4$ & 294 & Pt$^{+3}_2$Mo$^{+6}$B$^{-3}_4$ & 295 & Pt$^{+3}_2$Mo$^{+6}$N$^{-3}_4$ & 296 & Pt$^{+3}_2$Mo$^{+6}$P$^{-3}_4$ \\
		\hline
		297 & Pt$^{+3}_2$Mo$^{+6}$Sb$^{-3}_4$ & 298 & Pt$^{+3}_2$W$^{+6}$As$^{-3}_4$ & 299 & Pt$^{+3}_2$W$^{+6}$B$^{-3}_4$ & 300 & Pt$^{+3}_2$W$^{+6}$N$^{-3}_4$ \\
		\hline
		301 & Pt$^{+3}_2$W$^{+6}$P$^{-3}_4$ & 302 & Pt$^{+3}_2$W$^{+6}$Sb$^{-3}_4$ & 303 & Re$^{+3}_2$Mo$^{+6}$As$^{-3}_4$ & 304 & Re$^{+3}_2$Mo$^{+6}$B$^{-3}_4$ \\
		\hline
		305 & Re$^{+3}_2$Mo$^{+6}$N$^{-3}_4$ & 306 & Re$^{+3}_2$Mo$^{+6}$P$^{-3}_4$ & 307 & Re$^{+3}_2$Mo$^{+6}$Sb$^{-3}_4$ & 308 & Re$^{+3}_2$W$^{+6}$As$^{-3}_4$ \\
		\hline
		309 & Re$^{+3}_2$W$^{+6}$B$^{-3}_4$ & 310 & Re$^{+3}_2$W$^{+6}$N$^{-3}_4$ & 311 & Re$^{+3}_2$W$^{+6}$P$^{-3}_4$ & 312 & Re$^{+3}_2$W$^{+6}$Sb$^{-3}_4$ \\
		\hline
		313 & Rh$^{+3}_2$Mo$^{+6}$As$^{-3}_4$ & 314 & Rh$^{+3}_2$Mo$^{+6}$B$^{-3}_4$ & 315 & Rh$^{+3}_2$Mo$^{+6}$N$^{-3}_4$ & 316 & Rh$^{+3}_2$Mo$^{+6}$P$^{-3}_4$ \\
		\hline
		317 & Rh$^{+3}_2$Mo$^{+6}$Sb$^{-3}_4$ & 318 & Rh$^{+3}_2$W$^{+6}$As$^{-3}_4$ & 319 & Rh$^{+3}_2$W$^{+6}$B$^{-3}_4$ & 320 & Rh$^{+3}_2$W$^{+6}$N$^{-3}_4$ \\
		\hline
		321 & Rh$^{+3}_2$W$^{+6}$P$^{-3}_4$ & 322 & Rh$^{+3}_2$W$^{+6}$Sb$^{-3}_4$ & 323 & Ru$^{+3}_2$Mo$^{+6}$As$^{-3}_4$ & 324 & Ru$^{+3}_2$Mo$^{+6}$B$^{-3}_4$ \\
		\hline
		325 & Ru$^{+3}_2$Mo$^{+6}$N$^{-3}_4$ & 326 & Ru$^{+3}_2$Mo$^{+6}$P$^{-3}_4$ & 327 & Ru$^{+3}_2$Mo$^{+6}$Sb$^{-3}_4$ & 328 & Ru$^{+3}_2$W$^{+6}$As$^{-3}_4$ \\
		\hline
		329 & Ru$^{+3}_2$W$^{+6}$B$^{-3}_4$ & 330 & Ru$^{+3}_2$W$^{+6}$N$^{-3}_4$ & 331 & Ru$^{+3}_2$W$^{+6}$P$^{-3}_4$ & 332 & Ru$^{+3}_2$W$^{+6}$Sb$^{-3}_4$ \\
		\hline
		333 & Sc$^{+3}_2$Mo$^{+6}$As$^{-3}_4$ & 334 & Sc$^{+3}_2$Mo$^{+6}$B$^{-3}_4$ & 335 & Sc$^{+3}_2$Mo$^{+6}$N$^{-3}_4$ & 336 & Sc$^{+3}_2$Mo$^{+6}$P$^{-3}_4$ \\
		\hline
		337 & Sc$^{+3}_2$Mo$^{+6}$Sb$^{-3}_4$ & 338 & Sc$^{+3}_2$W$^{+6}$As$^{-3}_4$ & 339 & Sc$^{+3}_2$W$^{+6}$B$^{-3}_4$ & 340 & Sc$^{+3}_2$W$^{+6}$N$^{-3}_4$ \\
		\hline
		341 & Sc$^{+3}_2$W$^{+6}$P$^{-3}_4$ & 342 & Sc$^{+3}_2$W$^{+6}$Sb$^{-3}_4$ & 343 & Ta$^{+3}_2$Mo$^{+6}$As$^{-3}_4$ & 344 & Ta$^{+3}_2$Mo$^{+6}$B$^{-3}_4$ \\
		\hline
		345 & Ta$^{+3}_2$Mo$^{+6}$N$^{-3}_4$ & 346 & Ta$^{+3}_2$Mo$^{+6}$P$^{-3}_4$ & 347 & Ta$^{+3}_2$Mo$^{+6}$Sb$^{-3}_4$ & 348 & Ta$^{+3}_2$W$^{+6}$As$^{-3}_4$ \\
		\hline
		349 & Ta$^{+3}_2$W$^{+6}$B$^{-3}_4$ & 350 & Ta$^{+3}_2$W$^{+6}$N$^{-3}_4$ & 351 & Ta$^{+3}_2$W$^{+6}$P$^{-3}_4$ & 352 & Ta$^{+3}_2$W$^{+6}$Sb$^{-3}_4$ \\
		\hline
		353 & Tc$^{+3}_2$Mo$^{+6}$As$^{-3}_4$ & 354 & Tc$^{+3}_2$Mo$^{+6}$B$^{-3}_4$ & 355 & Tc$^{+3}_2$Mo$^{+6}$N$^{-3}_4$ & 356 & Tc$^{+3}_2$Mo$^{+6}$P$^{-3}_4$ \\
		\hline
		357 & Tc$^{+3}_2$Mo$^{+6}$Sb$^{-3}_4$ & 358 & Tc$^{+3}_2$W$^{+6}$As$^{-3}_4$ & 359 & Tc$^{+3}_2$W$^{+6}$B$^{-3}_4$ & 360 & Tc$^{+3}_2$W$^{+6}$N$^{-3}_4$ \\
		\hline
		361 & Tc$^{+3}_2$W$^{+6}$P$^{-3}_4$ & 362 & Tc$^{+3}_2$W$^{+6}$Sb$^{-3}_4$ & 363 & Ti$^{+3}_2$Mo$^{+6}$As$^{-3}_4$ & 364 & Ti$^{+3}_2$Mo$^{+6}$B$^{-3}_4$ \\
		\hline
		365 & Ti$^{+3}_2$Mo$^{+6}$N$^{-3}_4$ & 366 & Ti$^{+3}_2$Mo$^{+6}$P$^{-3}_4$ & 367 & Ti$^{+3}_2$Mo$^{+6}$Sb$^{-3}_4$ & 368 & Ti$^{+3}_2$W$^{+6}$As$^{-3}_4$ \\
		\hline
		369 & Ti$^{+3}_2$W$^{+6}$B$^{-3}_4$ & 370 & Ti$^{+3}_2$W$^{+6}$N$^{-3}_4$ & 371 & Ti$^{+3}_2$W$^{+6}$P$^{-3}_4$ & 372 & Ti$^{+3}_2$W$^{+6}$Sb$^{-3}_4$ \\
		\hline
		373 & V$^{+3}_2$Mo$^{+6}$As$^{-3}_4$ & 374 & V$^{+3}_2$Mo$^{+6}$B$^{-3}_4$ & 375 & V$^{+3}_2$Mo$^{+6}$N$^{-3}_4$ & 376 & V$^{+3}_2$Mo$^{+6}$P$^{-3}_4$ \\
		\hline
		377 & V$^{+3}_2$Mo$^{+6}$Sb$^{-3}_4$ & 378 & V$^{+3}_2$W$^{+6}$As$^{-3}_4$ & 379 & V$^{+3}_2$W$^{+6}$B$^{-3}_4$ & 380 & V$^{+3}_2$W$^{+6}$N$^{-3}_4$ \\
		\hline
		381 & V$^{+3}_2$W$^{+6}$P$^{-3}_4$ & 382 & V$^{+3}_2$W$^{+6}$Sb$^{-3}_4$ & 383 & Y$^{+3}_2$Mo$^{+6}$As$^{-3}_4$ & 384 & Y$^{+3}_2$Mo$^{+6}$B$^{-3}_4$ \\
		\hline
		385 & Y$^{+3}_2$Mo$^{+6}$N$^{-3}_4$ & 386 & Y$^{+3}_2$Mo$^{+6}$P$^{-3}_4$ & 387 & Y$^{+3}_2$Mo$^{+6}$Sb$^{-3}_4$ & 388 & Y$^{+3}_2$W$^{+6}$As$^{-3}_4$ \\
		\hline
		389 & Y$^{+3}_2$W$^{+6}$B$^{-3}_4$ & 390 & Y$^{+3}_2$W$^{+6}$N$^{-3}_4$ & 391 & Y$^{+3}_2$W$^{+6}$P$^{-3}_4$ & 392 & Y$^{+3}_2$W$^{+6}$Sb$^{-3}_4$ \\
		\hline
		393 & Zr$^{+3}_2$Mo$^{+6}$As$^{-3}_4$ & 394 & Zr$^{+3}_2$Mo$^{+6}$B$^{-3}_4$ & 395 & Zr$^{+3}_2$Mo$^{+6}$N$^{-3}_4$ & 396 & Zr$^{+3}_2$Mo$^{+6}$P$^{-3}_4$ \\
		\hline
		397 & Zr$^{+3}_2$Mo$^{+6}$Sb$^{-3}_4$ & 398 & Zr$^{+3}_2$W$^{+6}$As$^{-3}_4$ & 399 & Zr$^{+3}_2$W$^{+6}$B$^{-3}_4$ & 400 & Zr$^{+3}_2$W$^{+6}$N$^{-3}_4$ \\
		\hline
		401 & Zr$^{+3}_2$W$^{+6}$P$^{-3}_4$ & 402 & Zr$^{+3}_2$W$^{+6}$Sb$^{-3}_4$ & 403 & Cr$^{+5}_2$Mo$^{+6}$C$^{-4}_4$ & 404 & Cr$^{+5}_2$Mo$^{+6}$Si$^{-4}_4$ \\
		\hline
		405 & Cr$^{+5}_2$W$^{+6}$C$^{-4}_4$ & 406 & Cr$^{+5}_2$W$^{+6}$Si$^{-4}_4$ & 407 & Ir$^{+5}_2$Mo$^{+6}$C$^{-4}_4$ & 408 & Ir$^{+5}_2$Mo$^{+6}$Si$^{-4}_4$ \\
		\hline
		409 & Ir$^{+5}_2$W$^{+6}$C$^{-4}_4$ & 410 & Ir$^{+5}_2$W$^{+6}$Si$^{-4}_4$ & 411 & Mn$^{+5}_2$Mo$^{+6}$C$^{-4}_4$ & 412 & Mn$^{+5}_2$Mo$^{+6}$Si$^{-4}_4$ \\
		\hline
		413 & Mn$^{+5}_2$W$^{+6}$C$^{-4}_4$ & 414 & Mn$^{+5}_2$W$^{+6}$Si$^{-4}_4$ & 415 & Nb$^{+5}_2$Mo$^{+6}$C$^{-4}_4$ & 416 & Nb$^{+5}_2$Mo$^{+6}$Si$^{-4}_4$ \\
		\hline
		417 & Nb$^{+5}_2$W$^{+6}$C$^{-4}_4$ & 418 & Nb$^{+5}_2$W$^{+6}$Si$^{-4}_4$ & 419 & Os$^{+5}_2$Mo$^{+6}$C$^{-4}_4$ & 420 & Os$^{+5}_2$Mo$^{+6}$Si$^{-4}_4$ \\
		\hline
		421 & Os$^{+5}_2$W$^{+6}$C$^{-4}_4$ & 422 & Os$^{+5}_2$W$^{+6}$Si$^{-4}_4$ & 423 & Re$^{+5}_2$Mo$^{+6}$C$^{-4}_4$ & 424 & Re$^{+5}_2$Mo$^{+6}$Si$^{-4}_4$ \\
		\hline
		425 & Re$^{+5}_2$W$^{+6}$C$^{-4}_4$ & 426 & Re$^{+5}_2$W$^{+6}$Si$^{-4}_4$ & 427 & Ru$^{+5}_2$Mo$^{+6}$C$^{-4}_4$ & 428 & Ru$^{+5}_2$Mo$^{+6}$Si$^{-4}_4$ \\
		\hline
		429 & Ru$^{+5}_2$W$^{+6}$C$^{-4}_4$ & 430 & Ru$^{+5}_2$W$^{+6}$Si$^{-4}_4$ & 431 & Ta$^{+5}_2$Mo$^{+6}$C$^{-4}_4$ & 432 & Ta$^{+5}_2$Mo$^{+6}$Si$^{-4}_4$ \\
		\hline
		433 & Ta$^{+5}_2$W$^{+6}$C$^{-4}_4$ & 434 & Ta$^{+5}_2$W$^{+6}$Si$^{-4}_4$ & 435 & Tc$^{+5}_2$Mo$^{+6}$C$^{-4}_4$ & 436 & Tc$^{+5}_2$Mo$^{+6}$Si$^{-4}_4$ \\
		\hline
		437 & Tc$^{+5}_2$W$^{+6}$C$^{-4}_4$ & 438 & Tc$^{+5}_2$W$^{+6}$Si$^{-4}_4$ & 439 & V$^{+5}_2$Mo$^{+6}$C$^{-4}_4$ & 440 & V$^{+5}_2$Mo$^{+6}$Si$^{-4}_4$ \\
		\hline
		441 & V$^{+5}_2$W$^{+6}$C$^{-4}_4$ & 442 & V$^{+5}_2$W$^{+6}$Si$^{-4}_4$ \\
		\cline{1-4}
	\end{longtable}

	\normalsize
	
	\newpage
	
	% Page 2: Cr2WTe4
	\begin{flushleft}
		\large\bfseries System: Cr$_{2}$WTe$_{4}$
	\end{flushleft}
	
	\vspace{-0.2cm}
	
	\begin{minipage}[t]{0.5\textwidth}
		Lattice constants: $a$ = $b$ = 5.79 \AA \\
		Space Group: $P\bar{4}2m$ \\
		Band Gap: 0.294 eV \\
		HSE06 Band Gap: 0.756 eV
	\end{minipage}%
	\begin{minipage}[t]{0.5\textwidth}
		Magnetic Moment: 3.8 $\mu_{\text{B}}$ \\
		Formation Energy: $-$1.36 eV/f.u. \\
		Spin Polarization Ratio: 94\% \\
		Charge-to-Spin Conversion Efficiency: 94\%
	\end{minipage}
	
	\vspace{0.1cm}
	
	\begin{figure}[H]
		\centering
		\includegraphics[width=5.75in, height=4.88in]{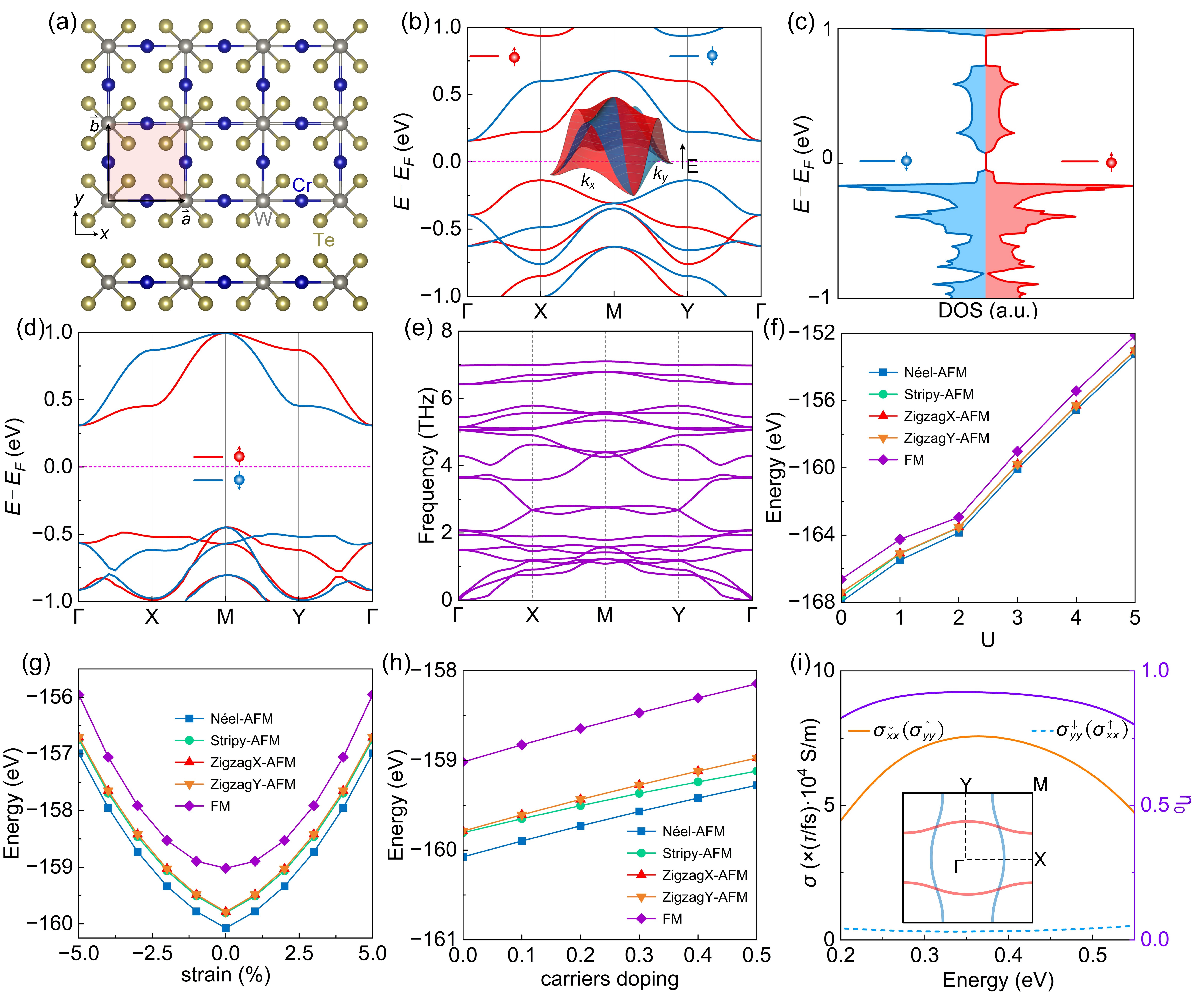}		
		\caption{Properties of the Cr$_2$WTe$_4$ monolayer. (a) Atomic structure (top and side views). (b) Band structure from GGA-PBE+$U$ calculations, with the tent state highlighted in the inset. (c) Density of states. (d) Band structure calculated with the HSE06 functional for comparison. (e) Phonon spectrum. Total energies of the FM, N\'eel-AFM, Stripy-AFM, ZigzagX-AFM and ZigzagY-AFM states under varying (f) Hubbard $U$ parameters, (g) biaxial strains from $-5\%$ to $+5\%$, and (h) carrier doping. (i) Conductivity and spin polarization efficiency as a function of energy for different spin channels and in-plane directions.}
		\label{fig:S5}	
	\end{figure}
	
	\newpage
	
	% Page 3: Mn2MoO4
	\begin{flushleft}
		\large\bfseries System: Mn$_{2}$MoO$_{4}$
	\end{flushleft}
	
	\vspace{-0.2cm}
	
	\begin{minipage}[t]{0.5\textwidth}
		Lattice constants: $a$ = $b$ = 5.09 \AA \\
		Space Group: $P\bar{4}2m$ \\
		Band Gap: 0.642 eV \\
		HSE06 Band Gap: 2.000 eV
	\end{minipage}%
	\begin{minipage}[t]{0.5\textwidth}
		Magnetic Moment: 4.1 $\mu_{\text{B}}$ \\
		Formation Energy: $-$13.2 eV$/$f$.$u$.$ \\
		Spin Polarization Ratio: 99\% \\
		Charge-to-Spin Conversion Efficiency: 99\%
	\end{minipage}
	
	\vspace{0.1cm}
	
	\begin{figure}[H]
		\centering
		\includegraphics[width=5.75in, height=4.88in]{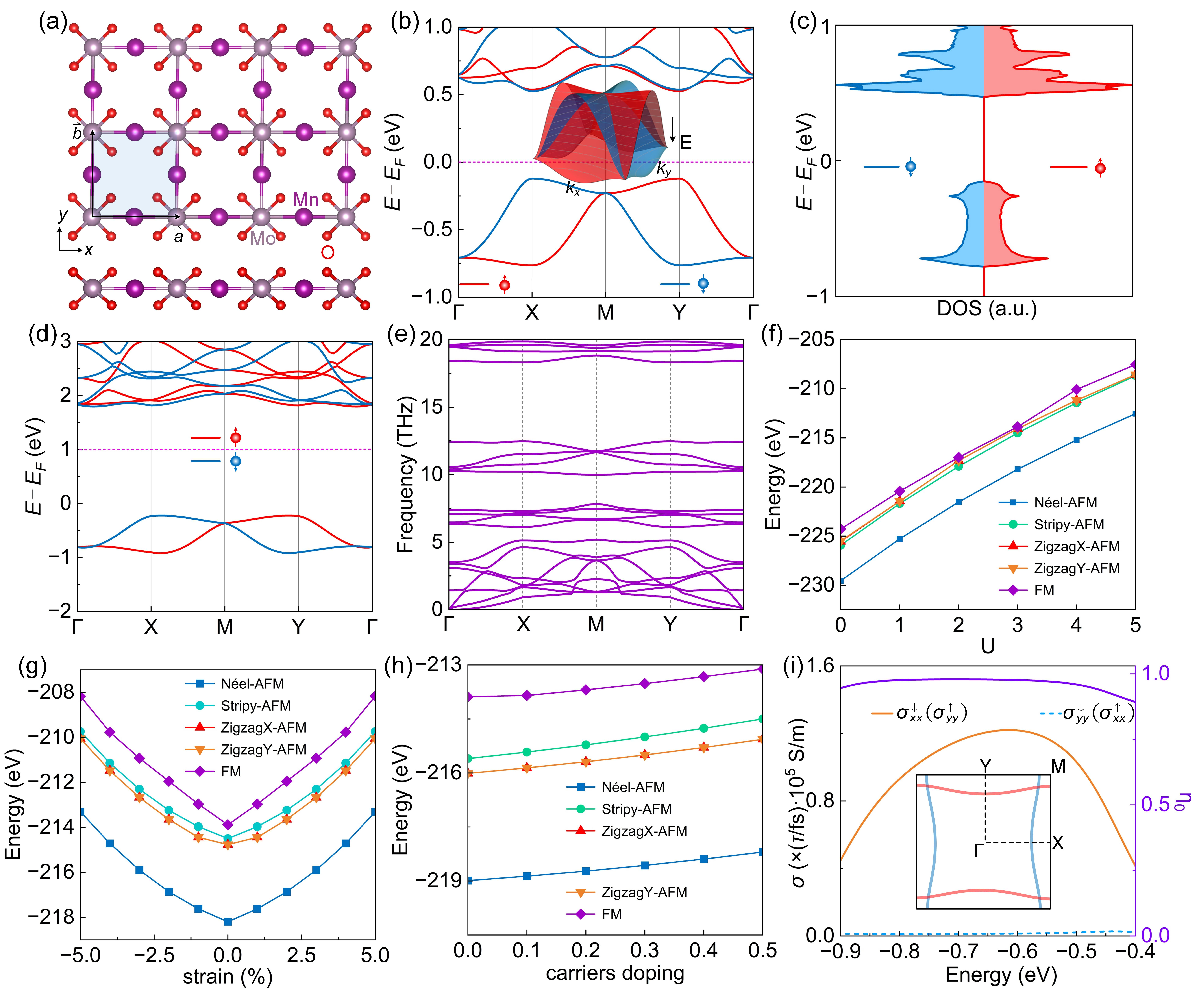}		
		\caption{Properties of the Mn$_{2}$MoO$_{4}$ monolayer. (a) Atomic structure (top and side views). (b) Band structure from GGA-PBE+$U$ calculations, with the tent state highlighted in the inset. (c) Density of states. (d) Band structure calculated with the HSE06 functional for comparison. (e) Phonon spectrum. Total energies of the FM, N\'eel-AFM, Stripy-AFM, ZigzagX-AFM and ZigzagY-AFM states under varying (f) Hubbard $U$ parameters, (g) biaxial strains from $-5\%$ to $+5\%$, and (h) carrier doping. (i) Conductivity and spin polarization efficiency as a function of energy for different spin channels and in-plane directions.}
		\label{fig:S6}	
	\end{figure}
	
	\newpage
	
	% Page 4: Mn2MoS4
	\begin{flushleft}
		\large\bfseries System: Mn$_{2}$MoS$_{4}$
	\end{flushleft}
	
	\vspace{-0.2cm}
	
	\begin{minipage}[t]{0.5\textwidth}
		Lattice constants: $a$ = $b$ = 5.43 \AA \\
		Space Group: $P\bar{4}2m$ \\
		Band Gap: 0.558 eV \\
		HSE06 Band Gap: 1.497 eV
	\end{minipage}%
	\begin{minipage}[t]{0.5\textwidth}
		Magnetic Moment: 3.9 $\mu_{\text{B}}$ \\
		Formation Energy: $-$13.5 eV$/$f$.$u$.$ \\
		Spin Polarization Ratio: 100\% \\
		Charge-to-Spin Conversion Efficiency: 100\%
	\end{minipage}
	
	\vspace{0.1cm}
	
	\begin{figure}[H]
		\centering
		\includegraphics[width=5.75in, height=4.88in]{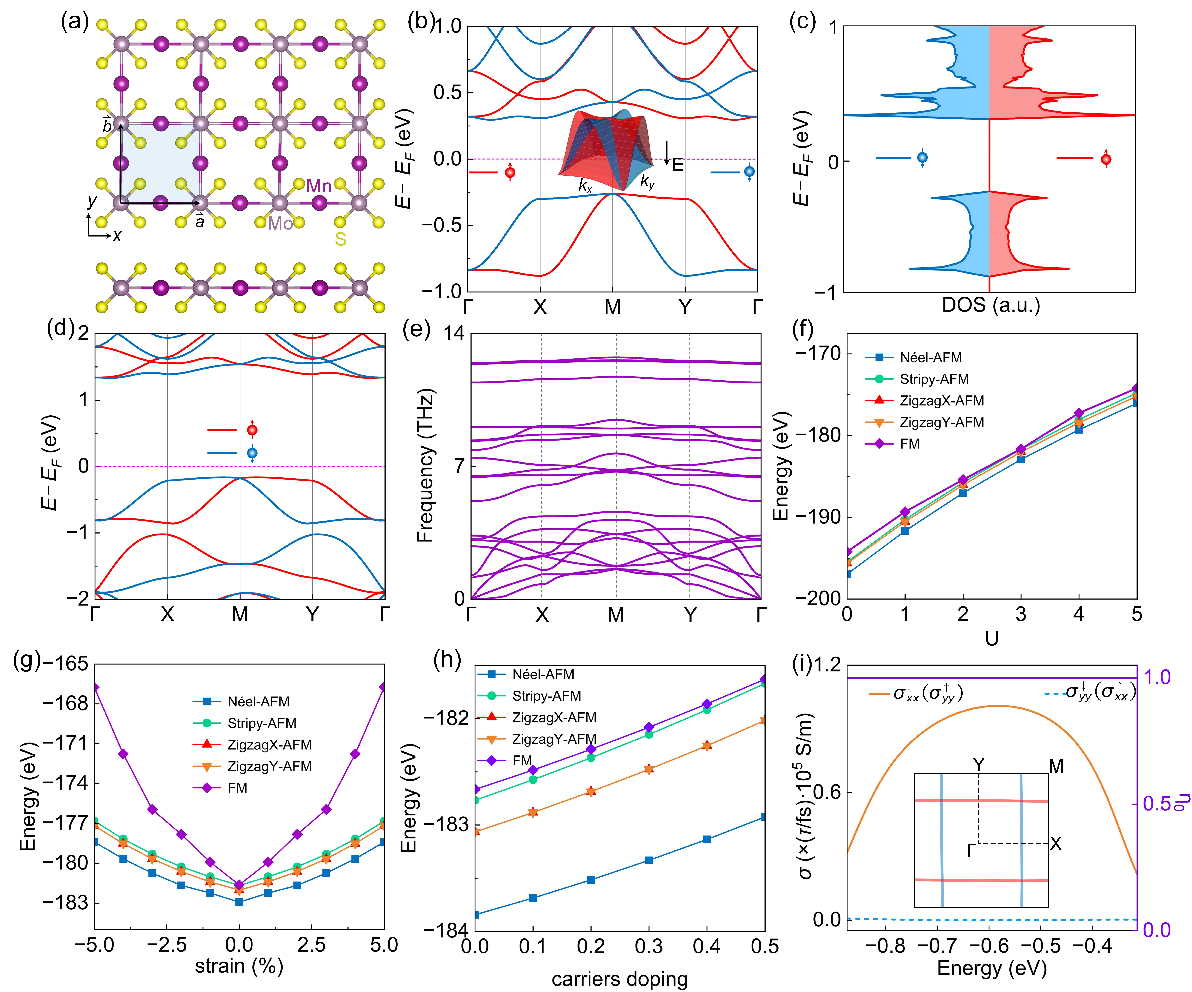}		
		\caption{Properties of the Mn$_{2}$MoS$_{4}$ monolayer. (a) Atomic structure (top and side views). (b) Band structure from GGA-PBE+$U$ calculations, with the tent state highlighted in the inset. (c) Density of states. (d) Band structure calculated with the HSE06 functional for comparison. (e) Phonon spectrum. Total energies of the FM, N\'eel-AFM, Stripy-AFM, ZigzagX-AFM and ZigzagY-AFM states under varying (f) Hubbard $U$ parameters, (g) biaxial strains from $-5\%$ to $+5\%$, and (h) carrier doping. (i) Conductivity and spin polarization efficiency as a function of energy for different spin channels and in-plane directions.}
		\label{fig:S7}
	\end{figure}
	
	\newpage
	
	% Page 5: Mn2MoSe4
	\begin{flushleft}
		\large\bfseries System: Mn$_{2}$MoSe$_{4}$
	\end{flushleft}
	
	\vspace{-0.2cm}
	
	\begin{minipage}[t]{0.5\textwidth}
		Lattice constants: $a$ = $b$ = 5.57 \AA \\
		Space Group: $P\bar{4}2m$ \\
		Band Gap: 0.590 eV \\
		HSE06 Band Gap: 1.504 eV
	\end{minipage}%
	\begin{minipage}[t]{0.5\textwidth}
		Magnetic Moment: 4.0 $\mu_{\text{B}}$ \\
		Formation Energy: $-$5.31 eV$/$f$.$u$.$ \\
		Spin Polarization Ratio: 99.5\% \\
		Charge-to-Spin Conversion Efficiency: 99.5\%
	\end{minipage}
	
	\vspace{0.1cm}
	
	\begin{figure}[H]
		\centering
		\includegraphics[width=5.75in, height=4.88in]{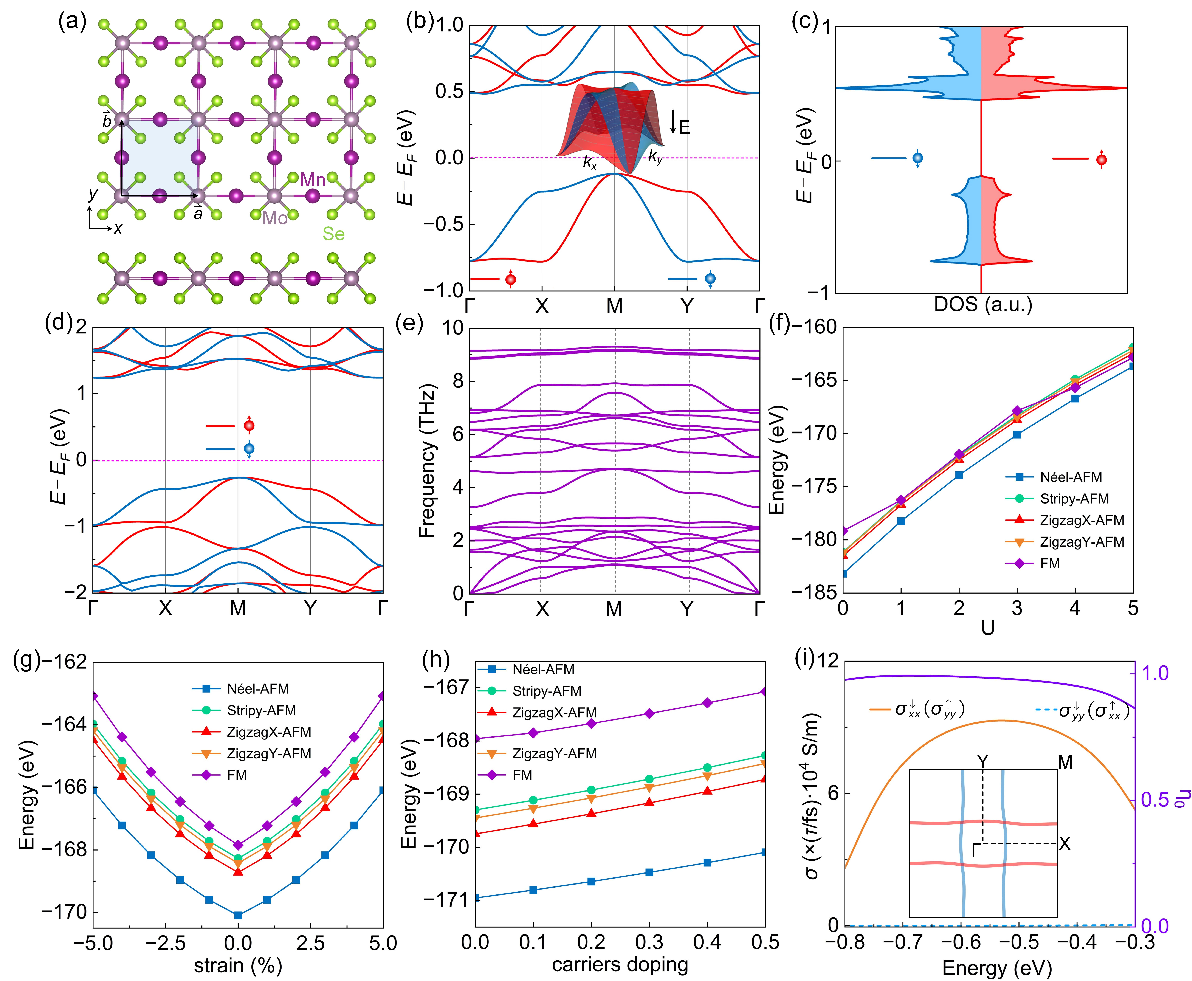}		
		\caption{Properties of the Mn$_{2}$MoSe$_{4}$ monolayer. (a) Atomic structure (top and side views). (b) Band structure from GGA-PBE+$U$ calculations, with the tent state highlighted in the inset. (c) Density of states. (d) Band structure calculated with the HSE06 functional for comparison. (e) Phonon spectrum. Total energies of the FM, N\'eel-AFM, Stripy-AFM, ZigzagX-AFM and ZigzagY-AFM states under varying (f) Hubbard $U$ parameters, (g) biaxial strains from $-5\%$ to $+5\%$, and (h) carrier doping. (i) Conductivity and spin polarization efficiency as a function of energy for different spin channels and in-plane directions.}
		\label{fig:S8}
	\end{figure}
	
	\newpage
	
	% Page 6: Mn2WO4
	\begin{flushleft}
		\large\bfseries System: Mn$_{2}$WO$_{4}$
	\end{flushleft}
	
	\vspace{-0.2cm}
	
	\begin{minipage}[t]{0.5\textwidth}
		Lattice constants: $a$ = $b$ = 5.13 \AA \\
		Space Group: $P\bar{4}2m$ \\
		Band Gap: 0.003 eV \\
		HSE06 Band Gap: 1.106 eV
	\end{minipage}%
	\begin{minipage}[t]{0.5\textwidth}
		Magnetic Moment: 4.1 $\mu_{\text{B}}$ \\
		Formation Energy: $-$12.6 eV$/$f$.$u$.$ \\
		Spin Polarization Ratio: 99\% \\
		Charge-to-Spin Conversion Efficiency: 99\%
	\end{minipage}
	
	\vspace{0.1cm}
	
	\begin{figure}[H]
		\centering
		\includegraphics[width=5.75in, height=4.88in]{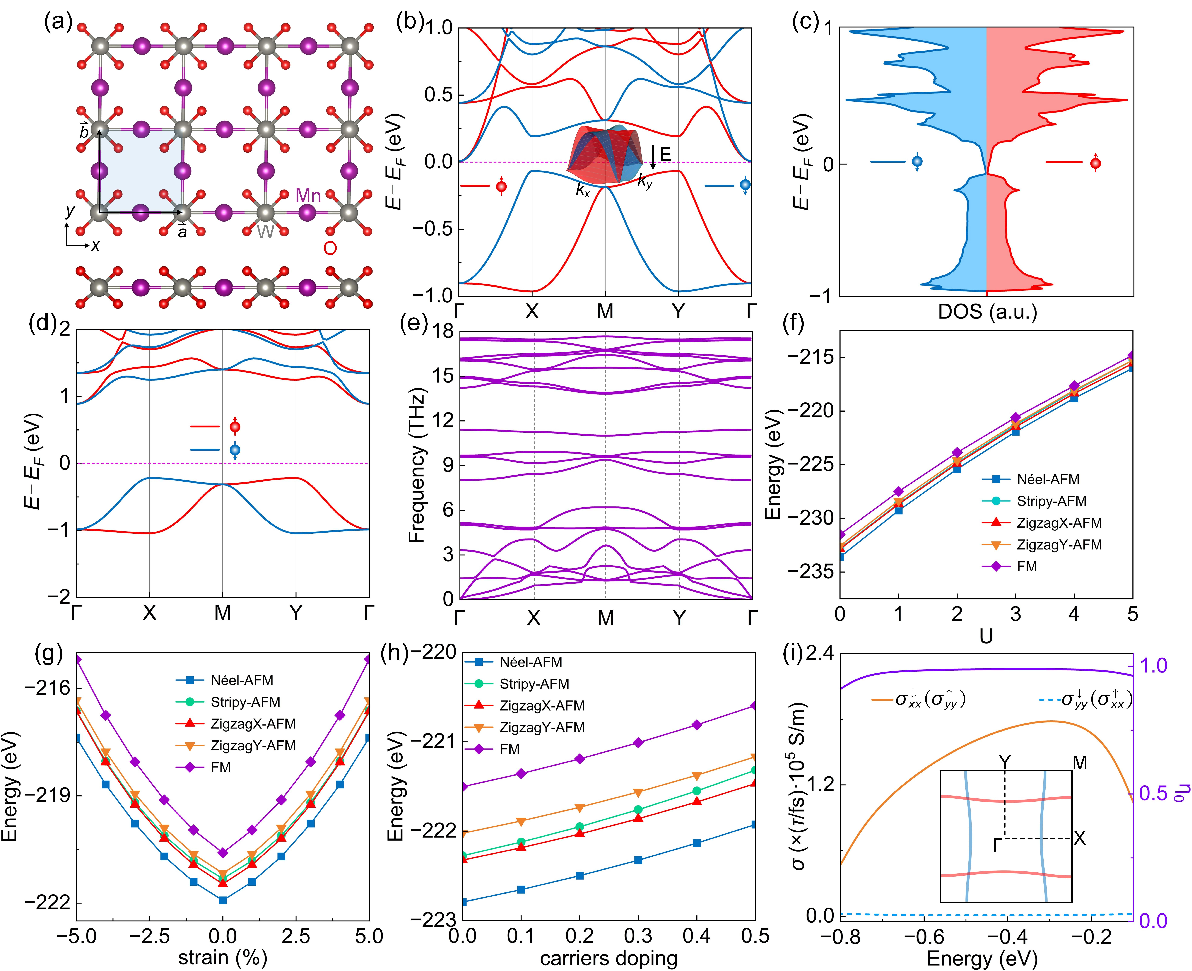}		
		\caption{Properties of the Mn$_{2}$WO$_{4}$ monolayer. (a) Atomic structure (top and side views). (b) Band structure from GGA-PBE+$U$ calculations, with the tent state highlighted in the inset. (c) Density of states. (d) Band structure calculated with the HSE06 functional for comparison. (e) Phonon spectrum. Total energies of the FM, N\'eel-AFM, Stripy-AFM, ZigzagX-AFM and ZigzagY-AFM states under varying (f) Hubbard $U$ parameters, (g) biaxial strains from $-5\%$ to $+5\%$, and (h) carrier doping. (i) Conductivity and spin polarization efficiency as a function of energy for different spin channels and in-plane directions.}
		\label{fig:S9}
	\end{figure}
	
	\newpage
	
	% Page 7: Mn2WS4
	\begin{flushleft}
		\large\bfseries System: Mn$_{2}$WS$_{4}$
	\end{flushleft}
	
	\vspace{-0.2cm}
	
	\begin{minipage}[t]{0.5\textwidth}
		Lattice constants: $a$ = $b$ = 5.45 \AA \\
		Space Group: $P\bar{4}2m$ \\
		Band Gap: 0.355 eV \\
		HSE06 Band Gap: 1.178 eV
	\end{minipage}%
	\begin{minipage}[t]{0.5\textwidth}
		Magnetic Moment: 3.9 $\mu_{\text{B}}$ \\
		Formation Energy: $-$13.1 eV$/$f$.$u$.$ \\
		Spin Polarization Ratio: 100\% \\
		Charge-to-Spin Conversion Efficiency: 100\%
	\end{minipage}

	\vspace{0.1cm}
	
	\begin{figure}[H]
		\centering
		\includegraphics[width=5.75in, height=4.88in]{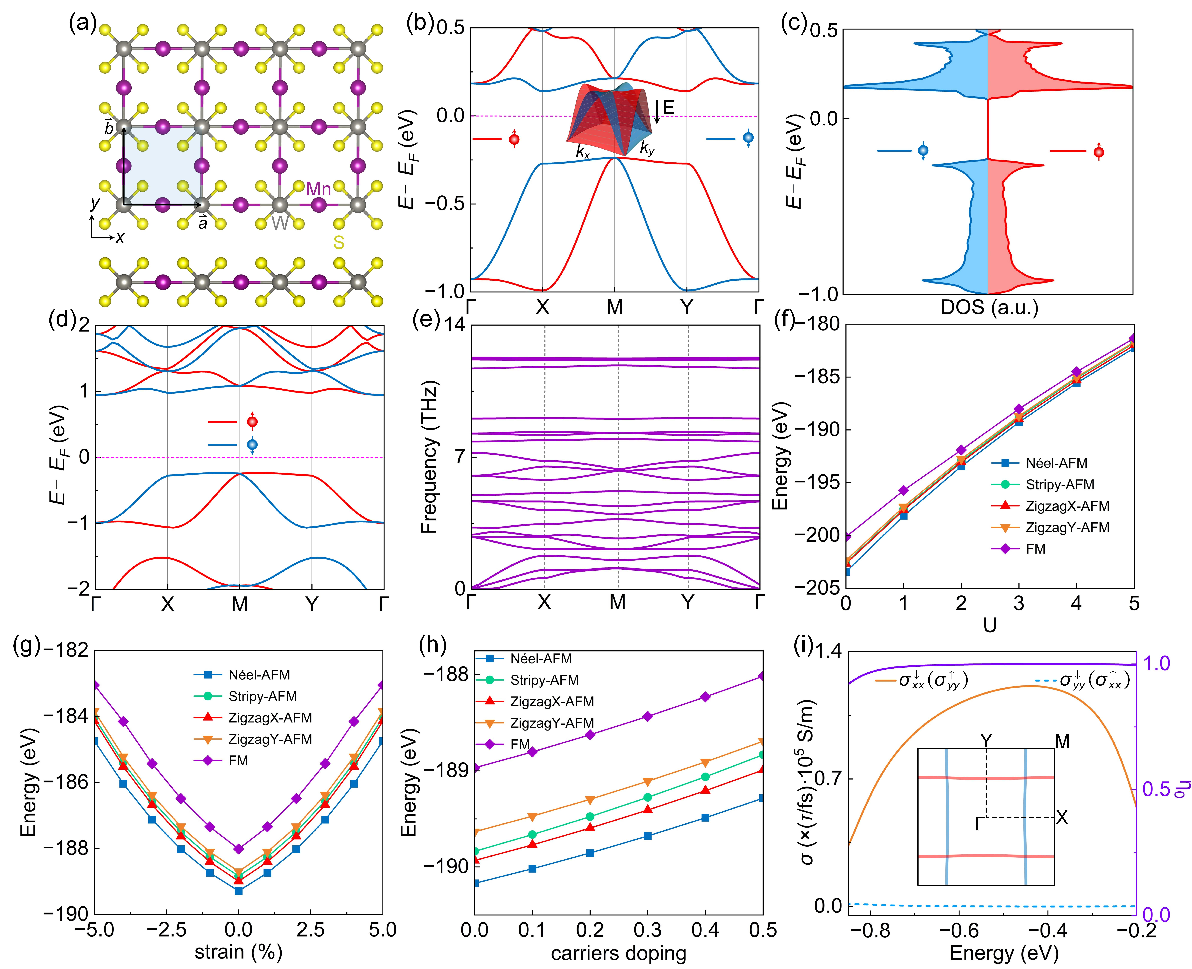}		
		\caption{Properties of the Mn$_{2}$WS$_{4}$ monolayer. (a) Atomic structure (top and side views). (b) Band structure from GGA-PBE+$U$ calculations, with the tent state highlighted in the inset. (c) Density of states. (d) Band structure calculated with the HSE06 functional for comparison. (e) Phonon spectrum. Total energies of the FM, N\'eel-AFM, Stripy-AFM, ZigzagX-AFM and ZigzagY-AFM states under varying (f) Hubbard $U$ parameters, (g) biaxial strains from $-5\%$ to $+5\%$, and (h) carrier doping. (i) Conductivity and spin polarization efficiency as a function of energy for different spin channels and in-plane directions.}
		\label{fig:S10}
	\end{figure}
	
	\newpage
	
	% Page 8: Mn2WSe4
	\begin{flushleft}
		\large\bfseries System: Mn$_{2}$WSe$_{4}$
	\end{flushleft}
	
	\vspace{-0.2cm}
	
	\begin{minipage}[t]{0.5\textwidth}
		Lattice constants: $a$ = $b$ = 5.13 \AA \\
		Space Group: $P\bar{4}2m$ \\
		Band Gap: 0.448 eV \\
		HSE06 Band Gap: 1.219 eV
	\end{minipage}%
	\begin{minipage}[t]{0.5\textwidth}
		Magnetic Moment: 4.0 $\mu_{\text{B}}$ \\
		Formation Energy: $-$4.78 eV$/$f$.$u$.$ \\
		Spin Polarization Ratio: 99.5\% \\
		Charge-to-Spin Conversion Efficiency: 99.5\%
	\end{minipage}
	
	\vspace{0.1cm}
	
	\begin{figure}[H]
		\centering
		\includegraphics[width=5.75in, height=4.88in]{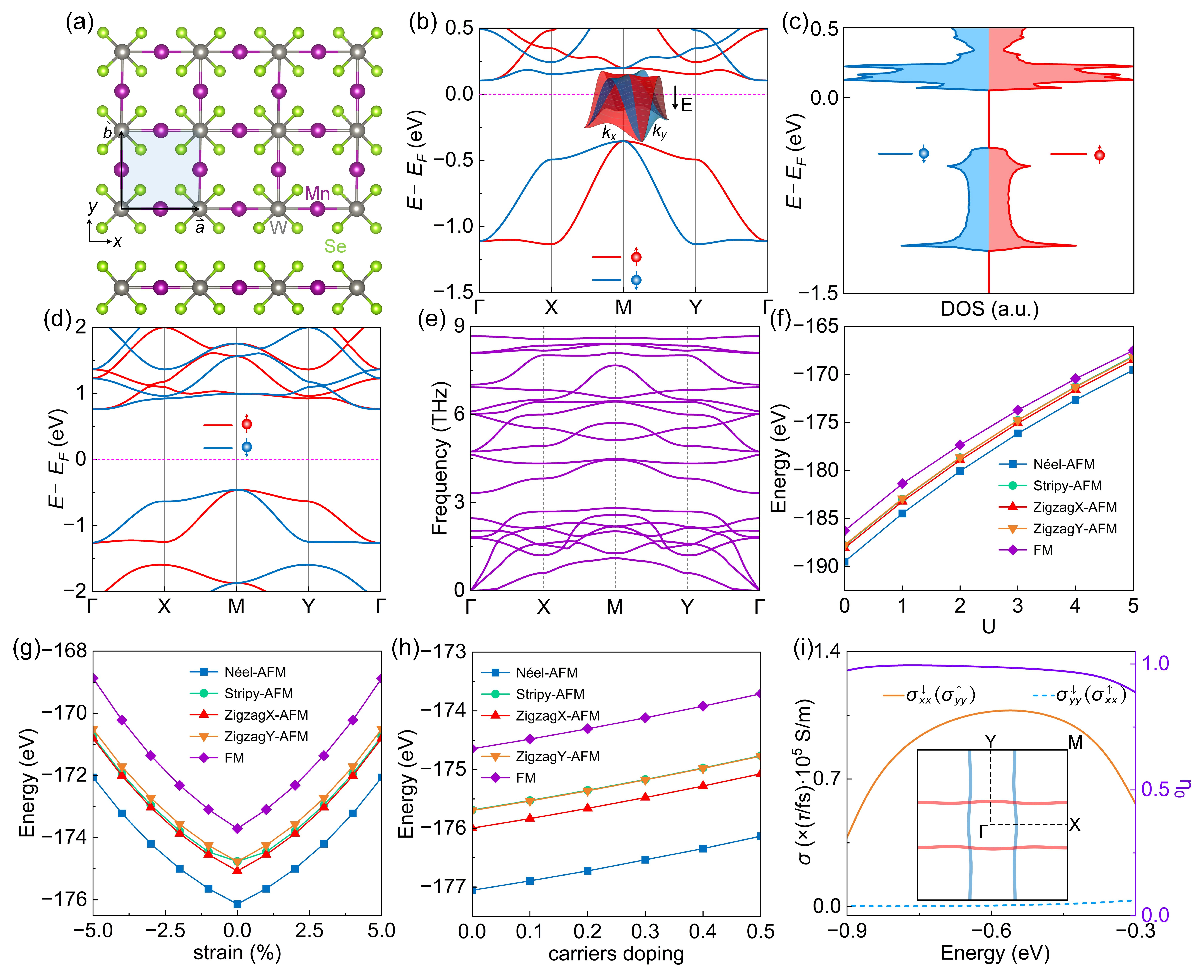}		
		\caption{Properties of the Mn$_{2}$WSe$_{4}$ monolayer. (a) Atomic structure (top and side views). (b) Band structure from GGA-PBE+$U$ calculations, with the tent state highlighted in the inset. (c) Density of states. (d) Band structure calculated with the HSE06 functional for comparison. (e) Phonon spectrum. Total energies of the FM, N\'eel-AFM, Stripy-AFM, ZigzagX-AFM and ZigzagY-AFM states under varying (f) Hubbard $U$ parameters, (g) biaxial strains from $-5\%$ to $+5\%$, and (h) carrier doping. (i) Conductivity and spin polarization efficiency as a function of energy for different spin channels and in-plane directions.}
		\label{fig:S11}
	\end{figure}

	\subsubsection*{{VI.2 Representative 2D SADL material: Cr$_2$WSe$_4$}}
	{\qquad Among the screened 2D candidates, monolayer Cr$_2$WSe$_4$ is selected as a representative SADL system because it exhibits a clear tent-state band structure and contains the heavy W element, providing a stringent test of SOC robustness. Below, we summarize its stability, effective-model characterization, and transport robustness.}
	
	{\qquad The stability of monolayer Cr$_2$WSe$_4$ is summarized in Fig.~\ref{fig:S12}(a--f) and Table~\ref{tab:S3}. The phonon spectrum shows no imaginary modes, and the \emph{ab initio} molecular dynamics simulation at 300~K retains the structural integrity, supporting dynamical and thermal stability [Figs.~\ref{fig:S12}(a,b)]. The elastic constants satisfy the 2D mechanical stability criteria, as listed in Table~\ref{tab:S3}. The Monte Carlo heat-capacity curve gives a N\'eel temperature of $T_N\approx593$~K [Fig.~\ref{fig:S12}(c)], while the total-energy comparisons in Figs.~\ref{fig:S12}(d--f) show that the N\'eel-AFM state remains the lowest-energy magnetic configuration under variations of the Hubbard $U$ parameter, biaxial strain, and carrier doping.}
	
	\begin{table}[htbp]
		\centering
		\caption{Elastic stiffness tensor components $C_{ij}$ of Cr$_{2}$WSe$_{4}$. All values are given in N/m, which satisfies the elastic stability criteria: $C_{11}C_{22}-C_{12}^2 > 0$ and $C_{66} > 0$.}
		\label{tab:S3}
		\begin{tabular*}{\textwidth}{@{\extracolsep{\fill}}cccc}
			\toprule
			& \multicolumn{3}{c}{Elastic Constants $C_{ij}$ (N/m)} \\
			\cmidrule{2-4}
			$i\backslash j$ & 1 & 2 & 6 \\
			\midrule
			1 & 47.3 & 29.6 & 0.0 \\
			2 & 29.6 & 47.3 & 0.0 \\
			6 & 0.0 & 0.0 & 20.4 \\
			\midrule
			\multicolumn{4}{l}{Stability verification:} \\
			\multicolumn{4}{l}{$C_{11}C_{22}-C_{12}^2 = (47.3)^2 - (29.6)^2 = 1361.1  > 0$} {$C_{66} = 20.4 > 0$} \\
			\bottomrule
		\end{tabular*}
	\end{table}
	
	{\qquad The low-energy electronic structure of Cr$_2$WSe$_4$ provides a quantitative connection to the anisotropic 2D $d$-wave model developed in Sec.~I. The minimal-model fit to the spin-resolved tent-state bands gives an effective transverse-to-longitudinal hopping ratio $\gamma=0.083$ [Fig.~\ref{fig:S12}(g)], well below the band-center strong-SADL tolerance $\gamma\lesssim0.318$. This places Cr$_2$WSe$_4$ deep inside the high-$\eta_0$ region of the analytical phase diagram (Fig.~1j). The same tent-state band character is retained at the HSE06 level [Fig.~\ref{fig:S12}(h)] and under carrier doping/gate tuning [Fig.~\ref{fig:S12}(i)].}

	\begin{figure}[H]
		\centering
		\includegraphics[width=0.9\textwidth]{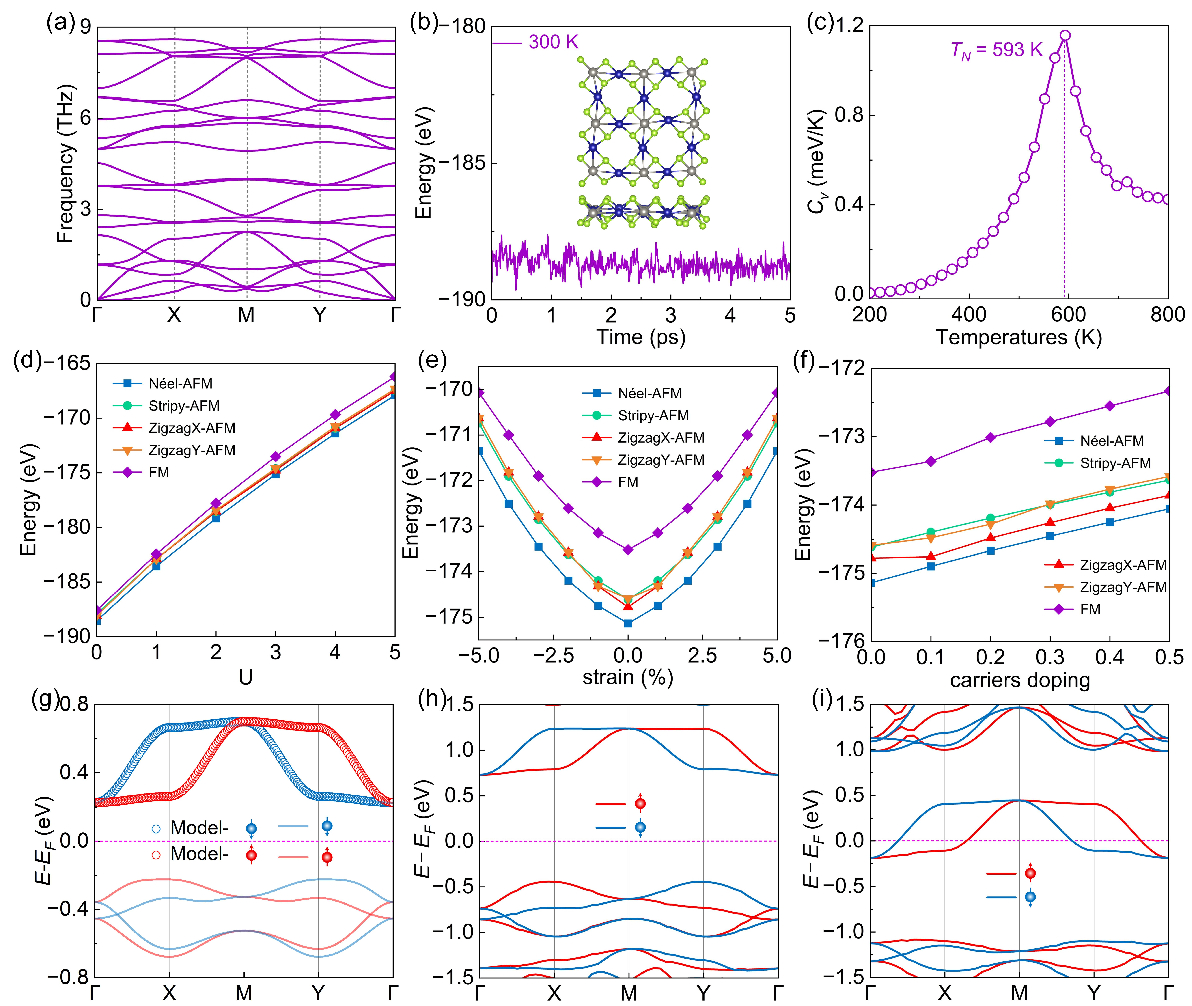}
		\caption{Assessment of the stability and electronic structure of the Cr$_2$WSe$_4$ monolayer. (a) Phonon spectrum. (b) Molecular dynamics simulation at 300 K. (c) Heat capacity ($C_{v}$) as a function of temperature from Monte Carlo simulations. Total energy comparisons for the FM, N\'eel-AFM, Stripy-AFM, ZigzagX-AFM, and ZigzagY-AFM states as a function of (d) the Hubbard $U$ parameter, (e) biaxial strain ($-$5\% to $+$5\%), and (f) carrier doping. {Electronic-structure validation: (g) finite-anisotropy model fit to the low-energy tent-state bands, (h) HSE06 band structure, and (i) electron-doped PBE$+U$ band structure.}}
		\label{fig:S12}
	\end{figure}
	
	{\qquad Including SOC provides a direct test of whether the SADL-relevant tent-state framework and open-line geometry are retained in the presence of SOC-induced spin mixing. As shown in Fig.~\ref{fig:S13}(a--c), the overall tent-state band morphology and the corresponding open Fermi-contour geometry remain clearly identifiable after SOC is included. The main SOC-induced changes are local: small hybridization gaps appear near band crossings, and spin is no longer a strictly conserved quantum number. These local gaps, however, are confined to limited portions of the open contours and do not disrupt the extended quasi-1D Fermi-line segments, which retain strong spin character and pronounced velocity anisotropy. This provides the basis for the largely preserved open-line transport contrast in the SOC-included electronic structure.} 	
	
	{\qquad The band-structure robustness discussed above is then quantified at the transport level using Kubo calculations based on the SOC-included Wannier Hamiltonian. The $\mathcal{T}$-odd component identifies the altermagnetic SADL-related response, whereas the $\mathcal{T}$-even component isolates the conventional SOC-driven spin Hall background. With SOC included, the dominant spin conductivities change only weakly, and the normalized conversion efficiency remains above $90\%$ in the relevant SADL energy window [Fig.~\ref{fig:S13}(d)]. At a representative energy of $E=0.47$~eV, the $\mathcal{T}$-odd SADL efficiency remains above $90\%$ for disorder broadening up to $\Gamma=83$~meV [Fig.~\ref{fig:S13}(e)]. In contrast, the $\mathcal{T}$-even SOC-driven spin Hall efficiency, defined as the ratio of the transverse SOC spin Hall conductivity to the primary longitudinal conductivity, stays below $1\%$ over the same range [Fig.~\ref{fig:S13}(f)]. This contrast supports that the high charge-to-spin conversion in Cr$_2$WSe$_4$ is dominated by the nonrelativistic altermagnetic SADL response rather than by a conventional SOC spin Hall background.}

	% Fig. S13
	\begin{figure}[H]
		\centering
		\includegraphics[width=0.95\textwidth]{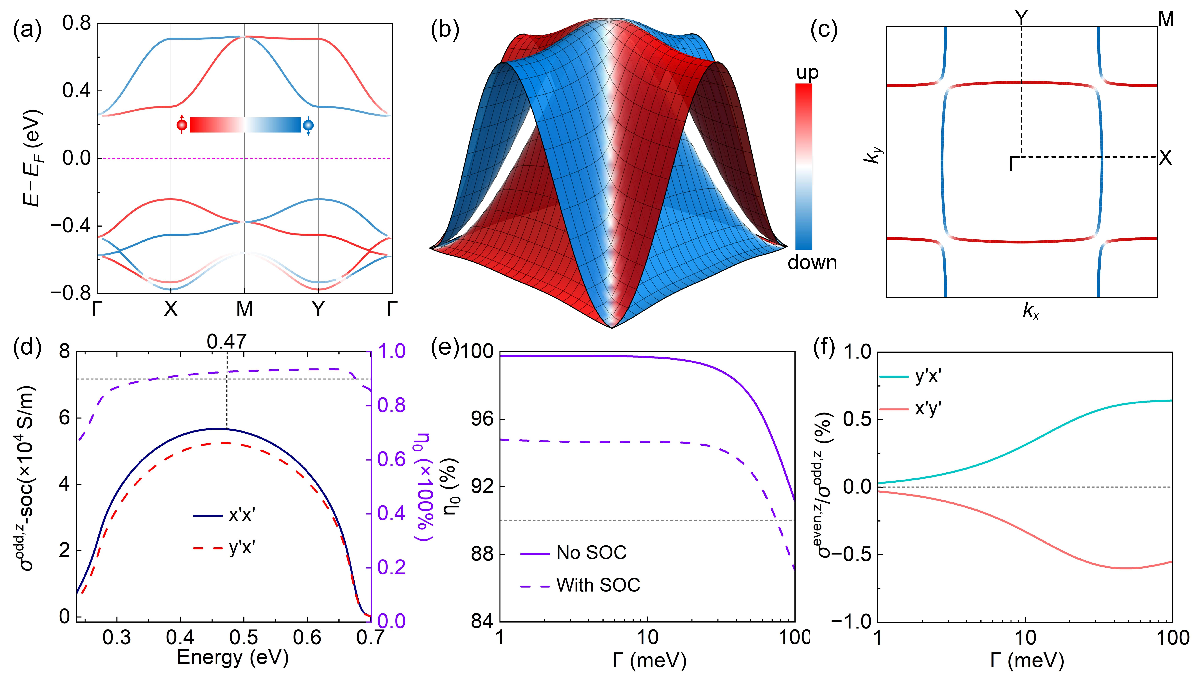}
		\caption{{
				SOC-included electronic structure and Kubo analysis for monolayer Cr$_2$WSe$_4$.
				(a) PBE$+U$+SOC band structure.
				(b,c) SOC-included tent-state band structure and constant-energy contours in the SADL regime.
				(d) Energy-dependent $\mathcal{T}$-odd longitudinal charge conductivity, transverse spin conductivity, and the resulting charge-to-spin conversion efficiency.
				(e) $\mathcal{T}$-odd SADL efficiency at the representative energy $E=0.47$~eV as a function of disorder broadening $\Gamma$, compared with the non-SOC result.
				(f) $\mathcal{T}$-even SOC-driven spin Hall efficiency as a function of $\Gamma$.}
		}
		\label{fig:S13}
	\end{figure}

	{\qquad Because the Kubo analysis separates the dominant $\mathcal{T}$-odd SADL response from the much smaller $\mathcal{T}$-even SOC background, it also points to experimentally distinguishable symmetry signatures. Upon reversal of the N\'eel vector, the $\mathcal{T}$-odd SADL-induced transverse spin current is expected to change sign, whereas the $\mathcal{T}$-even SOC-driven spin Hall background should remain unchanged. Comparing opposite magnetic domains, or measurements before and after electrical switching of the N\'eel vector, can therefore help isolate the SADL contribution. A complementary experimental check is provided by the angular dependence: the SADL transverse response follows $\theta_{\mathrm{SH}}=\eta_0|\sin2\theta|$ and vanishes for principal-axis driving, $\theta=0^\circ$ or $90^\circ$. A residual transverse spin signal in this geometry can then provide an estimate of the conventional SOC-related background.}
	
	{\qquad The CRTA and constant-$\Gamma$ Kubo calculations above establish the band-structure and SOC-robustness baselines, but they do not include state-dependent scattering anisotropy. To test this effect, we further performed momentum-dependent (MD) scattering calculations for Cr$_2$WSe$_4$ using the AMSET protocol described in Sec.~V. The calculations include ADP, POP, and IMP scattering, generating momentum- and band-dependent relaxation times. As a benchmark for the same protocol, Fig.~\ref{fig:S14}(a) shows that the calculated GaN mobility reproduces the experimental temperature-dependent trend~\textcolor{blue}{[\hyperlink{SM-ref-amset}{13}]} when all these scattering mechanisms are included.}
	
	% Fig. S14
	\begin{figure}[H]
		\centering
		\includegraphics[width=0.68\textwidth]{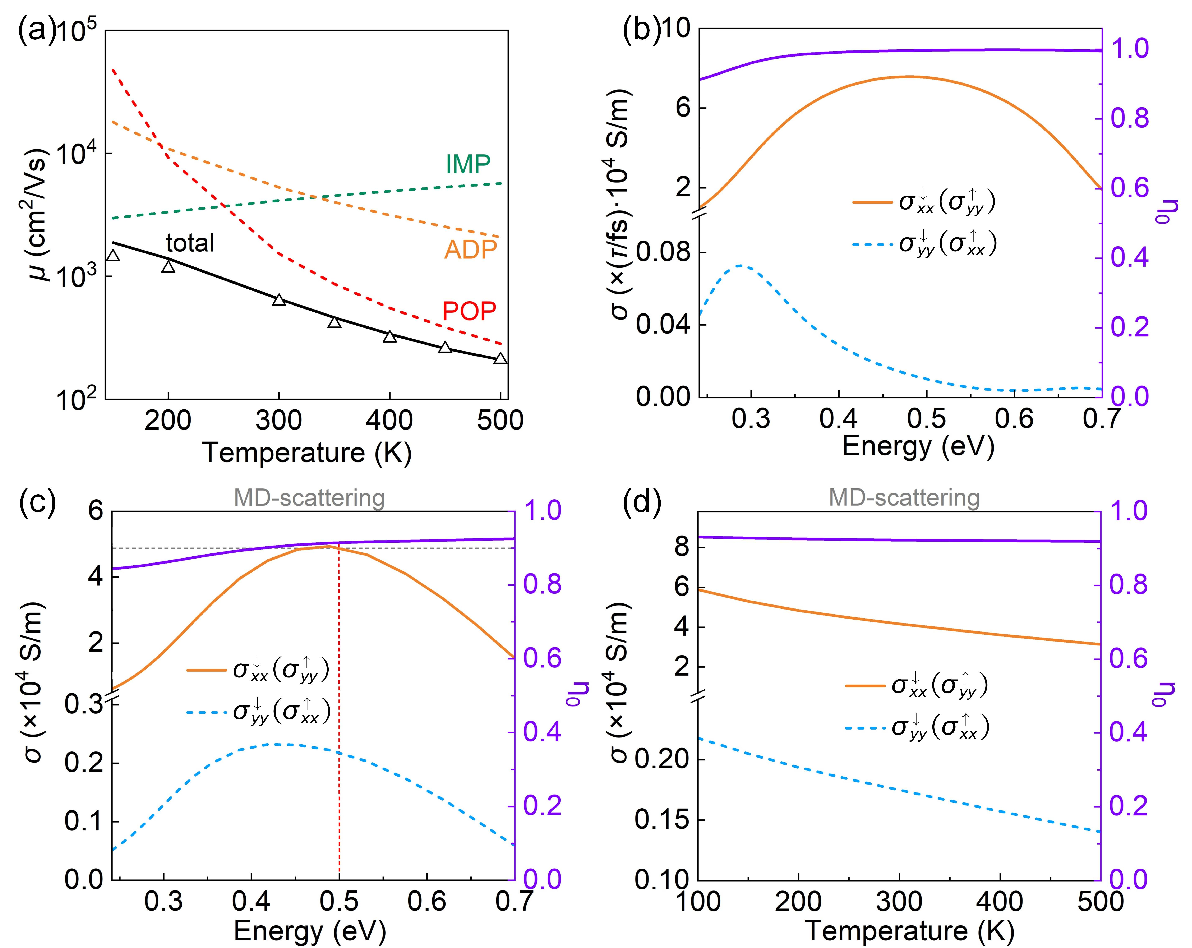}
		\caption{{
				Momentum-dependent scattering calculations.
				(a) Benchmark mobility calculation for the GaN semiconductor: open triangles denote experimental data, the black line denotes the calculated total mobility including IMP, ADP, and POP scattering, and dashed lines denote the calculated mobilities when only one scattering mechanism is included.
				(b) CRTA spin-resolved conductivities and axial spin polarization $\eta_0$ for Cr$_2$WSe$_4$.
				(c) Energy-dependent spin-resolved conductivities and $\eta_0$ for Cr$_2$WSe$_4$ with momentum- and band-dependent relaxation times, including IMP, ADP, and POP scattering.
				(d) Temperature-dependent spin-resolved conductivities and $\eta_0$ at a representative energy $E=0.5$~eV in the SADL regime, including the same scattering mechanisms.}
		}
		\label{fig:S14}
	\end{figure}

	{\qquad The MD scattering results provide a transport-level test of the dynamical picture underlying SADL. In the CRTA baseline, the large $\eta_0$ reflects the velocity contrast between the dominant and weak spin-transport channels [Fig.~\ref{fig:S14}(b)]. After replacing the single constant lifetime by momentum- and band-dependent relaxation times, the absolute conductivities are renormalized, but the contrast between $\sigma_{\mathrm{maj}}$ and $\sigma_{\mathrm{min}}$ remains large (the corresponding axial polarization $\eta_0 > 90\% $) over a broad SADL energy window and across the temperature range considered [Figs.~\ref{fig:S14}(c,d)]. If realistic scattering efficiently connected the separated opposite-velocity states on the open Fermi lines or strongly enhanced leakage into the weak channel, the ratio $\sigma_{\mathrm{min}}/\sigma_{\mathrm{maj}}$ would increase and $\eta_0$ would be substantially reduced. Instead, $\eta_0$ remains high, supporting the picture that the finite momentum separation of open Fermi lines suppresses efficient velocity-reversing scattering between separated branches. Therefore, MD scattering renormalizes the absolute transport scale but does not erase the spin-channel contrast that underlies SADL.}
	
	% Section VII		
	\newpage
	\begin{center}
		\large\bfseries {VII. Realization of SADL in Three-Dimensional Materials} 
	\end{center}
	
	\qquad From a pool of 3415 AFM materials in the Materials Project database~\textcolor{blue}{[\hyperlink{SM-ref-Horton2025}{14}, \hyperlink{SM-ref-Jain2013}{15}]}, we selected compounds with tetragonal $(a = b \neq c)$ and cubic $(a = b = c)$ crystal symmetry that host the structural motif depicted in Fig. 1a. 
	These compounds underwent a hierarchical computational screening process, summarized in Fig.~\ref{fig:S15}. First, we performed structural optimization and calculated their AFM band structures using the GGA-PBE$+U$ method. Subsequently, candidates demonstrating tent states were selected for validation through more accurate HSE06 calculations. Finally, systematic calculations of the transport properties were conducted for the resulting candidates.
	\begin{figure}[H]
		\centering
		\includegraphics[width=0.9\textwidth]{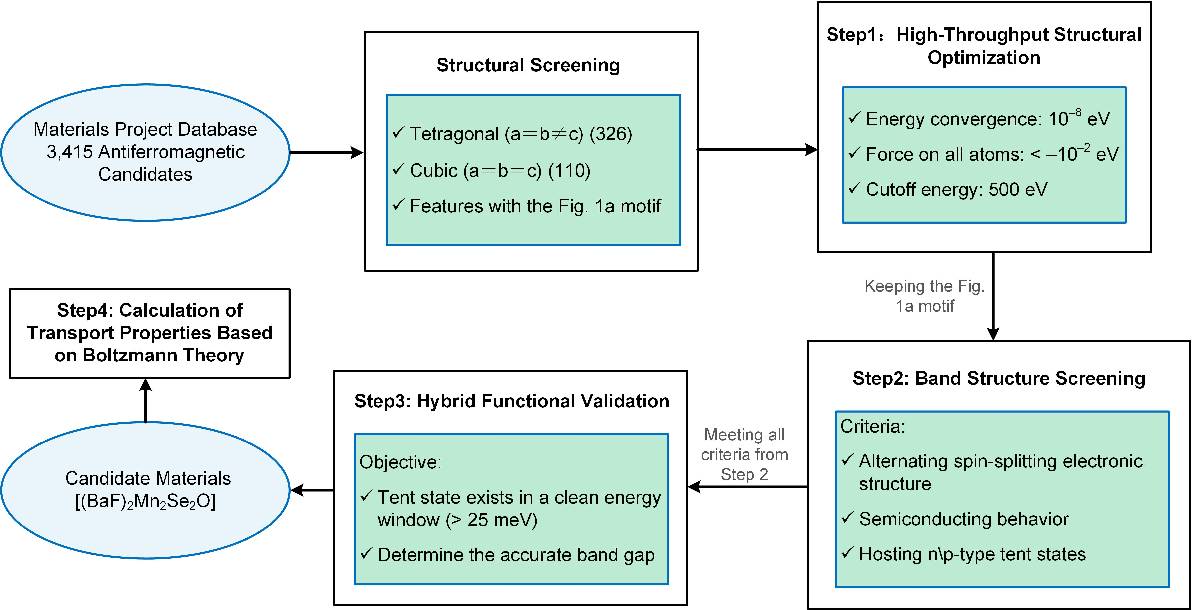}
		\caption{Computational screening workflow for identifying 3D tent-state materials.}
		\label{fig:S15}
	\end{figure}	
	
	\newpage
	{\qquad Among the six 3D altermagnetic candidates with quasi-2D transport characteristics (Fig.~\ref{fig:S16}), the synthesized layered compound (BaF)$_2$Mn$_2$Se$_2$O most closely satisfies the full set of SADL criteria. Its weak out-of-plane dispersion allows the SADL-relevant band-edge states to retain predominantly 2D character. We therefore focus on this experimentally synthesized candidate to examine its electronic structure, transport robustness, and possible experimental routes for detecting SADL signatures.}
	\begin{figure}[H]
		\centering
		\includegraphics[width=0.9\textwidth]{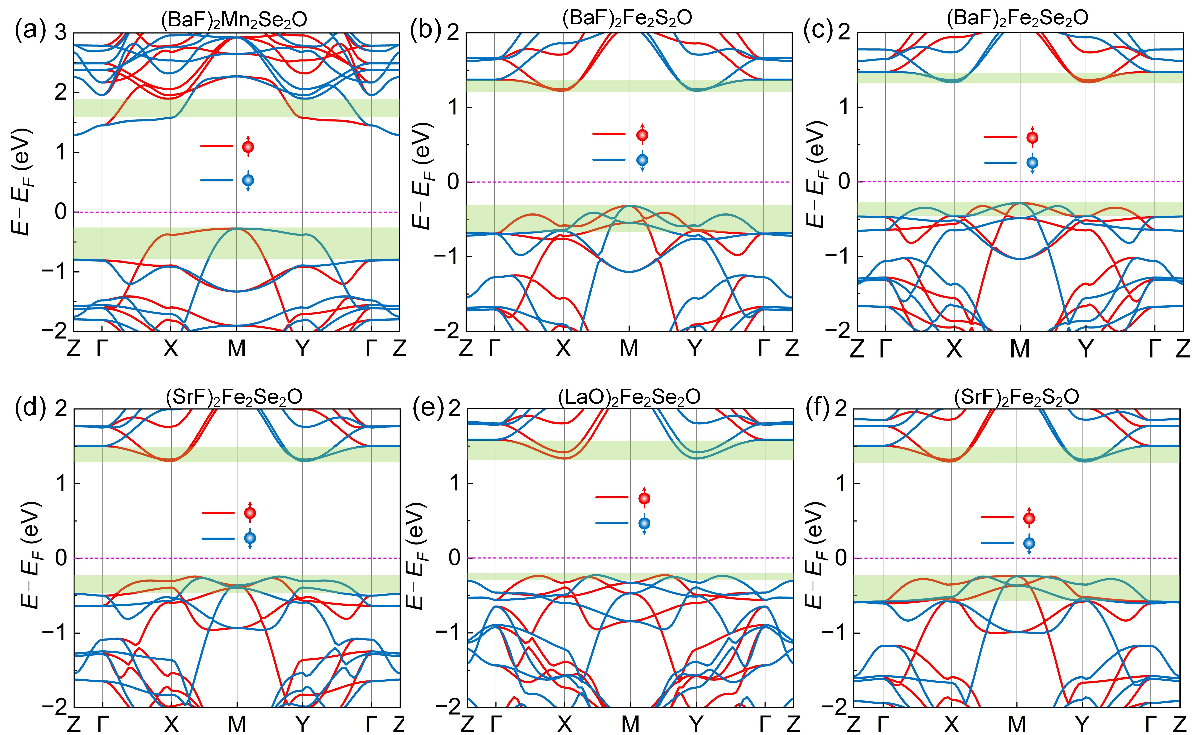}
		\caption{Electronic band structures for the six 3D altermagnets with quasi-2D dispersion in the highlighted energy windows.}
		\label{fig:S16}
	\end{figure}
	
	{\qquad As shown in Figs.~\ref{fig:S17}(a,b), the valence- and conduction-band edges of (BaF)$_2$Mn$_2$Se$_2$O exhibit clear 2D tent-state features. Although additional bands and hybridization effects distort the deeper-energy regions, these tent-state branches remain sufficiently isolated over sizable clean energy windows where the SADL physics can be identified. To quantify the low-energy dispersion within these SADL-relevant windows, we fitted the spin-resolved branches using the anisotropic 2D $d$-wave model of Sec.~I [Fig.~\ref{fig:S17}(c)]. The fitting gives $\gamma=0.108$ for the conduction-band window and $\gamma=0.042$ for the valence-band window, placing both within the strong-SADL region (Fig.~1j). The same tent-state features remain identifiable when SOC is included [Fig.~\ref{fig:S17}(d)], at the HSE06 level [Fig.~\ref{fig:S17}(e)], and under carrier doping [Fig.~\ref{fig:S17}(f, g)].}
	
	\begin{figure}[H]
		\centering
		\includegraphics[width=0.9\textwidth]{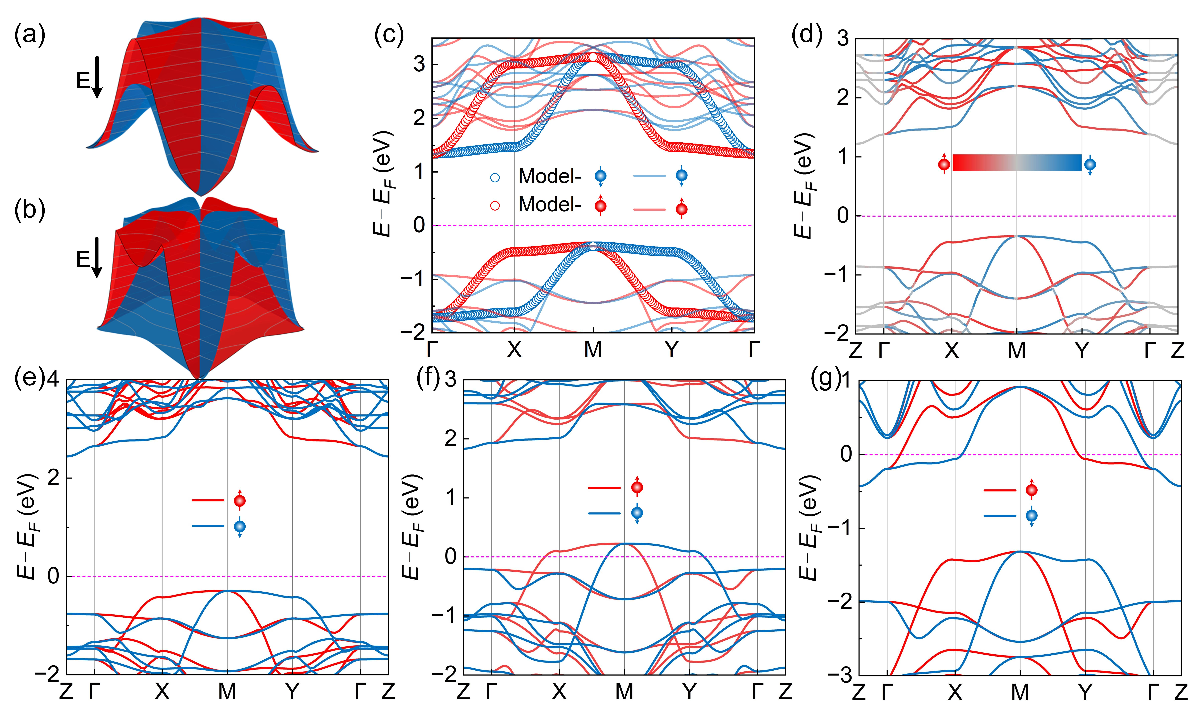}
		\caption{
			Electronic band structure of (BaF)$_2$Mn$_2$Se$_2$O.
			Schematic illustrations of the tent states for (a) $n$-type and (b) $p$-type doping at the conduction and valence band edges, respectively.
			{(c) Minimal-model fit to the spin-resolved low-energy bands in the SADL windows, where open circles and solid lines denote the fitted model bands and first-principles bands, respectively.}
			Band structures calculated {(d)} at the GGA-PBE{$+U$} level including spin-orbit coupling (SOC) and {(e)} with the HSE06 method.
			Band structures for {(f)} hole-doped and {(g)} electron-doped systems.
			Black arrows on the 3D band diagrams indicate the direction of increasing energy.}
		\label{fig:S17}
	\end{figure}
	
	{\qquad Considering that the layered compound (BaF)$_2$Mn$_2$Se$_2$O has already been synthesized~\textcolor{blue}{[\hyperlink{SM-ref-Liu2011}{16}]}, we propose it as an experimentally accessible platform for testing the SADL signatures predicted above. Two complementary routes are particularly relevant: momentum-space spectroscopy of the spin-resolved open Fermi lines, and spin-transport measurements of the angular, carrier-density, and temperature dependence of the charge-to-spin conversion response.}
	
	{\qquad First, high-resolution angle-resolved photoemission spectroscopy (ARPES) and spin-resolved ARPES (SARPES) can probe the static SADL fingerprints in momentum space. The layered structure of (BaF)$_2$Mn$_2$Se$_2$O is favorable for clean cleavage, and its calculated electronic structure shows pronounced quasi-2D character with weak $k_z$ dispersion. Photon-energy-dependent ARPES can therefore be used to select the relevant out-of-plane momentum plane, such as the $k_z=0$ plane, and then map the in-plane constant-energy contours. Conventional ARPES can test whether quasi-1D open Fermi lines appear in the SADL energy window, while SARPES can determine their spin texture and examine whether the two sets of open lines carry spin polarizations locked to orthogonal crystalline directions. The feasibility of this route is supported by recent ARPES/SARPES studies of altermagnetic systems, including spin-polarized dispersions in semiconducting MnTe~\textcolor{blue}{[\hyperlink{SM-ref-Krempasky2024Nature}{17}]} and momentum-dependent $d$-wave spin textures in layered KV$_2$Se$_2$O~\textcolor{blue}{[\hyperlink{SM-ref-Jiang2025NatPhys}{18}]}.}

	{\qquad Second, the dynamical spin-current response can be tested by transport measurements after tuning the Fermi level into the SADL regime, for example by chemical doping, ionic-liquid or electrochemical gating, or surface dosing. As derived from the macroscopic conductivity tensor in Sec.~IV, the SADL-induced transverse charge-to-spin conversion efficiency should follow $\theta_{\mathrm{SH}}=\eta_0|\sin2\theta|$, reaching a maximum for diagonal in-plane driving and vanishing for principal-axis driving. This angular dependence would provide an indirect macroscopic signature of SADL and its underlying $d$-wave spin-group symmetry. In addition, the momentum-dependent scattering results in Fig.~\ref{fig:S18} predict that the conversion efficiency remains high over broad energy and temperature ranges, providing an indirect route to test the dynamical robustness of SADL. Experimentally, this can be examined by measuring $\eta_0$ or $\theta_{\mathrm{SH}}$ as a function of carrier density and temperature (in particular, the sizable valence-band SADL window of (BaF)$_2$Mn$_2$Se$_2$O makes hole-side tuning especially suitable for testing carrier-density robustness). The spin-current response may be detected using established schemes such as spin-torque ferromagnetic resonance, harmonic Hall measurements, or nonlocal spin-current detection in suitable heterostructures, with related spin-torque and spin-current measurements already demonstrated in the altermagnet RuO$_2$~\textcolor{blue}{[\hyperlink{SM-ref-Bai2022PRL}{19}, \hyperlink{SM-ref-Karube2022PRL}{20}]}.}

	\begin{figure}[H]
		\centering
		\includegraphics[width=0.78\textwidth]{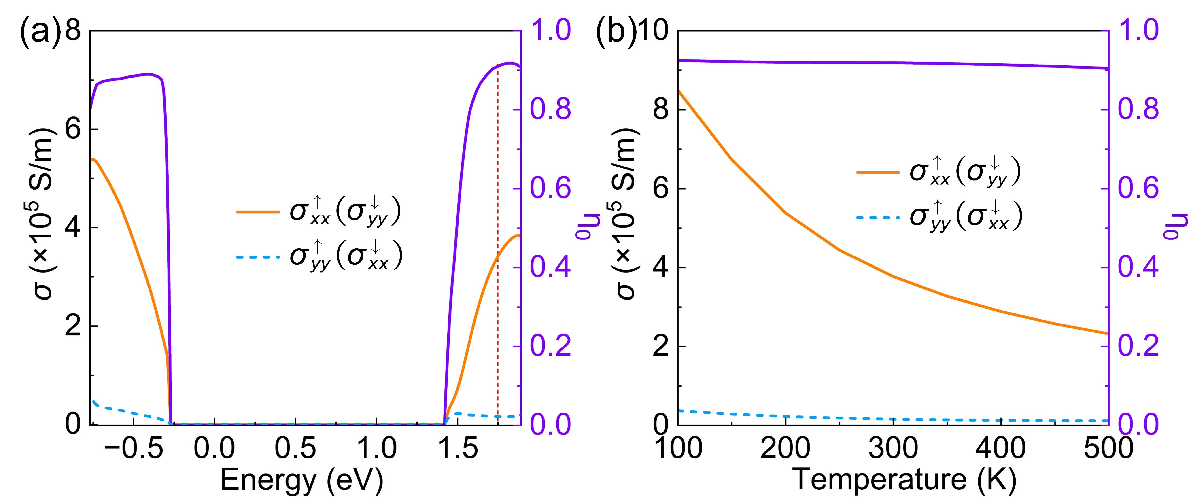}
		\caption{{
				AMSET-based momentum-dependent scattering calculations for (BaF)$_2$Mn$_2$Se$_2$O, including IMP, ADP, and POP scattering.
				(a) Energy-dependent spin-resolved conductivities and axial spin polarization $\eta_0$.
				(b) Temperature-dependent spin-resolved conductivities and $\eta_0$ at a representative SADL-regime energy of 1.75 eV.
		}}
		\label{fig:S18}
	\end{figure}
	
	\clearpage
	\begin{center}\large\bfseries References\end{center}
\begin{enumerate}[label={[\arabic*]},leftmargin=*,labelsep=0.5em,itemsep=3pt,parsep=0pt,topsep=6pt] 
		\item \hypertarget{SM-ref-Pizzi2014}{} G. Pizzi, D. Volja, B. Kozinsky, M. Fornari, and N. Marzari, BoltzWann: A Code for the Evaluation of Thermoelectric and Electronic Transport Properties with a Maximally Localized Wannier Functions Basis, \href{https://doi.org/10.1016/j.cpc.2013.09.015}{\journaltext{Comput. Phys. Commun. \boldfirstnumber{185, 422 (2014)}.}} 
		\item \hypertarget{SM-ref-Ziman1972}{} J. M. Ziman, \textit{Principles of the Theory of Solids} \journaltext{(Cambridge University Press, Cambridge, England, 1972).} 
		
		\item \hypertarget{SM-ref-vasp}{} G. Kresse and J. Furthm\"uller, Efficient Iterative Schemes for Ab Initio Total-Energy Calculations Using a Plane-Wave Basis Set, \href{https://doi.org/10.1103/PhysRevB.54.11169}{\journaltext{Phys. Rev. B \boldfirstnumber{54, 11169 (1996)}.}}
		
		\item \hypertarget{SM-ref-paw}{} G. Kresse and D. Joubert, From Ultrasoft Pseudopotentials to the Projector Augmented-Wave Method, \href{https://doi.org/10.1103/PhysRevB.59.1758}{\journaltext{Phys. Rev. B \boldfirstnumber{59, 1758 (1999)}.}} 
		
		\item \hypertarget{SM-ref-pbe}{} J. P. Perdew, K. Burke, and M. Ernzerhof, Generalized Gradient Approximation Made Simple, \href{https://doi.org/10.1103/PhysRevLett.77.3865}{\journaltext{Phys. Rev. Lett. \boldfirstnumber{77, 3865 (1996)}.}} 
		
		\item \hypertarget{SM-ref-monkhorst}{} H. J. Monkhorst and J. D. Pack, Special Points for Brillouin-Zone Integrations, \href{https://doi.org/10.1103/PhysRevB.13.5188}{\journaltext{Phys. Rev. B \boldfirstnumber{13, 5188 (1976)}.}} 
		
		\item \hypertarget{SM-ref-hse06}{} J. Heyd, G. E. Scuseria, and M. Ernzerhof, Hybrid Functionals Based on a Screened Coulomb Potential, \href{https://doi.org/10.1063/1.1564060}{\journaltext{J. Chem. Phys. \boldfirstnumber{118, 8207 (2003)}.}} 
		
		\item \hypertarget{SM-ref-wannier90}{} A. A. Mostofi, J. R. Yates, Y.-S. Lee, I. Souza, D. Vanderbilt, and N. Marzari, Wannier90: A Tool for Obtaining Maximally-Localised Wannier Functions, \href{https://doi.org/10.1016/j.cpc.2007.11.016}{\journaltext{Comput. Phys. Commun. \boldfirstnumber{178, 685 (2008)}.}} 
		
		\item \hypertarget{SM-ref-wannier90update}{} A. A. Mostofi, J. R. Yates, G. Pizzi, Y.-S. Lee, I. Souza, D. Vanderbilt, and N. Marzari, An Updated Version of Wannier90: A Tool for Obtaining Maximally-Localised Wannier Functions, \href{https://doi.org/10.1016/j.cpc.2014.05.003}{\journaltext{Comput. Phys. Commun. \boldfirstnumber{185, 2309 (2014)}.}} 
		
		\item \hypertarget{SM-ref-Smejkal2021}{} R. Gonz\'alez-Hern\'andez, L. \v{S}mejkal, K. V\'yborn\'y, Y. Yahagi, J. Sinova, T. Jungwirth, and J. \v{Z}elezn\'y, Efficient Electrical Spin Splitter Based on Nonrelativistic Collinear Antiferromagnetism, \href{https://doi.org/10.1103/PhysRevLett.126.127701}{\journaltext{Phys. Rev. Lett. \boldfirstnumber{126, 127701 (2021)}.}} 
		
		\item \hypertarget{SM-ref-Freimuth2014PRB}{} F. Freimuth, S. Bl\"ugel, and Y. Mokrousov, Spin-Orbit Torques in Co/Pt(111) and Mn/W(001) Magnetic Bilayers from First Principles, \href{https://doi.org/10.1103/PhysRevB.90.174423}{\journaltext{Phys. Rev. B \boldfirstnumber{90, 174423 (2014)}.}} \item \hypertarget{SM-ref-ZeleznyWannierLR}{} J. \v{Z}elezn\'y, Wannier Linear Response, \href{https://bitbucket.org/zeleznyj/wannier-linear-response/wiki/Home}{\journaltext{https://bitbucket.org/zeleznyj/linear-response-symmetry.}} 
		
		\item \hypertarget{SM-ref-amset}{} A. M. Ganose, J. Park, A. Faghaninia, R. Woods-Robinson, K. A. Persson, and A. Jain, Efficient Calculation of Carrier Scattering Rates from First Principles, \href{https://doi.org/10.1038/s41467-021-22440-5}{\journaltext{Nat. Commun. \boldfirstnumber{12, 2222 (2021)}.}}
		
		\item \hypertarget{SM-ref-Horton2025}{} M. K. Horton, P. Huck, R. X. Yang, J. M. Munro, S. Dwaraknath, A. M. Ganose, R. S. Kingsbury, M. Wen, J. X. Shen, T. S. Mathis, A. D. Kaplan, K. Berket, J. Riebesell, J. George, A. S. Rosen, E. W. C. Spotte-Smith, M. J. McDermott, O. A. Cohen, A. Dunn, M. C. Kuner, G.-M. Rignanese, G. Petretto, D. Waroquiers, S. M. Griffin, J. B. Neaton, D. C. Chrzan, M. Asta, G. Hautier, S. Cholia, G. Ceder, S. P. Ong, A. Jain, and K. A. Persson, Accelerated Data-Driven Materials Science with the Materials Project, \href{https://doi.org/10.1038/s41563-025-02272-0}{\journaltext{Nat. Mater. \boldfirstnumber{24, 1522 (2025)}.}}
		
		\item \hypertarget{SM-ref-Jain2013}{} A. Jain, S. P. Ong, G. Hautier, W. Chen, W. D. Richards, S. Dacek, S. Cholia, D. Gunter, D. Skinner, G. Ceder, and K. A. Persson, Commentary: The Materials Project: A Materials Genome Approach to Accelerating Materials Innovation, \href{https://doi.org/10.1063/1.4812323}{\journaltext{APL Mater. \boldfirstnumber{1, 011002 (2013)}.}}		
		
		\item \hypertarget{SM-ref-Liu2011}{} R. H. Liu, J. S. Zhang, P. Cheng, X. G. Luo, J. J. Ying, Y. J. Yan, M. Zhang, A. F. Wang, Z. J. Xiang, G. J. Ye, and X. H. Chen, {S}tructural and {M}agnetic {P}roperties of the {L}ayered {M}anganese {O}xychalcogenides ({L}a{O})$_2${M}n$_2${S}e$_2${O} and ({B}a{F})$_2${M}n$_2${S}e$_2${O}, \href{https://doi.org/10.1103/PhysRevB.83.174450}{\journaltext{Phys. Rev. B \boldfirstnumber{83, 174450 (2011)}.}}
		
		\item \hypertarget{SM-ref-Krempasky2024Nature}{} J. Krempask\'y, L. \v{S}mejkal, S. W. D'Souza, M. Hajlaoui, G. Springholz, K. Uhl\'i\v{r}ov\'a, F. Alarab, P. C. Constantinou, V. Strocov, D. Usanov, W. R. Pudelko, R. Gonz\'alez-Hern\'andez, A. Birk Hellenes, Z. Jansa, H. Reichlov\'a, Z. \v{S}ob\'a\v{n}, R. D. Gonzalez Betancourt, P. Wadley, J. Sinova, D. Kriegner, J. Min\'ar, J. H. Dil, and T. Jungwirth, Altermagnetic Lifting of Kramers Spin Degeneracy, \href{https://doi.org/10.1038/s41586-023-06907-7}{\journaltext{Nature \boldfirstnumber{626, 517 (2024)}.}}
		
		\item \hypertarget{SM-ref-Jiang2025NatPhys}{} B. Jiang, M. Hu, J. Bai, Z. Song, C. Mu, G. Qu, W. Li, W. Zhu, H. Pi, Z. Wei, Y.-J. Sun, Y. Huang, X. Zheng, Y. Peng, L. He, S. Li, J. Luo, Z. Li, G. Chen, H. Li, H. Weng, and T. Qian, A Metallic Room-Temperature $d$-Wave Altermagnet, \href{https://doi.org/10.1038/s41567-025-02822-y}{\journaltext{Nat. Phys. \boldfirstnumber{21, 754 (2025)}.}}
		
		\item \hypertarget{SM-ref-Bai2022PRL}{} H. Bai, L. Han, X. Y. Feng, Y. J. Zhou, R. X. Su, Q. Wang, L. Y. Liao, W. X. Zhu, X. Z. Chen, F. Pan, X. L. Fan, and C. Song, Observation of Spin Splitting Torque in a Collinear Antiferromagnet $\mathrm{RuO}_2$, \href{https://doi.org/10.1103/PhysRevLett.128.197202}{\journaltext{Phys. Rev. Lett. \boldfirstnumber{128, 197202 (2022)}.}}
		
		\item \hypertarget{SM-ref-Karube2022PRL}{} S. Karube, T. Tanaka, D. Sugawara, N. Kadoguchi, M. Kohda, and J. Nitta, Observation of Spin-Splitter Torque in Collinear Antiferromagnetic $\mathrm{RuO}_2$, \href{https://doi.org/10.1103/PhysRevLett.129.137201}{\journaltext{Phys. Rev. Lett. \boldfirstnumber{129, 137201 (2022)}.}}		
	\end{enumerate}

\endgroup
\end{document}